# Exosphere-Mediated Migration of Volatile Species On Airless Bodies Across the Solar System

Jordan K. Steckloff*[1,2], David Goldstein[1], Laurence Trafton[4,1], Philip Varghese[1], Parvathy Prem[3]

*corresponding author: steckloff@utexas.edu

[1]University of Texas at Austin, Department of Aerospace Engineering and Engineering Mechanics, Aerospace Engineering Building, C0600, 2617 Wichita Street, Austin, TX 78712-1221 USA
[2]Planetary Science Institute, 1700 East Fort Lowell, Suite 106, Tucson, AZ 85719-2395, USA
[3]Johns Hopkins University Applied Physics Laboratory, 11100 Johns Hopkins Road, Laurel, MD, 20723, USA
[4]University of Texas at Austin, McDonald Observatory, 3640 Dark Sky Drive, Fort Davis, TX, 79734, USA

**Abstract**

Surface-bound exospheres facilitate volatile migration across the surfaces of nearly airless bodies. However, such transport requires that the body can both form and retain an exosphere. To form a sublimation exosphere requires the surface of a body to be sufficiently warm for surface volatiles to sublime; to retain an exosphere, the ballistic escape and photodestruction rates and other loss mechanisms must be sufficiently low. Here we construct a simple free molecular model of exospheres formed by volatile desorption/sublimation. We consider the conditions for forming and retaining exospheres for common volatile species across the Solar System, and explore how three processes (desorption/sublimation, ballistic loss, and photodestruction) shape exospheric dynamics on airless bodies. Our model finds that the $CO_2$ exosphere of Callisto is too dense to be sustained by impact-delivered volatiles, but could be maintained by only ~ 7 hectares of exposed $CO_2$ ice. We predict the peak surface locations of Callisto's $CO_2$ exosphere along with other Galilean moons, which could be tested by *JUICE* observations. Our model finds that to maintain Iapetus' two-tone appearance, its dark Cassini Regio likely has unresolved exposures of water ice, perhaps in sub-resolution impact craters, that amount to up to ~0.06% of its surface. In the Uranian system, we find that the $CO_2$ deposits on Ariel, Umbriel, Titania, and Oberon are unlikely to have been delivered via impacts, but are consistent with both a magnetospheric origin or sourced endogenously. We suggest that exosphere-mediated volatile transport could produce these moons' leading/trailing $CO_2$ asymmetries, and may be a seasonal equinox feature that could be largely erased by volatile migration during the Uranian solstices. We calculate that ~2.4 – 6.4 mm thick layer of $CO_2$ could migrate about the surface of Uranus' large moons during a seasonal cycle.

## 1. Introduction

The majority of objects in the Solar System lack thick atmospheres. Aside from Mercury, the other three terrestrial planets, along with Titan, Triton, and Pluto are known to possess thick, collisional atmospheres (surface pressures of ~1 – $10^7$ Pa) that facilitate surface modifications over time. Io is a notable intermediate case in which its dayside atmosphere is collisional while its nightside and polar atmospheres may be in the transitional or free molecular flow regime (*Walker et al., 2012*). Nevertheless, many nominally "airless" bodies of the Solar System possess, perhaps temporarily, highly rarefied surface-bound exospheres that mediate transportation of volatile species about their surfaces and systems. Indeed, exospheric dynamics can deliver water to polar cold traps on Mercury (*Butler, 1997; Paige et al., 2013; Ernst et al., 2018*), the Moon (*Arnold, 1979; Ong et al., 2010; Stewart et al., 2011; Prem et al., 2015; 2018*), and Ceres (*Tu et al., 2014; Schorghofer et al., 2017; Landis et al., 2019*). Exospheric dynamics are argued to transfer nitrogen and methane from above the exobase of Pluto to Charon (*Trafton et al., 1987; Whipple et al., 1989; Tucker et al., 2014; Hoey et al., 2017a; 2017b*), and alter the dynamics and evolution of plumes on Io (*McDoniel et al., 2017*). To understand the extent to which exospheric/atmospheric dynamics evolve the surfaces of airless bodies in the solar system, one must understand the conditions required for an object to host an exosphere.

We consider here the case of quasi-steady, surface bound sublimation/desorption exospheres to explore how three processes (desorption/sublimation, ballistic loss, and photodestruction) affect exosphere shape and structure, and impact the exosphere's dynamical behavior and lifetime. We exclude transient situations from the scope of this work (e.g., impact-generated exospheres that are not retained or volatiles emitted via vents or plumes). Although the exospheric loss mechanisms described here would apply to these situations, the creation of exospheres through these mechanisms are interesting special cases best examined on a case by case basis. We also restrict our analysis to bodies large enough to be spherical (or approximately spherical) to avoid the complicated surface temperature distributions of more complex shapes. They may orbit the Sun or be satellites of a planet, and have surfaces that are icy, rocky (refractory), or both. We further exclude from consideration plasma or sputtering processes – these are indeed important processes that are responsible for the exospheres above icy satellites in the Jovian and Saturnian systems. Nevertheless, these thermal processes can coexist with sputtering processes , which have been well discussed elsewhere (*e.g., Johnson et al., 1981; Shi et al., 1995; Vorburger et al., 2015*).

To experience quasi-steady, sublimation-driven exospheric dynamics, an object must be warm enough to form an exosphere from surface volatiles, and have high enough surface gravity to retain it. Furthermore, exospheric molecules must survive long enough against photodestruction to build up an exosphere. We neglect loss to permanent cold traps: that process is generally weak and is, in fact, an outcome of our discussion. Each of these three processes (sublimation, thermal escape, and photodissociation) have characteristic timescales, length-scales, and/or flux rates that define their contribution to exosphere formation/destruction, exospheric structure, and exosphere-mediated volatile transport. Here we explore the interplay between these three processes across the Solar System for four of the most common volatiles currently present on the surfaces of airless bodies: $H_2O$, $CO_2$, CO, and $N_2$. We treat each species independently, even though their evolution may be coupled (e.g., through their effect on albedo or thermodynamic properties of intimately mixed frozen volatiles). We focus on a comparison of these three characteristic timescales/fluxes, to identify conditions under which these processes dominate to form/prevent exosphere formation. We then compare the behavior of these volatiles with observations of airless bodies in the inner Solar System, Asteroid Belt, and on satellites of the giant planets.

## 2. Methods

We focus on the behavior of two endmember cases that describe the body surface composition and exospheric shape. Bodies can have a surface that is primarily refractory with a small amount of a volatile species aloft or adsorbed, forming a finite reservoir of volatiles (e.g., the Moon and Mercury with water or argon). Alternatively, the body's surface may be primarily composed of frozen volatiles, forming an effectively infinite volatile reservoir (e.g., the water-ice surfaces of Callisto, Ganymede, Rhea, and Dione). Although the microphysical interactions that cause surfaces and exospheres to exchange molecules are physically similar in principle, differences in the macroscopic structure of these two end-member situations require different modeling approaches that account for their finite/non-finite volatile reservoirs. We should note that a single body may experience both of these cases simultaneously for different volatile species (e.g., Callisto for $CO_2$ and $H_2O$).

Exospheric gases on airless bodies such as the Moon are continually lost through thermal escape, charge exchange, photodestruction, photoionization, solar wind scavenging and sputtering, and migration to cold traps, along with other possible mechanisms. Thus, these exospheres require sources to replenish themselves, such as solar wind (for $H_2O$; *Thomas, 1974; Stern, 1999; Jones et al. 2020*), volatile rich impactors (e.g., comet nuclei or CI chondrites), or from endogenous releases of volatiles (e.g., hydrothermal or volcanic activity; *Vu et al., 2017; Needham and Kring, 2017; Palomba et al., 2019; Raponi et al., 2019; Schenk et al., 2019; Thomas et al., 2019*). For the ice-rich surface case, the surface itself also contributes exospheric molecules, and can dominate the production through sublimation or sputtering. All of these cases produce a dynamic exosphere, the evolution and decay of which we analyze in this work. Of the various processes that act to source/destroy exospheres, we focus on a subset of the steady (rather than episodic) processes that affect all airless bodies across the Solar System: thermally-driven desorption/sublimation and photodestruction. As indicated above, we exclude from our consideration exospheric components that are produced from sputtering processes, such as $O_2$ on outer icy moons (*e.g., Teolis et al., 2010*). Instead, we will focus exclusively on the sublimation/desorption produced exospheric components of the common volatiles $H_2O$, $CO_2$, CO, and $N_2$. We note that many of these species can also be produced by sputtering in the outer Solar System (*e.g., Vorburger et al., 2015*); here we consider only the thermally produced component of these species' exospheres, neglecting their non-thermally produced components. It should be noted that, due to the nearly collisionless nature of exospheres, an object can simultaneously host multiple quasi-independent exospheres of different species and/or source processes.

| Assumption/ Consideration | Applies to: | Does not apply to: |
|---|---|---|
| Surface-bound exosphere | "Airless" objects | Objects with collisional atmospheres (e.g., Pluto, Triton, Mars, Venus) |
| Non-sputtering exospheric source | Distant objects from the Sun, satellites outside strong magnetic fields | Near Sun objects or satellites orbiting in strong magnetic fields |
| Negligible nighttime desorption | Objects with strong day/night temperature variations, negligible internal heat flux | Objects with high heat flux and/or little day/night variation (e.g., Sedna at periapse) |

**Table 1:** *A list of physical assumptions/considerations concerning this model***.** Objects that meet these criteria should experience thermally driven exospheric dynamics that are well described by this model. Objects that fail to meet some (or all) of these criteria may experience deviations from model predictions, which can become quite significant. Nevertheless, the analysis here could describe the sublimative components of these respective atmospheres.

For the refractory surface case (finite volatile reservoir), we focus on the microphysical behavior of a single, typical molecule experiencing each of these three processes. One of the timescales of this endmember refractory case is how long a typical molecule would remain condensed on the object's illuminated surface before desorbing (the "residence time", $\tau_{res}$). Because these volatile molecules may condense on the surface at night and sublimate at dawn, they begin to desorb to the exosphere once their surface location rotates into daylight past the dawn terminator. However, if the supply of volatiles is limited (finite), the surface will quickly deplete itself of volatiles, resulting in an exosphere that stretches across the illuminated hemisphere, but with a peak density that is centered on the longitude at which adsorbed/condensed molecules preferentially desorb/sublimate (typically on/near the dawn terminator; *e.g., Hodges, 1975; Stern, 1999; Schorghofer, 2014*). The longitudinal location of this morning exosphere can be computed by comparing the residence time of the molecule with the time elapsed since dawn. Since lines of latitude experience different temperature profiles throughout the day, the shape of this morning exosphere does not necessarily lie along a particular line of longitude. To first order, the longitudinal width of this morning exosphere is comparable to the average ballistic hop distance of these desorbed/sublimed molecules.

Non-condensing/adsorbing gases behave differently and are instead present across the surface of the body. Exospheric (volumetric) densities of these species (e.g., lunar helium) peak with cold nighttime temperatures (*Hodges, 1975, Benna et al., 2015*), due to gases becoming denser at colder temperatures and preferentially collecting near areas of colder temperatures[1]. Our analysis neglects such non-condensing species, focusing instead on volatile-driven

[1] Even in the free molecular limit, colder temperatures equate with slower gas molecules. This leads to molecules taking longer to leave areas of colder temperatures, resulting in an accumulation of molecules and higher molecular density

exospheres, and their interactions with the surface. Further, we do not separately include the effects of thermal creep (thermal diffusion; *e.g., Sone, 2007*), which requires a strong temperature gradient along the surface to move molecules and is therefore only potentially applicable near the terminator.

For the other endmember, the icy surface case (infinite reservoir), we consider the sublimation flux of molecules from the surface, to determine exospheric density at the surface as a function of subsolar latitude, and the expected lifetime of an ejected molecule in the exosphere (prior to ballistic escape or photodestruction). Such bodies effectively have a semi-infinite reservoir of surface volatiles, thus the sublimation flux of molecules out of/into the surface is sensitive to the surface temperature. The surface temperature itself is determined by balancing solar heating with radiative and sublimative cooling. This neglects internal heat flux, horizontal heat conduction within the surface, and other radiative heating terms. Such effects are generally very minor perturbations on the surface temperature of an airless body; this radiative equilibrium approach generally provides a good estimate of surface temperature.

In practice, this leads to two cases for each volatile: a cooler temperature case in which the surface primarily cools through thermal radiation, and a warmer case in which the surface loses heat primarily through sublimation cooling. The transition between these two regimes can be thought of as defining the sublimation temperature (*Steckloff et al., 2015*). The warmer, sublimation cooled case is beyond the scope of this work as such cases are highly collisional atmospheres, rather than rarefied exospheres (Figure 1).

We justify this limitation in scope using the model of *Steckloff et al., (2015*) to compute the mean free path near the nucleus of cometary atmospheres for three common volatiles as a function of heliocentric distance. This model assumes a very low albedo object composed purely of the volatile ice under consideration. We compare this mean free path to two characteristic length scales: the size of the nucleus and an approximation for the atmospheric scale height ($RT/(m_{mol}g_{surf})$, where $T$ is the surface temperature at the subsolar point, $R$ is the ideal gas constant, $m_{mol}$ is the molar mass of the atmospheric molecules, and $g_{surf}$ is the gravitational acceleration at the surface) of a comet the size of 67P/Churyumov-Gerasimenko, which is representative of a typical Jupiter Family Comet. We use the smaller of these two length scales to compute the Knudsen number at the surface of cometary atmospheres. Figure 1 shows that, for all locations where the body is predominantly cooled by sublimation, the atmosphere is highly collisional and many orders of magnitude too dense to be treated as an exosphere, and thus outside the scope of this work.

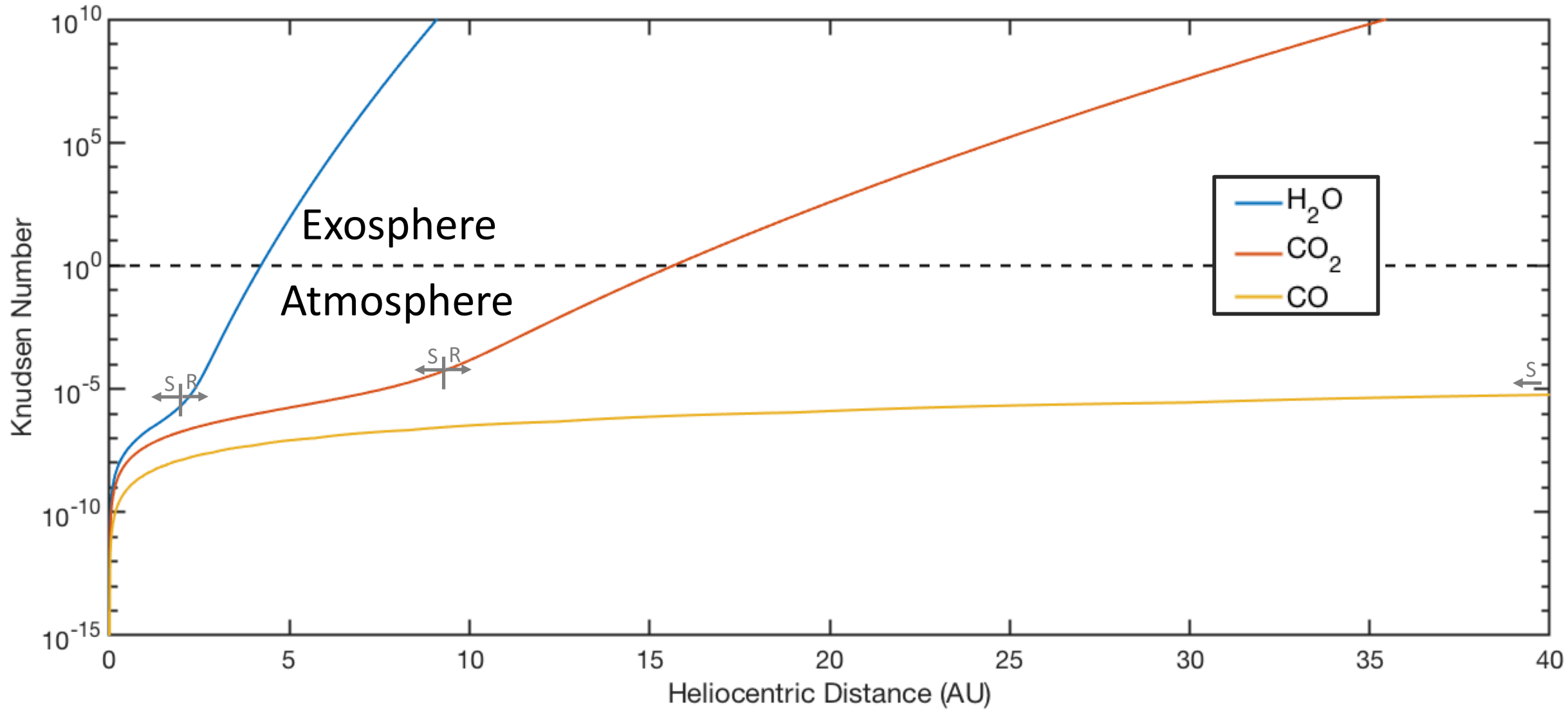

**Figure 1:** *The Knudsen Number of cometary atmospheres at the surface for the sublimation-cooled case.* The Knudsen Number is the ratio of the mean free path to a fundamental length-scale of the situation (e.g., atmosphere scale height or size of the body). Grey lines denote approximate location of transition from the sublimation-dominated cooling (S) to radiation-dominated cooling (R) regimes. Sublimation-dominated cooling has very low Knudsen numbers, characteristic of a collision-dominated atmosphere, and are therefore outside the scope of this work.

Ultimately, we will compare these sublimation dynamics with photodestruction and ballistic escape processes, both of which remove molecules from the exosphere. The photodestruction timescale depends on the flux of solar UV photons with energies that break molecular bonds; this flux depends on heliocentric distance. The ballistic escape timescale depends on the ability of the body to retain molecules gravitationally, which itself depends on the speeds of exospheric molecules leaving the surface, and mass and radius of the body. Finally, the refractory surface (finite reservoir) case may form morning exospheres, the location and structure of which depend on the diurnal timescale of the host body. Note that, by restricting ourselves to collisionless dynamics, we are explicitly neglecting organized transport processes among the gas molecules themselves. For all cases, we assume zero spin-pole obliquity (i.e., no axial tilt).

*2.1 Refractory Surface (finite reservoir) - Residence Time of a Volatile Ice*

To form an exosphere from surface volatiles, the illuminated surface of an object must be sufficiently warm such that volatiles are likely to sublimate/desorb from the surface into the gas phase. This process is described by the residence time of the volatile, which specifies the expected time that a molecule will remain adsorbed/condensed on the surface before desorbing/subliming as a function of temperature; equivalently, this is the inverse of the probability of desorbing/sublimating per unit time (*Wilmoth, 1973*). The residence time ($\tau_{res}$) was first derived by *Frenkel (1924)* using the molecular theory considerations of *Langmuir* (*1913; 1916*) to derive the expected time that a typical molecule remains adsorbed on a surface as a function of temperature:

$$\tau_{res} = \tau_0 e^{\frac{u_O}{k_B T}} \quad (1)$$

where $\tau_0$ is the lattice oscillation period of a condensed molecule for motion perpendicular to the surface, $u_0$ is the energy released upon condensation (molecular binding energy, sometimes called the "activation energy"), $k_B$ is Boltzmann's constant, and $T$ is the temperature of the surface. We note that the lattice oscillation periods and binding energies for the volatiles considered here (Table 2) have little variation, up to a factor of a few. Nevertheless, since $u_0$ appears in the exponent, small variations in this binding energy can lead to relatively large changes in the residence time $\tau_{res}$. As with previous modeling efforts (*Stewart et al. 2011; Prem et al, 2015*), we assume that volatiles exist in sufficient quantities such that most surface-bound molecules bind to other molecules of the same species (or alternatively, that there exists a sufficient, firmly bound monolayer of volatiles, that most molecules are adsorbing/desorbing from the second layer or above). Thus, we consider the binding energy and lattice oscillation of volatiles bound to their own chemical species. Note that some recent work suggests that only a small fraction of the surface of the Moon may have a monolayer of water on its surface (*Hendrix et al. 2019*), however this does not rule out that molecules preferentially bind to molecules of the same species. Furthermore, water adsorbed onto minerals may have a different binding energy and oscillation period; this is excluded out of necessity to keep this modeling effort practicable.

| **Volatile Species** | **Oscillation Period ($\tau_0$)** | **Binding Energy ($u_0$)** | **Photodestruction/photoionization Timescale ($\tau_{photo}$) at 1 AU** |
|---|---|---|---|
| $H_2O$ | $5.0\times10^{-13}$ s [a] | $6.64\pm0.02\times10^{-20}$ J (unannealed) [a]<br>$7.00\pm0.07\times10^{-20}$ J (annealed) [a] | $8.3\times10^{4}$ s [e,f] |
| $CO_2$ | $2.9\times10^{-12}$ s [b] | $3.71\pm0.07\times10^{-20}$ J [b] | $4.95\times10^{5}$ s ($X\ ^1\Sigma^+$)[e,f] |
| CO | $5.0\times10^{-13}$ s [a] | $\approx1.3\times10^{-20}$ J (on CO) [a]<br>$\approx 4.7\times10^{-20}$ J (on $H_2O$) [a] | $1.33\times10^{6}$ s [e,f] |
| $N_2$ | $6.8\times10^{-13}$ s [c] | $\approx1.22\times10^{-20}$ J [d] | $9.73\times10^{5}$ s [e,f] |

**Table 2:** Oscillation periods and molecular binding energies for $H_2O$, $CO_2$, CO, and $N_2$. [a]*Sandford and Allamandola, 1988* [b]*Sandford and Allamandola, 1990* [c]*Sangala et al., 2011* [d]*Shakeel et al., 2018* [e]*Huebner et al., 1992* [f]*Huebner and Mukherjee, 2015*

The residence time is highly sensitive to temperature, with sufficiently cold surfaces on the night sides and high latitudes of bodies leading to residence times many orders of magnitude longer than that body's diurnal timescale. This temperature depends on the specific species: around ~130 K for $H_2O$, ~70 K for $CO_2$, and ~23 K for both CO and $N_2$. This condition generally holds for the bodies and volatiles considered in this work. Thus, volatiles that strike the night side effectively adsorb onto the surface until sunlight sufficiently warms the surface during daylight. We compute the longitude at which these adsorbed molecules will typically desorb from the surface into the exosphere ($\phi_{desorb(\theta)}$) by accounting for the position dependence of the surface temperature with the following model based on Arnold (*1979*)

$$T_{(\theta,\phi)}(t) = \begin{cases} T_{min} + \left(T_{peak} - T_{min}\right)\left[\cos\theta\cos(\phi - \phi_o(t))\right]^{\frac{1}{4}} & \text{if } |\phi - \phi_o| \leq 90^{\circ} \\ T_{min} & \text{if } |\phi - \phi_o| > 90^{\circ} \end{cases} \quad (2)$$

where $T_{peak}$ and $T_{min}$ are the peak and minimum surface temperatures respectively, $\theta$ is latitude, $\phi$ is longitude, and $\phi_o(t)$ is the subsolar longitude, which changes as the body rotates:

$$\phi - \phi_o(t) = 360^{\circ}\frac{t}{P_{rot}} - 90^{\circ}, \quad (3)$$

Where $t = 0$ corresponds to dawn at that longitude. We integrate the cumulative residence time a parcel of the surface has experienced since crossing the dawn terminator, divided by the elapsed time since crossing the dawn terminator, to compute the integrated scaled desorption time $\hat{t}_{desorb(\theta,\phi)}$ (see Appendix A). Where the time $\hat{t}_{desorb(\theta,\phi)} = 1$ indicates the location where adsorbed surface volatiles are most likely to desorb into the exosphere. The latitudinally dependent time $t^{*}_{(\theta)}$ to reach $\hat{t}_{desorb(\theta,\phi)} = 1$ gives a corresponding longitudinal displacement $\Delta\phi_{desorb(\theta)}$ from the dawn terminator at which surface volatiles will most likely desorb (See Appendix A for details). Although actual surface temperatures of airless bodies can diverge from this simple equation (or be better approximated by different surface functional fits), we find that the ultimate properties of the exosphere are weakly sensitive to the exact surface temperature model chosen, allowing this surface temperature model to be sufficiently accurate for all bodies considered (see Appendix E). Indeed, the only property considered here that is sensitive to variations in the surface temperature model is the average sublimative mass flux, which can vary by nearly a factor of two. However, average sublimative flux does not apply to the refractory surface case considered here, rather it only applies to the icy surface case, which is described in the next subsection.

For some Solar System bodies, condensed/adsorbed volatiles will sublime/desorb shortly after dawn, forming an exosphere centered along the dawn terminator (*Stewart et al., 2011; Schorghofer, 2014; Prem et al., 2018*). If the residence of a volatile is longer than the duration of daylight ($t^{*}_{(\theta)} > P_{rot}/2$), a given molecule is unlikely to desorb/sublime during a diurnal cycle. Nevertheless, such bodies would still form a (more tenuous) exosphere that is broadly distributed across the illuminated surface, with peak concentrations associated with peak surface temperatures (as in the icy surface case, discussed in the next subsection). In either case, the density or pressure of the exosphere depends on the distribution of volatiles over the surface. A computation of such densities/pressures is beyond the scope of this work (but could of course be readily simulated with a Monte Carlo model).

*2.2 Icy Surface (infinite reservoir) – Surface Vapor Pressure*

For a surface dominated by a volatile ice (or for surfaces sufficiently cool that adsorbed molecules are unlikely to desorb during a diurnal cycle), and to the extent that subliming molecules are retained long enough for at least a few molecular hops, an exosphere will form with a surface density dictated by sublimative (i.e., solid-vapor) equilibrium. As a result, this infinite reservoir of volatiles requires a modification to the numerical methods outlined for the finite reservoir case to account for this temperature-dependent molecular flux. For this case, we weight the contribution of each surface area element by the normalized value of this sublimation mass flux.

We modify the sublimation model of Steckloff et al., (2015) to compute the sublimation mass flux ($\dot{m}$) as a function of longitude ($\phi$) and latitude ($\theta$) dependent temperature ($T_{(\theta,\phi)}$)

$$\dot{m} = \alpha_{(T)}\sqrt{\frac{m_{mol}}{2\pi R T_{(\theta,\phi)}}}P_{vap\,(\theta,\phi)} \tag{4}$$

$$P_{vap\,(\theta,\phi)} = P_{ref}e^{\frac{\lambda}{R}\left(\frac{1}{T_{ref}}-\frac{1}{T_{(\theta,\phi)}}\right)} \tag{5}$$

where $\alpha_{(T)}$ is the temperature dependent sublimation coefficient of the volatile (typically close to unity), $m_{mol}$ is the molar mass of the species, $R$ is the universal gas constant, $P_{vap\,(\theta,\phi)}$ is the saturation vapor pressure at the surface, $\lambda$ is the latent heat of sublimation of the species (assumed to be constant), and $P_{ref}$ is an empirically determined vapor pressure at a specific temperature $T_{ref}$. This model assumes a transparent atmosphere, such that radiative exchange between the surface and exospheric molecules is negligible.

We use these computed relations to calculate the molecular fluxes ($\dot{m}$) across the illuminated surface of the body. The molecular flux is important, because (unlike the finite reservoir case) all parts of the illuminated surface in the infinite reservoir case contribute different amounts of volatile molecules to the exosphere; the molecular flux for each area element therefore allows each area element's effect on the exosphere to be weighted by its contribution. In practice, this simply requires that we weight the finite-reservoir calculations by the molecular flux for each surface area element; our finite reservoir calculations (described next) can thus be easily modified to describe the infinite reservoir case.

*2.3 Fraction of Molecules that Escape Ballistically*

In addition to being warm enough to release molecules, an airless body can only possess a bound or quasi-bound exosphere if it is massive enough to gravitationally retain those exospheric molecules for a time (long relative to the diurnal cycle). The velocity distribution of atmospheric molecules in thermal equilibrium is described by a Maxwell-Boltzmann distribution, which peaks at a speed defined by the temperature of the gas. Thus, even if the mean thermal velocity of the gas is below the escape speed, there will still be molecules in the high speed tail of the distribution with sufficient speed to escape to space. If the atmosphere is thick, such molecules will most likely collide with another molecule before escaping. However, such up-going fast moving molecules near the exobase (where the mean free path of the molecules is comparable to the atmospheric scale height) will likely escape before colliding (Jeans escape). Surface-bound exospheres experience a loss process similar to Jeans escape, except that sufficiently fast molecules escape directly from the surface.

To compute the rate at which molecules escape from a surface-bound exosphere, we consider the fraction of molecules emitted above the escape speed for each surface area element, neglecting any additional velocity contribution that results from the rotation of the body. It is well known from basic kinetic theory that the velocity distribution function of molecules leaving a surface element in thermal equilibrium with a gas at temperature $T$ is not a half-Maxwellian but rather is biased to higher speeds. This outgoing flux is, via time-reversal symmetry, equal to the incident flux of molecules striking the wall of a box containing a gas at temperature $T$ in thermodynamic equilibrium. Thus, the speed distribution function for molecules leaving the surface is (*Armand, 1977; Hodges and Mahaffy, 2016*)

$$F_{(v)}dv = \frac{1}{2}\left(\frac{m}{k_B T}\right)^2 v^3 e^{-\frac{mv^2}{2k_B T}}dv\ . \tag{6}$$

which is the classic Armand distribution for molecular speeds of desorbing molecules. Using integration by parts to integrate this function from $v = v_0$ to $v = \infty$, we obtain:

$$f_{lost} \equiv f_{(v>v_0)} = (1 + \beta^2 v_0^2) e^{-\beta^2 v_0^2} , \quad (7)$$

where $f_{(v>v_0)}$ is the fraction of molecules emitted from a surface area element that have a speed greater than $v_0$ and $\beta = \sqrt{m/2k_B T}$. Setting $v_0$ equal to the speed at which a molecule permanently leaves an isolated body provides the fraction of the molecules that a surface element emits that never come back. In the isolated two-body problem, this speed is the surface escape speed. However, outside of a body's Hill sphere, other bodies gravitationally dominate that region of space. Thus, the speed at which a particle will likely never return to the body is the speed required to rise to the Hill radius. A particle that just meets this threshold will slow to zero speed at the Hill radius ($R_H$). Thus, we can solve for the critical launch speed ($v_H$) through conservation of specific gravitational potential and kinetic energy:

$$\frac{v_H^2}{2} - \frac{GM}{R} = -\frac{GM}{R_H} \quad (8)$$

with the Hill radius ($R_H$) determined by:

$$R_H \approx a_b \sqrt[3]{\frac{M}{3M_P}} \quad (9)$$

where $a_b$ is the semimajor axis of the body's orbit around its primary, and $M_P$ is the mass of the primary. Solving for the speed required to reach the Hill radius ($v_H$) and thus $v_0$:

$$v_0 = v_H = \sqrt[2]{2GM\left(\frac{1}{R} - \frac{1}{a_b}\sqrt[3]{\frac{3M_P}{M}}\right)}. \quad (10)$$

Finally, to compute the fraction of molecules lost from across the illuminated hemisphere of the body (as opposed to a single surface area element), we integrate over the portion of the surface emitting molecules (assume nightside to have negligible sublimation flux). Because the icy surface and finite reservoir case have different sublimation flux distributions on their surfaces, we need to treat the two endmember cases differently (see Appendix B for the following derivations). For the icy surface (infinite reservoir) case,

$$f_{lost_global} = \frac{\dot{m}_{ejected}}{\dot{m}_{total}} = \iint \frac{\dot{m}_{(\theta,\phi)}}{\dot{m}_{total}} \left(1 + \beta_{(\theta,\phi)}^2 v_0^2\right) e^{-\beta_{(\theta,\phi)}^2 v_0^2} \cos\theta \, d\theta \, d\phi. \quad (11)$$

We integrate this function numerically to compute the fraction of ejected molecules that are directly ejected on escape trajectories, along with the ejected mass flux.

For the refractory surface (finite reservoir) case, we consider the temperatures along the line of the morning exosphere. Because adsorbed/condensed volatiles are not present in an infinitely thick layer (but rather in a thin, possibly monolayer), we do not normalize by mass flux, resulting in an expression that computes the global fraction of molecules that are directly ejected to space along the path of the morning exosphere.

$$f_{lost_global} = \int \left(1 + \beta_{(\ell)}^2 v_0^2\right) e^{-\beta_{(\ell)}^2 v_0^2} \, d\ell \quad (12)$$

where path ($\ell$) is the set of latitudes ($\theta$) and longitudes ($\phi$) corresponding to location of the morning exosphere. This expression assumes that condensed/adsorbed molecules are uniformly distributed in latitude at the morning terminator, and emitted only at the location of the morning exosphere (itself a function of latitude). We note that this calculation ignores ballistic escape of molecules that experience multiple "hops" (i.e., do not immediately escape from the dawn exosphere, but rather reimpact a potentially warmer part of the surface where ballistic escape is likelier). This simplification would predominantly affect larger bodies (higher escape speeds) nearer the sun (warmer surface temperatures), such as the Moon and Mercury; smaller, colder bodies would be less affected by this simplification. In any case, this ballistic loss calculation represents a lower bound estimate of the ballistic loss from these bodies.

For cases of the refractory surface (finite reservoir) in which the volatile does not immediately sublimate during the day (the integrated residence time is not much shorter than a diurnal cycle), the resulting exosphere is tenuous, more widely distributed (rather than narrowly distributed along the dawn terminator; *e.g., Stern, 1999; Stewart et al., 2011; Prem et al., 2015*), with molecules ejected from the illuminated hemisphere in a manner that depends on surface temperature. This case best resembles the icy surface (infinite reservoir) case, without weighting by mass flux due to the limited amount of volatile present on the surface, but with weighting related to the inverse of residence time

$$f_{lost_global} = \iint \frac{G_{res}}{\tau_{res\,(\theta,\phi)}} \left(1 + \beta^2_{(\theta,\phi)} v_0^2\right) e^{-\beta^2_{(\theta,\phi)} v_0^2} \cos\theta \, d\theta \, d\phi. \qquad (13)$$

where

$$G_{res} = \iint \tau_{res\,(\theta,\phi)} \cos\theta \, d\theta \, d\phi. \qquad (14)$$

This situation cannot be approximated as a line integral, as there is no preferential longitude at which molecules sublimate/desorb into the exosphere. For both of these refractory surface (finite reservoir) cases, we numerically integrate to compute these quantities.

*2.4 Ballistic Hop Timescales and Hop Distances*

Even in the absence of photodestruction, surface sticking, and solar wind interactions, an exosphere density perturbation will decay due to the diffusion of fast-moving molecules. A characteristic lifetime of such a perturbation can be determined from the average duration of a ballistic hop of molecules moving slowly enough to be gravitationally retained by the body ($\tau_{ballistic}$), and from the fraction of molecules that escape at each hop ($f_{lost}$). Computing the typical ballistic timescale requires computing the duration of a ballistic hop as a function of ejection speed and angle, and normalizing by the probabilities that a molecule would have such a speed and ejection angle.

To compute the ballistic timescale for a given surface temperature, we compute the ballistic hop timescale for all ejection angles and speeds between zero and $v_H$ (the critical speed required to reach the body's Hill radius). If a ballistic hop timescale is longer than the photodestruction timescale, the particle is considered to be lost, and is not considered in this computation. We then weight the ballistic hop timescale for each combination of ejection angle and speed by their relative probabilities (Armand speed distribution and Lambertian angular distribution), and numerically integrate over parameter space (velocity component combinations with speeds less than $v_H$) to obtain the typical (average) duration of a ballistic hop as a function of temperature (See Appendix C for details of approach and ejection speed and angle distributions).

To complete this computation, we treat the two cases differently. For rocky surfaces (finite reservoirs), we complete this computation along the line of peak density of the day-side exosphere (near the morning terminator). For rocky surfaces in which there is no exosphere concentrated near the morning terminator, we integrate across the entire illuminated hemisphere. For icy surfaces (infinite reservoir), we also integrate across the entire illuminated surface, again weighting by sublimation mass flux. This results in an object's average ballistic hop duration ($\tau_{ballistic}$). We validated this approach against free-molecular Monte Carlo simulations, to ensure its accuracy and proper functioning (see Appendix D).

*2.5 Photodestruction Timescale*

Exposure to sunlight can destroy exospheric molecules. When a molecule absorbs a sufficiently energetic photon it dissociates. For an optically thin daytime exosphere, the probability of photodissociation is independent of position and the decay rate of molecules is described by

$$N_{(t)} = N_0 e^{-\frac{t}{\tau_{photo}}} \quad (15)$$

where $N_0$ is the initial number of gas molecules, $t$ is the elapsed time, and $\tau_{photo}$ is the photodestruction timescale.

*Huebner et al., (1992*) compiled an extensive list of photodissociation timescales for various photodestruction pathways for common molecules (including $H_2O$, $CO_2$, and CO) at the Earth's distance from the sun (1 AU). These timescales can be scaled to other heliocentric distances by an inverse square law. Thus, we compute the photodestruction timescale at a given heliocentric distance ($\tau_{photo}$) by scaling the timescales for dominant photodestruction pathways ($\tau_{photo\ 1\mathrm{AU}}$ ) by the square of the ratios of heliocentric distances:

$$\tau_{photo} = \tau_{photo\ 1\mathrm{AU}} \left(\frac{r_{body}}{r_{Earth}}\right)^2 \quad (16)$$

where $r_{body}$ is the heliocentric distance of the object under consideration and $r_{Earth}$ is the heliocentric distance of the Earth. This timescale is only applicable to illuminated molecules that are not stuck to the surface. Adsorbed/condensed molecules may also photodissociate (albeit at likely different rates than "free" molecules); nevertheless, we neglect this process of the purposes of our exosphere calculations.

We use the model of *Butler (1997)* to account for photodestruction. We use the computed ballistic hop duration to compute the fraction of molecules that would survive a ballistic hop for every combination of tangential and normal velocity components, and weight that combination's contribution to the average hop time and distance by this value. Simultaneously, we determine the fraction of ejected molecules that are lost to photodestruction.

*2.6 Expected Exospheric Residence Time*

The expected exospheric residence time ($\tau_{exo}$) measures the expected, cumulative amount of time that an adsorbed molecule could expect to persist in an exosphere, desorbed from the surface without escaping ballistically or being photodestroyed. Note that the expected exospheric residence time is an entirely distinct quantity from the residence time that a molecule spends adsorbed on the surface; they are similar in name only. To obtain this quantity ($\tau_{exo}$), we assume that emitted molecules that do not escape either remain on the sunlit side, or hop to the night side where they deposit on the surface, to be released again when their landing point rotates past the dawn terminator. These computations only consider the time a molecule spends in the gas phase, neglecting any time spent adsorbed/condensed on the surface (e.g., the amount of time free from the surface, in the exosphere).

We consider that during each ballistic hop (average duration of $\tau_{ballistic}$), only a fraction return to the surface ($1 - f_{lost}$) to be ejected again. The expected duration of its exospheric residency ($\tau_{exo}$) is the sum of an infinite geometric series

$$\tau_{exo} = \sum_{n=1}^{\infty}(1 - f_{lost})^n \tau_{ballistic} = \tau_{ballistic} \frac{(1-f_{lost})}{f_{lost}} . \quad (17)$$

While this simple model (i.e., equation 17) becomes less accurate and introduces errors when the ballistic hop distance is comparable to the size of the body, the size of these errors should be small. Large objects have higher gravity, which leads to smaller hop distances, while small bodies have larger hop distances but higher fractions of particles that escape. Both of these considerations prevent errors from growing very large, allowing the computed timescale $\tau_{exo}$ to represent the characteristic lifetime of a molecule in the exosphere reasonably well.

### *2.7 Expected Migration Distance*

Thus far, we have described the exospheric processes required to understand the free molecular motion of an adsorbed/condensed molecule as it leaves the surface to travel ballistically about the body. Through multiple such ballistic hops, a molecule may travel away from its point of origin, potentially migrating to polar cold traps, where the particle may subsequently remain frozen over the age of the Solar System, and thus contribute to polar ice deposits. Assuming that the migration process is approximately modeled by a two dimensional random walk on an infinite plane, with constant step size equal to the average ballistic hop distance ($d_{hop}$), the root mean square displacement after $N$ hops is

$$d_{net,RMS} = d_{hop}\sqrt{N}. \quad (18)$$

The real situation corresponds to variable hop distances and migration is occurring on the surface of sphere. Nevertheless, $d_{net,RMS}$ computed from the expected number of hops before the molecule escapes or is photodestroyed gives a reasonable estimate of the distance that molecules can be expected to migrate. If $d_{net,RMS}$ is comparable to the size of the body, then molecules stand a reasonable change of reaching a cold trap. However, if $d_{net,RMS}$ is small compared to the size of the object, then such exosphere-mediated volatile transport becomes unlikely.

## 3. Results and Discussion

These three processes (desorption/sublimation, ballistic motion/loss, and photodestruction) determine the shape, structure, and dynamics of exospheres. Furthermore, comparing the characteristic timescales, length scales, and loss fractions associated with these processes can determine what processes destroy an object's exosphere, and whether it can experience exosphere-mediated volatile migration to cold traps. Alternatively, loss fractions on smaller bodies (e.g., Amalthea, Hyperion, and Puck) are often so high (near 100%) that any exosphere that forms is effectively unbound, leading to rapid devolatilization. This inability to retain volatiles (and thus experience exosphere-mediated volatile transport) can also leave an imprint on the object's surface. We apply this model across the System, from the innermost planet Mercury to the moons of Uranus. We ignore major groups for which these model assumptions break down, such as active Centaurs and comet nuclei (which have collisional "atmospheres" that are simply streaming away), or trans-Neptunian objects (which have surface temperatures sufficiently cold to challenge the accuracy of model parameterizations and empirical inputs). In the rest of the Solar System inside the orbit of Neptune, we find that CO and $N_2$ are much too volatile to condense reliably on unilluminated hemispheres. Thus, only two volatiles in our model ($CO_2$ and $H_2O$) can be applied to airless bodies interior to Neptune (We neglect CO- and $N_2$-rich Triton, as it has a dense, collisional atmosphere). We consider physical phenomena observed on each body or within each system of moons, and find that such analysis can explain broad observational trends throughout the Solar System.

### *3.1 Water ($H_2O$) Exosphere in the Inner Solar System*

In the inner Solar System (inside the orbit of Jupiter), water tends to be the dominant volatile. The four largest airless bodies in the inner Solar System (Mercury, the Moon, Ceres, and Vesta) all have sufficiently warm surfaces that water desorbs from their surfaces within a few degrees of the dawn terminator (Fig. 2). Indeed, water-rich exospheric activity has been observed on Mercury (*Zurbuchen et al., 2008*), the Moon (*Halekas et al., 2015; Benna et al., 2019*), and Ceres (*A'Hearn et al., 1992; Küppers et al., 2014*). Therefore, the reason these bodies are effectively airless must be due to mechanisms that remove these molecules (here we consider photodissociation and exospheric escape). Considering only these two loss mechanisms, our calculations reveal that these four bodies break down into two groups: photodissociation-limited, and exospheric lifetime-limited.

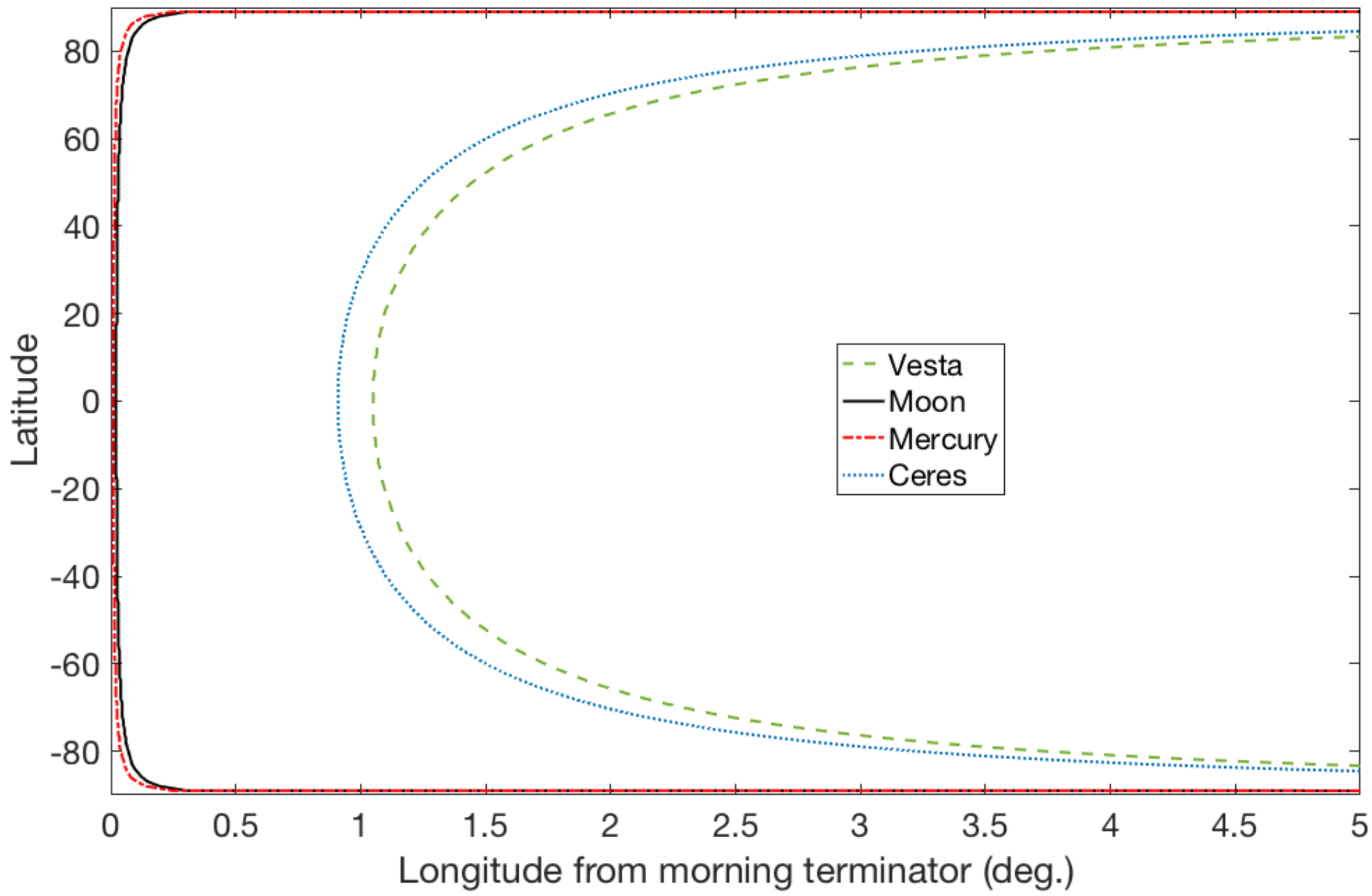

**Figure 2:** *The location of the morning $H_2O$ exosphere on Mercury, the Moon, Ceres, and Vesta.* We compute the longitude of water desorption as a function of latitude (morning terminator located at 0 degrees). Note that the resulting Dawn Atmospheric Enhancement (DAE) will be centered along these curves, with the width of the densest part of the DAE having a width on the order of the mean ballistic hop distance. Mercury and the Moon are closer to the Sun and rotate slower than Ceres and Vesta, resulting in morning exospheres centered much closer to the terminator. Results for the Moon agree closely with DSMC simulations (*Prem, 2017*)

Photodissociation-limited bodies are sufficiently close to the sun that their exospheric molecules break down before they have a chance to escape the gravitational influence of the body. Both Mercury and the Moon predominantly lose molecules through photodestruction, rather than ballistic escape, which is consistent with their respective photodestruction timescales (~3.8 and ~25 hours) being comparable to their expected water exospheric residence times (3.41 and 22.5 hours, respectively). By contrast, Ceres and Vesta predominantly lose molecules through ballistic escape, which is consistent with their photodestruction timescales (~8 days and ~5.8 days) being significantly longer than their expected exospheric residence time (8.17 and 2.01 hours, respectively). These residence times are based on how long a molecule would be expected to spend desorbed from the surface with a speed sufficiently low to keep it bound to the body. The time required to reach the Hill Radius and escape the body (assuming a molecule is ejected at a speed fast enough to escape the body) is typically on the order of months. Thus, even the molecules moving fast enough to escape Vesta and Ceres will most likely photodissociate prior to escaping their gravitational pull. In this sense, Ceres and Vesta behave similarly to comets in the Solar System, which eject water molecules that ultimately photodissociate prior to escaping, interact with the local electromagnetic field, and stream antisunward, forming an ion tail; this is consistent with the detected plasma bow shock at Ceres (*Jia et al., 2017*). However, these molecules are moving very fast (minimum speeds of five and two atmospheric scale heights per hour for Ceres and Vesta respectively), and generally leave the atmosphere/exosphere prior to photodissociating.

| | **Mercury** | **Moon** | **Ceres** | **Vesta** |
|---|---|---|---|---|
| **Object Radius** | 2440 km | 1737 km | 473 km | 263 km |
| **Avg. Bond Albedo** | 0.119[a] | 0.123±0.002[b] | 0.034±0.001[c] | 0.20±0.02[d] |
| **Heliocentric distance (AU)** | 0.39 | 1.0 | 2.77 | 2.36 |
| **Peak Surface Temperature (K)** | 700 [e] | 392[f] | 241 (est.) | 250 (est.) |
| **Minimum Surface Temperature (K)** | 100 [e] | 95[g] | 100 | 100 |
| **Surface escape speed (m/s)** | 4,250 | 2,380 | 510 | 360 |
| **Fraction of molecules lost (ballistically, per hop)** | $1.4\times10^{-52}$ % | $2.23\times10^{-15}$ % | 44.78% | 76.85% |
| **Fraction of molecules lost (photodestruction, per hop)** | 1.47 % | 0.53% | 3.38% | 2.04% |
| **Fraction of molecules lost (total, per hop)** | 1.47 % | 0.53% | 48.16% | 78.89% |
| **Expected Number of hops** | 67 | 187 | 1.1 | 0.27 |
| **Ballistic timescale ($H_2O$)** | 183 s | 443 s | 7.59 hours | 7.52 hours |
| **Expected Exospheric Residence Time ($H_2O$)** | 3.41 hours | 23.0 hours | 8.17 hours | 2.01 hours |
| **Ballistic hop distance ($H_2O$)** | 60 km | 140 km | 658 km | 400 km |
| **Expected Migration distance (RMS)** | 491 km | 1910 km | 690 km | 208 km |
| **Photodissociation timescale at heliocentric distance ($H_2O$)** | ~3.8 hours | ~25 hours | ~8.0 days | ~5.8 days |

**Table 3**: *Comparison of Thermally produced adsorption/desorption exospheric dynamics for $H_2O$ in the Inner Solar System.* Mercury, the Moon, Ceres, and Vesta are the four largest airless bodies in the inner Solar System, and the most capable of retaining an exosphere. Estimated peak surface temperatures are derived using albedo and heliocentric distance to solve radiative energy balance at sub-solar point. Refs: [a] Veverka et al., 1988 [b] Lane and Irvine, 1973 [c] Li et al., 2016 [d] Li et al., 2013b [e] Vasavada et al., 1999 [f] Hurley et al., 2015 [g] Williams et al. 2017

From these results, we can explore the ability of water molecules to migrate about a body. Our results show that the root mean square (RMS) migration distance is comparable to the size of the body (object radius) for the Moon, Ceres, and Vesta, suggesting that exosphere-mediated volatile transport can deliver to these bodies' cold traps significant amounts of water that evade ballistic escape and photodestruction. This is consistent with observations (*Feldman et al., 1998; 200; 2001; Lawrence et al., 2011; Mitrofanov et al., 2012; Hayne et al., 2015; Fisher et al., 2017; Li et al., 2018*) and previous modeling (e.g., *Butler, 1997; Crider & Vondrak, 2003; Stewart et al. 2011; Schorghofer, 2014; Prem et al. 2015, 2017; Schorghofer et al. 2017*) of water in the lunar cold traps, and in permanently shadowed craters on Ceres (*Tu et al. 2014; Platz et al., 2016; Schorghofer et al. 2017*). Vesta's high axial tilt produces illumination conditions that preclude any polar cold traps (*Stubbs & Wang, 2012*); thus there are no places for water to collect long term on Vesta, explaining its lack of polar water.

Curiously, Mercury possesses thick polar water ice deposits in its polar cold traps, in spite of our calculations and those of previous models (*Butler, 1997; Schorghofer et al. 2017; Prem et al., 2019*) suggesting that little water would survive long enough to migrate to these cold traps due to the dominance of photodestruction. This may result from our model's assumption that all water molecules are exposed to wavelengths of light that can cause photodestruction; however if sufficient quantities of water were to be injected into Mercury's exosphere all at once (e.g., though a comet impact), the exosphere could become optically thick, causing exospheric $H_2O$ molecules to shield other $H_2O$ molecules at lower altitudes, significantly reducing the overall effective rate of photodissociation (*Arnold, 1979; Prem et al., 2015*). Thus, two scenarios favor the migration of water to Mercury's polar water deposits. First, water in Mercury's cold traps could have been delivered all at once, in an impact sufficiently large to experience significant self-shielding from photodestruction. This is consistent with suggestions that the Hokusai Crater-forming impact delivered the majority of Mercury's water in a single, large event (*Ernst et al., 2018; Deutsch et al., 2019*). Alternatively, the water could originate locally relative to the cold traps from high latitude sources/impacts (both large impacts or micrometeorites), most likely from a latitude of ±70° (*Butler 1997*). This is consistent with Frantseva et al. (2022), who found that indeed water-rich impactors alone could account for the entire expected volume of water found in Mercury's cold traps. Solar wind interactions with hydroxyl groups in surface materials could also populate Mercury's cold traps with water over solar-system timescales (*Jones et al. 2020*), as could plasma flux and/or sputtering on the nightside, however this may be inconsistent with the relatively high "purity" of Mercury's water (*Ernst et al. 2018*).

Finally, our calculations can predict the general shape and structure of the $H_2O$ exospheres of these bodies. Water desorbs from all of their surfaces at longitudes near the dawn terminator, forming a dawn-centric water exosphere. On Mercury and the Moon, the relatively small ballistic hop distance (compared to the size of the body) results in a narrow "banana-shaped" exosphere along their dawn terminators. This phenomenon was previously described on the Moon (*Stern, 1999*) and from DSMC simulations (*Stewart et al., 2011; Prem et al., 2015*), however this work shows that an even narrower exosphere should exist on Mercury. By contrast, the relatively large hop distances of water on Ceres and Vesta, which are comparable to the size of the body, suggest that their $H_2O$ exospheres would encompass the entire body. However, the large ballistic escape rates suggest that were such exospheres to form, they would be very tenuous compared to thermally-generated exospheres formed from comparable surface conditions on either Mercury or the Moon.

### *3.2 The Jovian System*

The Galilean moons of Jupiter are large, planet-size moons, three of which (Europa, Ganymede, and Callisto) have surfaces composed primarily of water ice and/or hydrated minerals (*Carlson et al., 1996*). At this distance from the sun water sublimation is significantly reduced, such that water sublimes from across the illuminated hemispheres of these moons, rather than forming a morning $H_2O$ exosphere; this pattern was recently found on Ganymede, with an $H_2O$ sublimation exosphere centered on its subsolar point (*Roth et al. 2021*). Ganymede and Callisto also contain significant deposits of $CO_2$ (*McCord et al., 1997;1998*). We use our model to characterize the $H_2O$ and $CO_2$ exospheres of the Galilean and other jovian moons.

We find that the Galilean moons have sufficient mass that loss from the $H_2O$ and $CO_2$ exospheres is dominated by photodestruction, rather than ballistic escape, and that $CO_2$ on the Galilean moons is sufficiently volatile that condensed/adsorbed deposits would form a morning terminator exosphere (figure 3). Furthermore, the photodestruction timescale of $CO_2$ is an order of magnitude longer than $H_2O$. As a result, $CO_2$ has a significantly easier time populating an exosphere and experiencing exosphere-mediated volatile transport on these bodies, consistent with observations of Callisto's $CO_2$-dominated exosphere (*Carlson, 1999*).

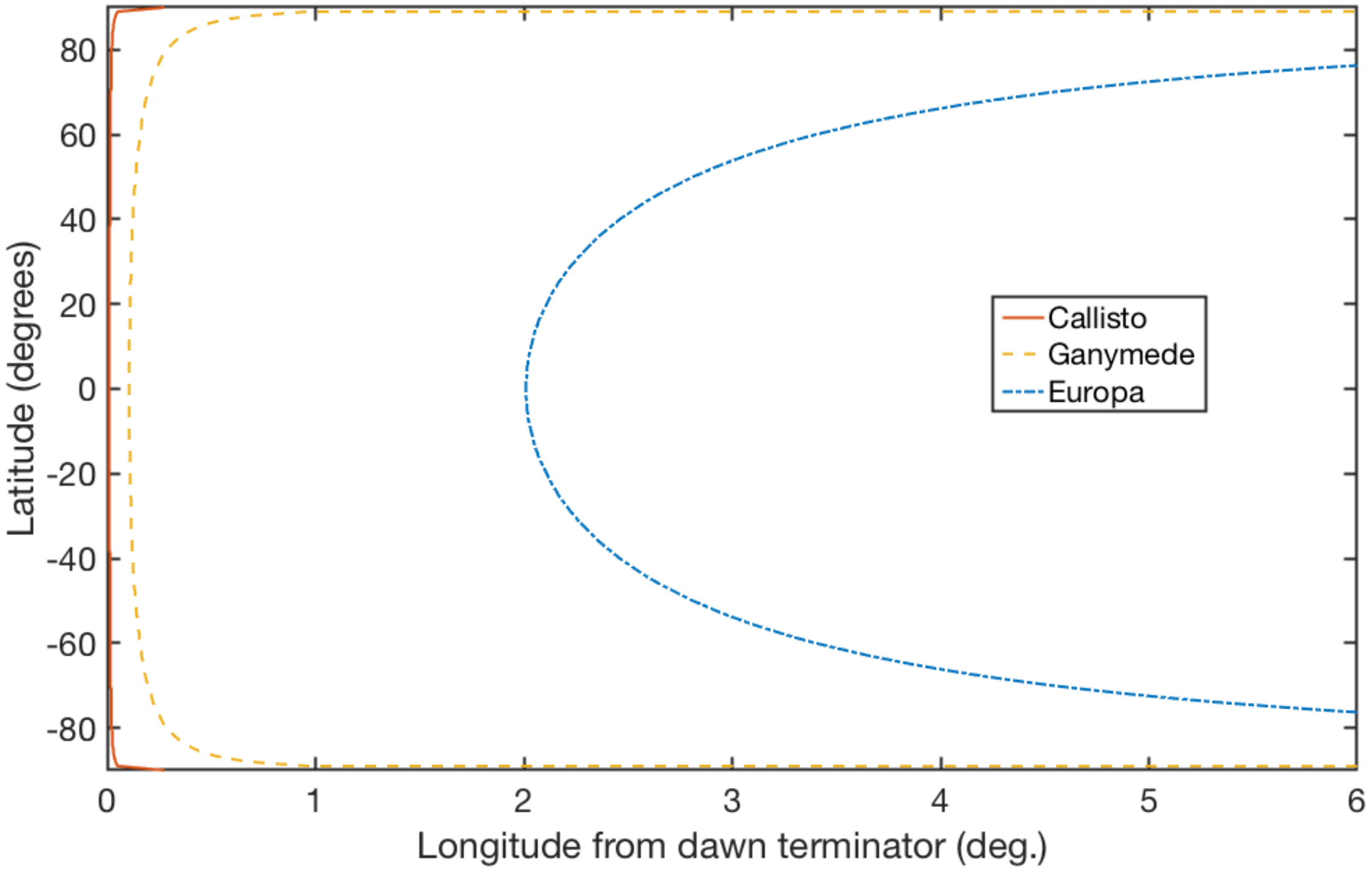

**Figure 3**: *The location of adsorption/desorption-produced $CO_2$ Exospheres on the Galilean Satellites.* Our model describes the location of peak $CO_2$ desorption (center of the Dawn Atmospheric Enhancement) near the morning terminators of Callisto, Ganymede, and Europa. Differences in the locations are due to different surface albedos/temperatures and rotation rates, but roughly correspond to a surface temperature of 88 K. Other $CO_2$ sources are not accounted for in this model, and could produce perturbations on these results.

Callisto's $CO_2$ exosphere was detected at an orbital phase of 75°, in a chord over 80-90°W, 2°N (*Carlson, 1999*). At this rotation phase, the subsolar point was located at 105°W, putting this chord though the early afternoon sky near the equator. This detection of $CO_2$ therefore missed Callisto's morning $CO_2$ terminator exosphere (see Fig. 3), which would have been significantly denser. Future spacecraft missions, such as ESA's upcoming *Jupiter Icy Moons Explorer (JUICE)* mission, should be able to detect this expected morning $CO_2$ sublimation enhancement in Callisto's atmosphere, and compare with its sputtering-induced exosphere (*e.g., Vorberger et al., 2015*). However, this requires that the sputtering and sublimation exospheres can be distinguished from one another; this should be possible due to their different shapes and structures, and has already been done for Ganymede's $H_2O$ exosphere (*Roth et al. 2021*).

Callisto's surface $CO_2$ distribution is not uniform. Callisto has bright and dark terrains (albedos of 0.8 and 0.2 respectively), with bright terrains associated with fresh crater rims and bright icy knobs; $CO_2$ is preferentially associated with these bright terrains; (*Moore et al., 1999; Hibbitts et al., 2002; Howard & Moore, 2008)*. $CO_2$ can be liberated through sputtering processes (*Vorberger et al., 2015*), or mass wasting of $CO_2$-rich "bedrock" (*Moore et al., 1999; Howard and Moore, 2008*), raising the question of whether these bright terrains are erosional or depositional; previous modeling work found that these terrains are likely a combination of these two processes (*Howard and Moore, 2008*).

We use the results of our model to understand the interactions between Callisto's surface and exosphere, and the flux required to maintain the observed 150 K, $7.5\times10^{-12}$ bar $CO_2$ exosphere (*Carlson et al. 1999*). At this temperature, the mean of the magnitude of these molecules' velocities is ~270 m/s. Dividing the surface pressure by this velocity gives the mass flux colliding with the surface, $2.8\times10^{-9}$ kg/m$^2$ s, or $3.8\times10^{16}$ molecules are striking every square meter of the surface every second ($2.8\times10^{30}$ molecules/s, when integrated over the entire surface). Our modeling results reveal that every molecule experiences on average ~36,000 of these hops before leaving the system; thus, $7.8\times10^{25}$ $CO_2$ molecules/s (5.7 kg/s of $CO_2$) are required to maintain Callisto's $CO_2$ exosphere.

Hibbitts et al. (2002) noted that Callisto's $CO_2$ is correlated with fresh impact craters; suggesting a causal relationship in which impacts could either deliver $CO_2$ or dredge it up from subsurface reservoirs. Using the findings of Zahnle et al. (*2003*), one could expect impacts to deliver, on average, on the order of ~1,000 m$^3$ of material per year (although individual impact events can certainly deliver significantly more than this). However, our results find that maintaining Callisto's $CO_2$ exosphere requires ~110,000 m$^3$ of $CO_2$ ice per year. Thus, unless a very recent impact is responsible for Callisto's exosphere, the impact delivery mechanism is unlikely to provide major contributions to Callisto's $CO_2$ exosphere.

Similarly, the micrometeorite flux provides little contribution to Callisto's exosphere. Callisto's surface is typically on the order of ~1-4 Gyr old (*Greeley et al. 2000*). If we consider that the impact gardening results of Costello et al (*2021*) for Europa are applicable to Callisto and extrapolate out to ~1 Gyr, one would expect Callisto's ~1-10 meters of surface material to generally be devolatilized through impacts and therefore contribute little to Callisto's exosphere. Although the impactors themselves may be volatile rich, the cumulative size-frequency distribution of the impactor flux in the Jovian system at small sizes is rather shallow, with a power law index of –0.9 (*Zahnle et al. 2003*). This means that the volume of impact-delivered material is dominated by larger impactors, and therefore micrometeorite impacts contribute negligibly to Callisto's exosphere.

Hibbitts et al. (2002) also consider that Callisto's $CO_2$ can originate from subsurface deposits either via impact excavation (e.g., impact gardening) or via mass wasting from Callisto's bright spires. We can estimate the area of exposed $CO_2$ ice required to maintain Callisto's $CO_2$ exosphere; assuming that all of this $CO_2$ originates from a single patch of exposed $CO_2$ ice (and that the vast majority of solar energy striking this patch is used to overcome $CO_2$'s latent heat of sublimation of 27,200 J/mol), we find that a total area on the order of only ~7 hectares (~17 acres) of exposed $CO_2$ across Callisto is required to maintain this sublimative flux (assuming the $CO_2$ exosphere is sublimative in origin). Such deposits would rapidly recede below a thermal skin depth of the surface[2], requiring regular mass wasting events to keep such ice deposits exposed, as has been suggested for maintaining comet activity (Steckloff et al. 2016; Steckloff & Samarasinha, 2018). Thus, on the order of ~7 hectares of fresh $CO_2$ ice would need to be exposed annually to maintain Callisto's exosphere, consistent with mass-wasting suggested by previous work (*Moore et al., 1999; Hibbitts et al., 2002; Howard & Moore, 2008*).

As exospheric $CO_2$ molecules collide with the surface, they temporarily condense/adsorb onto the surface, and remain for a period of time dictated by the temperature (the residence time). The bright terrains of Callisto would have an average temperature on the illuminated hemisphere of 111 K, corresponding to a $CO_2$ residence time of 0.11 s. At a surface pressure of $7.5\times10^{-12}$ bar and gas temperature of 150 K (*Carlson et al. 1999*), $3.8\times10^{16}$ $CO_2$ molecules/m$^2$ s collide with the surface, resulting in $4\times10^{15}$ $CO_2$ molecules/m$^2$ stuck on the surface at any time. Using a diameter of $CO_2$ molecule of 0.33 nm, ~0.05% of the surface would be covered by $CO_2$ molecules. Converting this to weight percent by multiplying by the density ratio of $CO_2$ and the (presumed) water ice substrate, we estimate that the bright terrains should host ~0.07% $CO_2$ by weight, just through contact with the exosphere. This closely agrees with the observationally obtained estimates of Hibbitts et al. (*2002*) of ~0.07 – 0.16% $CO_2$ on these bright terrains, with the range a result of uncertainties regarding ice grain sizes. Thus, the observed bright terrains, rather than hosting a source of $CO_2$, most likely contain $CO_2$ due to thermochemical equilibrium with the exosphere. Our modeling therefore reveals that Callisto's exosphere cannot build up sizeable frost deposits of $CO_2$ as the deposition rate is too low to counteract sublimative loss. Rather, the bright spires and bright high elevations are likely the result of mass wasting, which exposes subsurface ices.

We repeat this analysis for Callisto's dark terrains, which would have an average temperature of 141 K, and a resulting residence time of 0.00055 s. This results in $2\times10^{13}$ $CO_2$ molecules/m$^2$ stuck on the surface at any time; ~$2\times10^{-4}$ % of the surface covered by $CO_2$ molecules, or a weight percent of ~$4\times10^{-4}$ %. This differs significantly from the estimates of Hibbitts et al. (*2002*) of ~$1\times10^{-2}$ %, which assumed the surface dust of Callisto to have comparable properties to the Moon and Mars. However, it is unclear if this difference is due to modeling assumptions (e.g., Hibbitts et al. (2002) assumptions of Callisto's dust properties, our model's neglect of porous surface regolith which can trap more $CO_2$), or if Callisto's dark regolith has an unexpectedly high concentration of $CO_2$.

Our model also predicts a significant sublimation-driven $H_2O$ exosphere on Callisto. Multiplying the integrated sublimative flux by the expected number of hops and Callisto surface gravity, we calculate a maximum $H_2O$ exospheric pressure on the illuminated hemisphere of $1.6\text{x}10^{-4}$ Pa ($1.5\text{x}10^{-9}$ bar); this corresponds to a total exospheric mass of up to $1.6\text{x}10^{35}$ molecules (~4.9 million metric tons of $H_2O$). This pressure assumes the surface is pure water ice, and would scale with the near surface $H_2O$ mole fraction. Water sublimation from Callisto's predominant dark terrains dominate sublimative flux to the exosphere ($1.2\text{x}10^{18}$ molecules/m$^2$s

[2] Using the reasonable range of thermal diffusivities for icy bodies of ~$10^{-8}$ – $10^{-6}$ m$^2$/s (*Steckloff et al. 2021*), Callisto's diurnal thermal skin depth is ~0.1 – 1 m

on average) relative to the bright terrains (2.3x$10^{13}$ molecules/$m^2$s on average). This flux causes the dark terrains to recede at a rate of up to ~300 nm/yr. Meanwhile, the resulting exospheric flux (i.e., molecules striking the surface from the exosphere; the reverse direction of sublimation from the surface) dominates the sublimative flux from the bright terrains, causing water ice deposits to grow at a rate of up to ~1 mm/yr. Thus, the bright spires and terrains of Callisto are likely built and maintained by $H_2O$ frosts. Our calculations suggest that these frosts can continue to form so long as the exospheric $H_2O$ condensation/adsorption flux dominates the sublimative $H_2O$ flux from these bright terrains (this should occur so long as the $H_2O$ mixing ratio of the dark terrains is at least ~0.01%). This result is consistent with previous work suggesting that these terrains are depositional frosts (*Moore et al. 1999; Howard and Moore, 2008*).

Altogether, a picture of Callisto emerges as an actively evolving world, similar in many ways to comet nuclei; subliming and perhaps sputtering $CO_2$ building up an exosphere that slowly leaks to space. Mass wasting events likely are crucial toward maintaining $CO_2$ sublimative activity in a manner similar to comet nuclei (*Steckloff et al. 2016; Steckloff & Samarasinha, 2018*), causing refractory materials to settle downslope. Like a comet, such activity will cause Callisto's surface to tend towards a relatively flat, low-topography surface (*Vincent et al. 2017; Steckloff et al. 2018*), however impacts can generate new topography. Unlike comet nuclei, Callisto is large enough to retain sublimed gases, facilitating exosphere-mediated volatile transport that can form frost deposits; it has been suggests that some such frosts are possible on comets (*Rubin et al. 2014*; *Sunshine et al. 2016*), however they are unstable over diurnal cycles.

On Ganymede, we find a global sublimative $H_2O$ rate of $1.51\times10^{30}$ $s^{-1}$, comparable to previous results of $7\times10^{29}$ $s^{-1}$ (*Plainaki et al., 2015*). This corresponds to an average sublimative flux of $3.46\times10^{16}$ $s^{-1}m^{-2}$. This dominates the expected $H_2O$ sputtering flux in *Marconi (2007)* of $1.8\times10^{12}$ $s^{-1}m^{-2}$ or total sputtering $H_2O$ production in *Plainaki et al. (2015)* of 8.0x$10^{27}$ $s^{-1}$ and *Vorburger et al. (2022)* of $2.6\times10^{29}$ $s^{-1}$. On Ganymede, the peak $H_2O$ sublimation flux at the subsolar point is $2.41\times10^{17}$ $s^{-1}m^{-2}$; dividing by the thermal velocity (mean of the magnitude) of the subliming molecules of 423 m/s produces a surface density of $5.7\times10^{14}$ $m^{-3}$, which is consistent with the previous results of *Marconi (2007)* of 7x$10^{14}$ $m^{-3}$ and of *Plainaki et al. (2015)* of ~$2.5\times10^{14}$ $m^{-3}$, but with a significantly simpler model that does not track collisions or use Monte Carlo methods. This demonstrates the power of our relatively simple numerical model of sublimation exospheres.

On Ganymede, the surface is sufficiently cold, that any $H_2O$ molecule that strikes the surface should immediately stick (a sublimation coefficient of 1; *Gundlach et al. 2011*). Thus, the total mass of the $H_2O$ sublimation exosphere can be estimated by multiplying the global water production rate by the ballistic timescale, resulting in a total thermal $H_2O$ exospheric size of $7.53\times10^{32}$ molecules of $H_2O$ (~22,500 metric tons of $H_2O$). Furthermore, the production of water group ions can be estimated by multiplying the global production rate by the photodestruction rate, or $3.35\times10^{26}$ water molecules destroyed into ions per second.

The Jovian system also contains numerous small, irregular satellites. Amalthea, one of the largest of these with a mean radius of only ~84 km, is nevertheless dwarfed by the Galilean moons. Because of its low escape speed nearly all released molecules should escape the body, with no exosphere-mediated volatile transport. As a result, Amalthea would be expected to consist of thick, devolatilized lag deposits with very low albedos, consistent with observations (*Veverka et al., 1981; Porco et al., 2003*). Jupiter's other irregular moons (such as Himalia) would follow similar patterns. Thus, exospheres would only be expected to be observed on the Galilean moons.

| | | Amalthea | Europa | Callisto (global) | Callisto (bright) | Ganymede | Himalia |
|---|---|---|---|---|---|---|---|
| | **Object Radius** | 83.5 km | 1561 km | 2410 km | | 2634 km | 79 km |
| | **Avg. Bond Albedo** | 0.015[a] | 0.68±0.05 [b] | 0.13[c] | 0.7[e] | 0.32 – 0.39[c] | 0.05±0.01[d] |
| | **Heliocentric distance (AU)** | 5.2 | | | | | |
| | **Peak Surface Temperature (K)** | 173 | 130 | 167[e] | 131† | 153-157 | 175 |
| | **Minimum Surface Temperature (K)** | 80 (assumed) | 50 | ~80[e] | ~50†† | 70 | 80 |
| | **Surface escape speed (m/s)** | 58* | 2025 | 2440 | | 2741 | ~84 |
| **$H_2O$** | **Fraction of molecules lost (ballistic escape, per hop)** | 99.98 % | $8.73\times10^{-13}$ % | $2.4\times10^{-14}$ % | $1.2\times10^{-18}$ % | $1.56\times10^{-20}$ % | 99.9 % |
| | **Fraction of molecules lost (photodestruction, per hop)** | $3.67\times10^{-6}$ % | 0.0229 % | 0.0275 % | 0.0240 % | 0.0222 % | $3.51\times10^{-3}$ % |
| | **Fraction of molecules lost (ballistic + photo, per hop)** | 99.98 % | 0.0229 % | 0.0275 % | 0.0240 % | 0.0222 % | 99.9 % |
| | **Expected number of hops** | $2.18\times10^{-4}$ | 4366 | 3635 | 4165 | 4504 | $9.32\times10^{-4}$ |
| | **Ballistic timescale** | 51 min | 513 s | 618 s | 539 | 499 s | 19.7 hr |
| | **Expected Exospheric Residence Time** | 0.67 s | 25.9 days | 26 days | 26 days | 26 days | 66 s |
| | **Ballistic hop distance** | 42.7 km | 149 km | 206 km | 161 km | 161 km | 130 km |
| | **Expected Migration distance (RMS)** | 0.63 km | 9845 km | 12,400 km | 10,400 km | 10,804 km | 3.96 km |
| | **Photodissociation timescale** | 28 days | | | | | |
| **$CO_2$** | **Fraction of molecules lost (ballistic loss, per hop)** | 99.6 % (99.9 %) | $5.49\times10^{-54}$ % ($4.82\times10^{-35}$ %) | $1.1\times10^{-72}$ % ($2.28\times10^{-38}$ %) | $4.0\times10^{-82}$ % ($1.49\times10^{-49}$ %) | $6.44\times10^{-96}$ % ($3.74\times10^{-54}$ %) | 98.37 % (99.4 %) |
| | **Fraction of molecules lost (photodestruction, per hop)** | $1.17\times10^{-5}$ % ($3.78\times10^{-6}$ %) | $1.83\times10^{-3}$ % ($2.26\times10^{-3}$ %) | $2.0\times10^{-3}$ % ($2.76\times10^{-3}$ %) | $1.9\times10^{-3}$ % ($2.43\times10^{-3}$ %) | $1.7\times10^{-3}$ % ($2.26\times10^{-3}$ %) | 0.0112 % (0.00402 %) |
| | **Fraction of molecules lost (ballistic + photo. Per hop)** | 99.6 % (99.9 %) | $1.83\times10^{-3}$ % ($2.26\times10^{-3}$ %) | $2.0\times10^{-3}$ % ($2.76\times10^{-3}$ %) | $1.9\times10^{-3}$ % ($2.43\times10^{-3}$ %) | $1.7\times10^{-3}$ % ($2.26\times10^{-3}$ %) | 98.38 % (99.4 %) |
| | **Expected number of hops** | 0.004 (0.001) | 54,600 (44,200) | 50,000 (36,200) | 52,700 (41,200) | 58,800 (44,300) | 0.0165 (0.00571) |
| | **Ballistic timescale** | 51 min (51 min) | 245 s (303 s) | 270 s (369 s) | 253 s (325 s) | 229 s (302 s) | 25.0 hrs (25.6 hrs) |
| | **Expected Exospheric Residence Time** | 12 s (4.1 s) | 155 days (155 days) | 156 days (155 days) | 155 days (155 days) | 156 days (155 days) | 24.7 mins (8.76 mins) |
| | **Ballistic hop distance** | 42.6 km (42.6 km) | 37.9 km (56.8 km) | 43.8 km (79.9 km) | 38.8 km (62.6 km) | 36.5 km (62.7 km) | 129 km (130 km) |
| | **Expected Migration distance (RMS)** | 2.7 km (1.55 km) | 8,800 km (11,900 km) | 9,800 km (15,200 km) | 8,900 km (12,700 km) | 8,800 km (13,200 km) | 16 km (9.8 km) |
| | **Photodissociation timescale** | 155 days | | | | | |

**Table 4**: *Comparison of thermally produced adsorption/desorption exospheric dynamics for $H_2O$ and $CO_2$ in the Jovian System.* Calculations for an infinite reservoir case of $CO_2$ (e.g., the entire illuminated surface can sublime) are listed in parentheses; all calculation for $H_2O$ are infinte reservoir case. These results reveal that the Galilean satellites are the only objects in the Jovian system capable of forming bound exospheres, and experiencing any significant exosphere-mediated volatile transport [a]*Veverka et al., 1981* [b]*Grundy et al., 2007* [c]*Squyres, 1980* [d]*Porco et al., 2003* [e]*Moore et al., 1999*

*escape speed assuming a sphere with equal mass and volume

†computed from bond albedo, assuming emissivity of 0.9 (*Spencer 1987a*)

††assumed from Europa, which has a similar albedo

Amalthea's leading hemisphere is ~26±10% brighter than the trailing hemisphere (*Simonelli et al., 2000*). This is thought to be due to micrometeorite impacts, which form Jupiter's Amalthea ring, with charged particles only playing a minor role (*Simonelli et al., 2000)*. Amalthea also has bright terrains in southern Gaea crater that is up to a factor of ~3 brighter than the rest of the moon (*Simonelli et al., 2000*). Because Amalthea cannot experience exosphere-driven migration of volatiles, this albedo variation is not akin to a systematic polar deposit. Instead, it is likely the result of the Gaea impact exposing volatile-rich materials, or some other process that preferentially brightens polar terrains.

### *3.3 The Saturnian System*

The Saturnian system is located near the outer distance limit where solar heat can mobilize water molecules such that albedo variations can lead to significant variations in the evolutionary behavior of $H_2O$ over Solar System timescales (*Spencer and Denk, 2012*). Darker terrains (lower albedos) are warmer, and therefore have shorter residence times, kicking off volatiles relatively more rapidly than lighter, higher albedo terrains. Because volatile frosts tend to produce higher albedos (and therefore longer residence times), surface ices preferentially condense on the bright terrains, creating a positive feedback loop that should congregate volatiles into distinct, well-defined terrains (*Spencer, 1987b*). Thus, Saturn's larger airless moons are prone to an inherent albedo-driven surface instability.

Iapetus, a two-toned world (*Cassini, 1677*), is a paradigm of this effect; Iapetus has a dark leading hemisphere (Cassini Regio) with a 0.04 bond albedo and a bright trailing hemisphere with a 0.39 bond albedo (*Spencer & Denk, 2010*). This results in residence times for water ice at the subsolar point of 173 hrs and 1.99 hrs for these light and dark terrains respectively, a two order of magnitude difference; water preferentially resides in the bright terrains. *Spencer and Denk (2010)* found that if dark exogenic material were to fall on an initially uniform Iapetus, sublimation dynamics would cause water to migrate away from this dark material, amplifying the albedo contrast. As dark, exogenic material preferentially deposits on Iapetus' leading hemisphere, the leading hemisphere darkens due to infall and lag buildup, while the trailing hemisphere becomes brighter due to frost formation over ~2.4 Gyr (*Spencer & Denk, 2010*).

Deposition of exogenic material also occurs on Iapetus' trailing hemisphere, albeit at lower rates. However, deposition of water from Iapetus' leading hemisphere covers this terrain, keeping it bright[3]. Without this deposition of materials, Iapetus' trailing hemisphere would experience a net loss of ice, and be susceptible to the same runaway sublimation processes that made the leading hemisphere dark. Therefore, if Iapetus' albedo dichotomy is stable, then the leading hemisphere must continue to feed water to the bright hemisphere. This suggests that Cassini Regio on Iapetus is not ice-free, but rather contains exposures of water ice.

In equilibrium, equal amounts of water are migrating between the bright and dark terrains. This allows us to us to use our model to determine the minimum fraction of the surface of Iapetus' leading hemisphere that is exposed water ice. For this first-order estimate, we assume that the bright and dark terrains each occupy one hemisphere, and assume the surface to be pure water but with each terrain's respective albedos. We calculate that, when illuminated, the bright and dark terrains sublime $6.29\times10^{-4}$ kg/s and 1.00 kg/s respectively, a difference of a factor of 1590; these sublimative fluxes are likely accurate to within a factor of two (see Appendix E). Our model therefore suggests that, if water is not subliming and diffusing through the dark materials on Iapetus' leading hemisphere, then on the order of ~0.06% of the dark terrain must consist of water ice exposures at the surface. Given that the dark surface material is only a few ~10s of cm thick (*Le Gall et al., 2019*; consistent with the impact gardening depth; *Spencer & Denk, 2010*), it is plausible that numerous small impact craters punch though this crust, creating this required exposure, but would be unresolved in *Cassini* images.

Since the formation of these albedo dichotomies is merely a result of water's albedo-dependent response in the saturnian system, such processes should also be expected around other bodies. For example, Hyperion, a small moon that shows similar surface ices and albedo contrasts to Iapetus (*Cruikshank et al., 2007; Howard et al., 2012*), has a heavily mottled appearance, with a bright surface containing crater-like pits with dark materials at the bottoms of these pits (*Thomas et al., 2007; Cruikshank et al., 2007*). These pits themselves are likely the result of $CO_2$ sublimation (*Howard et al. 2012*). Sublimation rates of $H_2O$ are two orders of magnitude greater on these dark pit bottoms (~11.5 m/Gyr) compared to the brighter terrains (~0.15 m/Gyr), suggesting that sublimation can accelerate the formation of a dark lag deposit at these crater bottoms (*Howard et al., 2012*). Thus, Hyperion faces a similar albedo-sublimation instability, in which craters provide a topographic pit that forms an ever darker bottom relative to the brighter surrounding terrains. Since such dark materials triggered the formation of Iapetus' two-toned appearance (*Spencer & Denk, 2010*), why then does Hyperion not show consolidated, dominant dark and bright terrains?

---

[3] *Spencer & Denk (2010)* consider the effects of impact gardening on mixing this dark material with underlying ice, which they estimate can mix the top ~10 cm over ~Gyr timescales. However, they estimate that sublimation dominates impact gardening (even on Iapetus' bright terrains) on timescales longer than ~100 Myr.

This difference is most likely due is the different dynamics of $H_2O$ in the exospheres of Iapetus and Hyperion. Our modeling finds that, whereas only 23.4% of water molecules leaving Iapetus' dark terrains escape the system (allowing a typical water molecule to travel around the body), 99.9% of water molecules escape from Hyperion; such loss rates preclude exosphere-mediated volatile transport. As a result, Hyperion cannot experience Iapetus' depositional brightening and formation of a single dominant bright hemisphere. Furthermore, Hyperion's chaotic rotation state (*Wisdom et al., 1984*) prevents deposition of exogenic materials from preferentially depositing on a single hemisphere, helping to prevent the formation of a single, consolidated dark terrain. Similarly, ice-rich (*Owen et al., 1999*) Phoebe is too small to experience exospheric-driven volatile migration. Thus, Phoebe likely experiences the same localized albedo evolution as Hyperion, which is qualitatively consistent with its observed dark surface containing small patches of brighter materials in recently exposed terrains (*Owen et al., 1999; Porco et al., 2005*).

| | | Rhea | Hyperion | Iapetus (bright) | Iapetus (dark) | Phoebe |
|---|---|---|---|---|---|---|
| | **Object Radius** | 764.3 km | 135 km | 734.5 km | 734.5 km | 106.5 km |
| | **Avg. Bond Albedo** | 0.42–0.48[a] | 0.05-0.33[b] | 0.39 [c] | 0.04[c] | 0.1[d] |
| | **Heliocentric distance (AU)** | 9.6 | | | | |
| | **Peak Surface Temperature (K)** | ~103[f] | 127[g] | 113[e] | 129[e] | 112 |
| | **Minimum Surface Temperature (K)** | ~49[f] | 58[g] | ~47[f] | ~45[f] | ~40[f] |
| | **Surface escape speed (m/s)** | 635 | 74* | 573 | 573 | 102* |
| **$H_2O$** | **Mass loss rate (kg/s)** | $2.18\times10^{-6}$ | 0.0174 | $6.29\times10^{-4}$ | 1.00 | $7.06\times10^{-6}$ |
| | **Fraction of molecules lost (ballistic, per hop)** | 7.2% | 99.9% | 17.2 % | 23.1 % | 99.5 % |
| | **Fraction of molecules lost (photo, per hop)** | 0.096 % | $6.96\times10^{-4}$ % | 0.31 % | 0.34 % | $8.55\times10^{-3}$ % |
| | **Fraction of molecules lost (ballistic + photo, per hop)** | 7.3 % | 99.9 % | 17.5 % | 23.4 % | 99.5 % |
| | **Expected number of hops** | 12.7 | 0.0011 | 4.70 | 3.27 | 0.00481 |
| | **Ballistic timescale** | 2.29 hrs | 14.9 hrs | 8.02 hrs | 9.28 hrs | 33.25 hrs |
| | **Expected Exospheric Residence Time** | 29.0 hrs | 58 s | 37.7 hrs | 30.3 hrs | 9.59 min |
| | **Ballistic hop distance** | 758 km | 218 km | 880 km | 925 km | 175 km |
| | **Expected Migration distance (RMS)** | 2,700 km | 7.17 km | 1,908 km | 1,672 km | 12.1 km |
| | **Photodissociation timescale** | ~97 days | | | | |
| **$CO_2$** | **Fraction of molecules lost (ballistic, per hop)** | 0.00192 % (0.0290 %) | 98.6 % (99.4 %) | 0.0116 % (0.334 %) | 0.0138 % (0.774 %) | 95.7 % (97.3 %) |
| | **Fraction of molecules lost (photo, per hop)** | 0.00385 % (0.00520 %) | $1.51\times10^{-3}$ % ($6.81\times10^{-4}$ %) | 0.00509 % (0.0110 %) | 0.00522 % (0.0150 %) | 0.0134 % ($8.64\times10^{-3}$ %) |
| | **Fraction of molecules lost (ballistic + photo, per hop)** | 0.00577 % (0.0342 %) | 98.6 % (99.4 %) | 0.0167 % (0.345 %) | 0.0190 % 0.789 %) | 95.7 % (97.3 %) |
| | **Expected number of hops** | 17,300 (2930) | 0.015 (0.00643) | 5990 (289) | 5262 (126) | 0.0448 (0.0278) |
| | **Ballistic timescale** | 29.3 min (39.6 min) | 14.7 hrs (14.8 hrs) | 38.7 min (1.40 hrs) | 39.7 min (1.92 hrs) | 39.1 hrs (40.0 hrs) |
| | **Expected Exospheric Residence Time** | 352 days (80.4 days) | 12.8 min (5.73 min) | 161 days (16.9 days) | 145 days (10.1 days) | 1.75 hrs (1.11 hrs) |
| | **Ballistic hop distance** | 232.9 km (318 km) | 216 km (217 km) | 273.4 km (440 km) | 279 km (507 km) | 173 km (174 km) |
| | **Expected Migration distance (RMS)** | 30,700 km (17,200 km) | 26 km (17.4 km) | 21,200 km (7,480 km) | 20,200 km (5,680 km) | 36.6 km (28.9 km) |
| | **Photodissociation timescale** | 528 days | | | | |

**Table 5**: *Comparison of thermally produced adsorption/desorption-drive exospheric dynamics for $H_2O$ and $CO_2$ in the Saturnian System.* Water is assumed to be infinite reservoir case. $CO_2$ is presented as finite reservoir case, with infinite reservoir results presented in parentheses. These four airless bodies in the Saturnian system show that the large satellites (Rhea, Iapetus) have escape speeds much greater than the speed of water molecules, allowing the retention of an $H_2O$ exosphere. The smaller moons (Hyperion and Phoebe) have escape speeds significantly less than the speed of water molecules, and consequently cannot retain an exosphere or experience exospheric volatile migration.

Refs: [a]*Pitman et al., 2010* [b]*Howard et al., 2012* [c]*Spencer and Denk, 2010* [d]*Flasar et al., 2005* [e]*Spencer & Denk 2009* [f]*Howett et al., 2010* [g]*Howard et al., 2012* [h] *Flasar et al. 2005*
†computed from bond albedo, assuming emissivity of 0.9
††assumed from average values of other moons

Saturn's synchronously rotating moons with sizes comparable to Iapetus (e.g., Dione, Tethys, and Rhea) should also be capable of forming dominant dark and bright hemispheres via exosphere-mediated volatile transport. Although no $H_2O$-rich exospheres have been detected in the Saturnian system, Rhea and Dione nevertheless possess exospheres composed of $CO_2$ and $O_2$ (*Teolis et al., 2010; Tokar et al., 2012; Teolis & Waite, 2016*), and all three of these moons exhibit leading/trailing albedo dichotomies, with Dione in particular having the largest such dichotomy in the Solar System outside of Iapetus (*Buratti et al., 1998; Spencer & Denk, 2010*). However, these albedo dichotomies are much less extreme than that of Iapetus. Whereas Iapetus' bright terrain has an albedo ~10 times higher than its dark terrains, Dione's brighter hemisphere is only 1.45 times brighter (*Buratti et al., 1998; Spencer & Denk, 2010*). Given that these inner mid-sized moons (represented by Rhea in Table 5) have comparable ballistic hop distances, exospheric residence times, and RMS hop distances, why then do they not exhibit an albedo dichotomy as stark as Iapetus?

Iapetus, Rhea, Dione, and Tethys all rotate synchronously with diurnal/orbital periods of 79 days, 4.5 days, 2.7 days, and 1.9 days respectively. Spencer and Denk (2010) suggest that these differences of at least a factor of 20 lead to dramatic differences in the temperature of the surface and near-surface. This is consistent with the results of Howett et al. (*2010*), which demonstrate the importance of subsurface heat conduction on surface temperature. Thus, it appears that inherently lower temperatures on these inner icy moons, and resulting reduction in sublimative mass flux, precludes the formation of albedo dichotomies over the age of the Solar System.

Indeed, our modeling of the saturnian system is consistent with this hypothesis. We find that the sublimative water flux from the *bright* (colder) terrain on Iapetus is nevertheless ~150 times greater than Rhea. Because the timescale of albedo dichotomy formation should be linearly proportional to the sublimative flux rate, an albedo dichotomy on Rhea that is comparable to Iapetus could only form over the age of the Solar system if the observed dichotomy on Iapetus formed in less than ~30 Myr. This is at odds with numerical simulations, which show dichotomy formation timescales that are on the order of ~100 – 1000 Myr (*Spencer and Denk, 2010*). Thus, if currently observed conditions of Rhea and Iapetus are representative of their long-term evolution, surface temperatures alone explain why Iapetus formed an albedo dichotomy while no other icy moon of Saturn did.

Nevertheless, if the currently observed structure and active processes are *not* representative of the long-term state of the Saturnian system (i.e., that the Saturnian system behaved differently in the past), then there are two alternative possibilities that we cannot rule out. The first is that the inner mid-sized moons may not have a large albedo dichotomy because they are themselves very young (~100 Myr). The modeling of *Spencer and Denk* (*2010*) suggest that, if Rhea and Dione were to experience the same exosphere-mediated transport of volatiles as Iapetus, they would require on the order of a few ~100 Myr to achieve their current leading-trailing albedo dichotomies. This is consistent with the ages of large craters on Dione, which are younger than 1.4 Gyr (*López-Oquendo et al., 2019*). Furthermore, these timescales are consistent with ring composition arguments, which suggest that these moons formed from a disrupted Titan-sized moon on the order of ~100 Myr ago (*Canup, 2010*), and dynamical and tidal arguments, which suggest that these inner moons are ~100 Myr old (*Ćuk et al., 2016*). Additionally, studies of tidal dissipation (*Lainey et al., 2017*) and viscous spreading of rings (*Charnoz et al., 2011*) suggest young ages for these moons.

Alternatively, we propose the possibility that Rhea, Dione, and Tethys may have indeed once formed Iapetus-like albedo dichotomies that were later erased by resurfacing and the resulting changes in surface thermal properties. Whereas Iapetus is experiencing deposition of dark, devolatilized dust, the inner mid-sized moons orbit within Saturn's E-ring, which is composed of very pure water ice grains ejected from the Tiger Stripes of Enceladus (*Hedman et al., 2012*). The heat driving the activity in the Tiger Stripes is thought to be episodic or recent (*O'Neill & Nimmo, 2010; Spencer & Nimmo, 2013; Nakajima et al. 2019*); implying that the E-ring itself is also recent or episodically active. The grains ejected into the E-ring have very high albedos, producing very bright, pure surfaces (*Buratti et al., 1998*) upon deposition, which could prevent albedo dichotomy formation, with sputtering processes on their trailing hemispheres darkening their trailing hemispheres (*Schenk et al., 2011*). However, the spectra of the darker materials on Dione resembles the dark materials on Iapetus (*Clark et al., 2008*), suggesting that these moons experienced the deposition of comparable exogenic materials as Iapetus. Furthermore, the *dark* trailing hemispheres of Rhea and Dione have bond albedos of 0.42±0.10 and 0.37±0.08 respectively (*Pitman et al., 2010*), which are comparable to the bond albedo of the *bright* trailing hemisphere of Iapetus (*Spencer and Denk, 2010*). Thus, any albedo dichotomies similar to Iapetus that formed on these inner icy moons would have been recently erased by the onset of Enceladus' plume activity and the resulting formation of the E-ring. This alternative is consistent with the young ~1-100 Myr age of Enceladus' active Tiger Stripe terrain (*Porco et al., 2006*) and the possible ~100 Myr age of Enceladus itself (*Ćuk et al., 2016*).

### *3.4 Carbon Dioxide ($CO_2$) Exospheres in the Uranian System*

Significant, optically thick deposits of $CO_2$ have been discovered on the large moons of Uranus: Ariel, Umbriel, Titania, and Oberon (*Grundy et al., 2003; 2006; Cartwright et al., 2015*). This $CO_2$ is not a monolayer, but rather a relatively pure deposit of $CO_2$ ice (*Grundy et al., 2006*). Our results suggest that $CO_2$ is moderately volatile in the Uranian system, and that these pure deposits of $CO_2$ are consistent with exosphere-mediated volatile transport. Because the Uranian system has high rotation pole obliquities near 90 degrees, seasonal volatile transport is dominated by volatile migration from the summer hemisphere to the winter hemisphere. We can estimate the maximum thickness ($l_{frost}$) of the $CO_2$ frost layer than can be mobilized from one hemisphere to the other during a seasonal cycle by multiplying the sublimative flux to the exosphere ($\Phi_{sub}$) by the number of hops required for a particle to migrate by some characteristic distance (i.e., the radius of the moon) as given by equation 18, divided by the density of the ice ($\rho_{ice}$). Because the number of expected hops (or equivalently, the expected migration distance compared to the size of the moon) is so large, we can neglect the small fraction of molecules that would be lost during this migration process. Thus

$$l_{frost} = \frac{\Phi_{sub}}{\rho_{ice}} \left(\frac{d_{hop}}{R_{moon}}\right)^2 \quad (19)$$

We find that the depth of the $CO_2$ frost layer that can migrate during a single seasonal cycle is only ~2.4 – 6.4 mm thick on Umbriel, Ariel, Titania and Oberon; these results are consistent with the frost layer being optically thick (e.g., *Grundy et al. 2003; 2006*).

Given that $CO_2$ has an exospheric residence on these moons much shorter than a seasonal cycle (see Table 6), this suggests either a steady supply of exogenous $CO_2$ to the Uranian moons (e.g. sputtering or impacts), or a high initial abundance of $CO_2$. We can use our results to explore if an endogeneous origin of $CO_2$ in the Uranian system is reasonable. If we integrate over the surface to see how much material can sublimatively escape the large Uranian moons, we find that an average of 112 kg/m$^2$ of $CO_2$ ice would sublime from Umbriel, which is comparable to the flux of ~85 kg/m$^2$ that results from integrating under the curve in figure 6 from *Sori et al. (2017)*. Multiplying this number by the fraction of molecules lost per hop (Table 6), and dividing by the density of $CO_2$ ice (~1600 kg/m$^3$), we find that a layer of $CO_2$ 160 microns thick could be lost from the surface of Umbriel per orbit. Thus, over the age of the solar system, Umbriel would have lost at most a global layer of $CO_2$ 8700 m thick. Similarly, Ariel, Oberon, and Titania would have lost, respectively, a layer of $CO_2$ 310 m, 51 m, and 23 m thick over the age of the solar system; these differences are attributable to their different surface gravities (i.e. escape speeds) and surface temperatures (Table 6).

If we assume that the bulk compositions of these moons are comparable to comet nuclei, these are plausible amounts. $CO_2$ is the second most abundant cometary volatile ice, with a typical abundance relative to water on the order of ~10% (*Bockelée-Morvan et al. 2004*). Furthermore, the ice/rock ratio of comet nuclei is on the order of unity (*Choukroun et al. 2020*). Thus, when considering the amount of material required to source these lost quantities of pure $CO_2$, Umbriel, Ariel, Oberon, and Titania would require (respectively) up to ~87 km, ~3 km, ~0.5, and ~0.2 of the topmost material to have become devolatilized in $CO_2$. However, if even deeper layers of material have been sufficiently thermally processed to release $CO_2$, then $CO_2$ can nevertheless build up on the surface, consistent with observations. If we follow the thermal processing model of Steckloff et al. (*2021*) and assume thermal diffusivities of ~$10^{-8}$ m$^2$/s (comet-like) to ~$10^{-6}$ m$^2$/s (rock-like), we estimate that, over the age of the Solar System, on the order of the topmost 40 – 400 km of material would have been sufficiently thermally processed to reach the average diurnal temperature, allowing its $CO_2$ to be released to the exosphere. Thus, sufficient material on these moons may have been sufficiently thermally processed for the observed $CO_2$ in the Uranian system to have been endogenously sourced.

Alternatively, the observed $CO_2$ on the Uranian moons could be exogenic. The external impactor flux into the Uranian system is dominated by icy centaurs and ecliptic comets (*Shoemaker & Wolfe, 1982; Zahnle et al., 1998; Zahnle et al., 2003*), which contain abundant $H_2O$ and $CO_2$ (*Bockelée-Morvan et al., 2004*). Such impacts are quite rare (*Safrit et al. 2021*); using the results of Zahnle et al. (*2003*) and assuming a typical crater-to-impactor diameter ratio of 20, impacts into the large Uranian moons would deliver, on average on the order of ~20-100 m$^3$ of material per year. However, our results suggest that these moons could lose on the order of ~40,000 – 8 million m$^3$ of $CO_2$ ice per year. Furthermore, larger moons would generally collect slightly more exogenic $CO_2$ due to gravitational focusing. However, the observed trend is the opposite; the largest moons Titania and Oberon have significantly less $CO_2$ than the smaller moons Ariel and Umbriel (*Grundy et al., 2003; 2006; Cartwright et al., 2015*). Together, this suggests that impacts are unlikely to be a significant contributor to the observed $CO_2$ in the Uranian system.

Magnetospheric delivery of carbon to these moons and the resulting radiolytic sputtering, which can react to produce $CO_2$ (*Grundy et al., 2006*), naturally explains why the average integrated band depth associated with $CO_2$ decreases with increasing orbital radius of these moons (*Grundy et al., 2003; 2006; Cartwright et al., 2015*). Comparing the $CO_2$ loss rate from each moon (average sublimative flux integrated over the surface times the fraction of molecules lost via ballistic loss and photodestruction) provides an order-of-magnitude estimate of the relative $CO_2$ delivery flux required to maintain approximately equal $CO_2$ concentrations among these moons. From this, we find that Ariel and Umbriel require ~13× and ~370× greater $CO_2$ flux than Titania (Oberon requires ~2x greater flux than Titania). Given that Ariel and Umbriel are observed to have *greater* $CO_2$ concentrations (*Grundy et al., 2003; 2006; Cartwright et al., 2015*), these ratios are approximate lower bounds to the relative delivery fluxes. Nevertheless, this suggests that the $CO_2$ delivery among Uranus's major moons (if the observed $CO_2$ is exogeneous) decreases with increasing distance from the planet (otherwise, Ariel and Umbriel would have little detectable $CO_2$ relative to Titania and Oberon), which is consistent with a radiolytic sputtering origin (*Grundy et al. 2006*). Thus, our model is also consistent with a sputtering origin for the $CO_2$ in the Uranian system. Although, Voyager 2's non-detection of carbon ions in the Uranian magnetosphere (*Bridge et al., 1986; Krimigis et al., 1986*) is inconsistent with this mechanism, the Voyager 2 observations were not designed to detect $CO_2$. Ultimately, further study is needed to quantify the $CO_2$ sputtering flux in the Uranian system, which could determine whether an endogenous or sputtering $CO_2$ source is more plausible.

The Uranian system of moons lies in a region where residence times for $CO_2$ are comparable to rotation periods, and where surface albedo can profoundly affect $CO_2$ residence time (comparable to the Saturnian system for $H_2O$). Indeed, an Iapetus-like albedo dichotomy is diurnally stable so long as the bright terrain has a bond albedo of at least ~65-67%; which have been observed on other icy satellites such as Tethys (67 ± 11 %), Enceladus (81±4%; *Howett et al., 2010*), and Triton ($65^{+15}_{-10}$ %, though its atmosphere is not rarefied; *Nelson et al., 1990*). Thus, the moons of Uranus should form $CO_2$ exospheres from the $CO_2$ ice found on their surfaces (*Grundy et al., 2003, 2006; Cartwright et al., 2015*) that can experience Iapetus-like volatile migration and albedo dichotomies. Preferential impacts of exogenic material on these moons' leading hemispheres (*Zahnle et al., 2001*) deposit both water ice (*Grundy et al., 2006*) and red organic materials (*Cartwright et al., 2018*), which would largely stay in place due to the extremely cold temperatures, which inhibit their mobilization. Thus, exogenic impacts, and/or magnetic field impingement form natural processes that would break the symmetry of the system and favor the formation of leading/trailing albedo dichotomies. However, Voyager 2 (*Smith et al., 1986*) and earth-based observations (*e.g.*, *Karkoschka, 2001*) observed no such Iapetus-like geographic albedo trends on any large Uranian moon.

The likely reason for this appears to be the very high polar obliquity of the Uranian system, with the axis of rotation lying nearly in the plane of its orbit. As a result, Uranus's moons experience extreme seasonal heating variations that dominate the diurnal variations (*e.g., Sori et al., 2017*). We invert the RMS hop distance calculation (equation 18) to compute the number of hops required to reach each moon's antipode, and multiply by the ballistic hop timescale to estimate the time required to migrate this far. For Puck and Miranda, these timescales are 13 and 14 hrs, respectively, which is roughly an order of magnitude longer than their respective exospheric residence times. Thus, Puck and Miranda would lose their exospheric molecules before much migration could occur, leading to $CO_2$-free surfaces. However, $CO_2$ molecules on Ariel, Umbriel, Titania, and Oberon require timescales of roughly 24, 28, 71, and 59 hours, respectively, to migrate to the antipode of their initial position. These are more than an order of magnitude less than their exospheric residence times, and also shorter than each moon's rotation period. Thus, during the equinoxes (when the Uranian system experiences more typical diurnal day/night cycles), $CO_2$ molecules can migrate about each moon's surface in a single diurnal cycle; a leading/trailing dichotomy could set itself up over time.

In fact, the IRTF SpeX instrument detected a small, statistically significant leading/trailing hemisphere geometric albedo dichotomies on Ariel and Umbriel, with the trailing hemisphere brighter (and therefore cooler) than the leading hemispheres (*Cartwright et al. 2015*). This coincides with a significant leading/trailing asymmetry in their surface $CO_2$ distributions, with higher concentrations on their trailing hemispheres (*Grundy et al., 2003; 2006; Cartwright et al., 2015*), consistent with exosphere-mediated volatile transport to the colder hemisphere. The mechanism breaking the leading-trailing degeneracy is presently unknown, but may be due to preferential deposition of darker materials on their leading hemisphere and/or magnetospheric deposition/creation of bright materials through sputtering on their trailing hemispheres. Regardless, despite Titania and Oberon exhibiting no significant geometric albedo dichotomy, they too exhibit similar (albeit smaller) leading/trailing asymmetries in their surface $CO_2$ distributions (*Cartwright et al. 2015*).

Such a leading/trailing asymmetry appears inconsistent with our previous analysis, given that such distributions would be expected to be erased around the summer solstice. However, it is plausible that the observed leading/trailing asymmetry in the distribution of $CO_2$ on the Uranian moons is a seasonal effect that forms around the equinoxes, and which may not be fully erased during the solstices. Indeed, the observations that detected this distribution were made in 2002 (*Grundy et al., 2003*), 2001 – 2005 (*Grundy et al., 2006*), and 2000 – 2013 (*Cartwright et al., 2015*), which fall within the equivalent of one Uranus "month" of the 2007 equinox. The intense heating near the summer solstice (which lasts on the order of a decade) would rapidly cause any $CO_2$ on the illuminated hemispheres to migrate to the winter hemisphere, effectively erasing any buildup of $CO_2$ on a leading/trailing hemisphere that occurred since the previous summer. Thus, this asymmetry may begin to change as Uranus approaches its solstice in 2028, when pole-to-pole transport could erase this distribution. However, this leading/trailing asymmetry would persist in the winter hemisphere until summer, and possibly even strengthen (depending on mechanism). Indeed, these moons of Uranus may always exhibit some level of leading/trailing asymmetry in $CO_2$ whose amplitude varies seasonally, even if the dominant mode of transport is always from the summer to the winter hemisphere. Alternatively, if the $CO_2$ deposits are thicker than what can be mobilized during a seasonal cycle, the low latitude regions of these moons may be able to build up a permanent $CO_2$ deposit, as the high pole obliquity of these moons make these low latitude regions secular cold traps (*Sori et al. 2018*). Although the winter hemisphere is effectively unobservable from Earth, future observations may be able to investigate this migration looking at abundances as the seasons change.

It may also be possible to form a polar (e.g., north/south) albedo dichotomy on these moons due to seasonal, pole-to-pole volatile migration. Similar seasonal pole-to-pole volatile migration has been proposed to occur within Triton's collisional atmosphere (*Trafton, 1984*), preventing the formation of longitudinal albedo features. Forming an albedo dichotomy would require some process to break the symmetry of the system and cause material to start to preferentially sublime from one terrain and condense on another; no such mechanism is currently known to create a polar albedo dichotomy. For leading/trailing dichotomies, preferential deposition of exogenic material on the leading hemisphere and magnetic field impingement can break this symmetry. Similarly, impingement of particles trapped in planetary magnetic fields can preferentially strike one hemisphere if the magnetic dipole moment rotation vector and orbital angular moment vector of the moon align, darkening the surface. Uranus' magnetic field diverges significantly from this ideal geometry; its magnetic field dipole moment vector is offset from the center of the planet by 0.3 Uranus radii, and differs from the spin axis vector by 58.6° (*Ness et al., 1986; Connerney et al., 1987*). The exact effects of sputtering by particles trapped in the magnetic field on the surfaces of these moons, and whether this is able to break the north/south symmetry of the system, is currently poorly understood. Ultimately, further Earth-based or in situ observations of these moons are needed about its seasonal cycle, to understand the delivery, evolution, and migration of $CO_2$ in the Uranian system.

| | **Puck** | **Miranda** | **Ariel** | **Umbriel** | **Titania** | **Oberon** |
|---|---|---|---|---|---|---|
| **Object Radius** | 81 km | 236 km | 579 km | 585 km | 788 km | 761 km |
| **Surface escape speed (m/s)** | 69 | 193 | 559 | 520 | 773 | 727 |
| **Avg. Bond Albedo** | 0.035 ±0.006[a] | 0.24 ±0.06[b] | 0.230 ±0.025[a] | 0.100 ±0.010[a] | 0.170 ±0.015[a] | 0.140 ±0.015[a] |
| **Heliocentric distance (AU)** | 19.2 | | | | | |
| **Peak Surface Temperature (K)** | 92* | 86±1[b] | 84±1[b] | 88[c] | 88* | 89* |
| **Minimum Surface Temperature (K)** | 22† | 22† | 22† | 22[c]† | 22† | 22† |
| **Fraction of molecules lost (ballistic loss, per hop)** | 99.1 % | 67.3 % | 0.0479 % | 0.230 % | $2.13\times10^{-5}$ % | $1.85\times10^{-4}$ % |
| **Fraction of molecules lost (photodestruction, per hop)** | $8.92\times10^{-6}$ % | $8.97\times10^{-4}$ % | $1.19\times10^{-3}$ % | $1.76\times10^{-3}$ % | $6.05\times10^{-4}$ % | $7.01\times10^{-4}$ % |
| **Fraction of molecules lost (ballistic + photo, per hop)** | 99.1% | 67.3 % | 0.0491 % | 0.231% | $6.26\times10^{-4}$ % | $8.86\times10^{-4}$ % |
| **Expected number of hops** | 0.0091 | 0.49 | 2040 | 431 | 160,000 | 113,000 |
| **Ballistic timescale** | 1.45 hr | 2.23 hr | 36.2 min | 54.0 min | 18.4 min | 21.3 min |
| **Expected Exospheric Residence Time** | 47 sec | 1.09 hr | 51.3 days | 16.2 days | 5.59 years | 4.6 years |
| **Ballistic hop distance** | 86 km | 291 km | 258 km | 331 km | 163 km | 186 km |
| **Expected Migration distance (RMS)** | 8.19 km | 203 km | 11,600 km | 6880 km | 65,200 km | 62,500 km |
| **Avg sublimative flux to exosphere per Uranus Orbit (kg/m^2)** | 568 | 47.2 | 19.1 | 112 | 112 | 171 |
| **Thickness of CO2 slab that could have been lost over age of SS** | 81 km (size limited) | 236 km (size limited) | 310 m | 8700 m | 23 m | 51 m |
| **Estimated mobile seasonal frost layer thickness (mm)** | – | – | 2.4 | 22 | 3.0 | 6.4 |
| **Photodissociation timescale** | 5.8 yrs | | | | | |

**Table 6**: *Comparison thermally driven adsorption/desorption-drive exospheric dynamics for $CO_2$ in the Uranian System.* [a]*Karkoschka, 2001* [b]*Hanel et al., 1986* [c]*Sori et al., 2017*

*Peak surface temperature computed from radiative equilibrium using bond albedo

†mimimum temperature found to have minimal impact on computed values for $CO_2$, thus assumed to be seasonally uniform across moons

††assumed from average values of other moons

*escape speed assuming a sphere with equal mass and volume

## 4. Conclusions

We constructed and validated an efficient model of thermally produced, sublimation exospheres to investigate exosphere-mediated volatile migration in the Solar System. We use this model to estimate the general shape, structure, and dynamics of sublimating exospheric gases on airless bodies. We found that this model suggests that Mercury's high photodestruction rate and short ballistic hop distances make it more difficult for water to migrate to the polar cold traps. This is consistent with previous works, which argue that the water in Mercury's polar cold traps must have either been delivered in impacts close to the cold traps (*Butler, 1997*) and/or delivered in a single large impact (*Ernst et al., 2018; Deutsch et al., 2019*) at relatively low speed (Ernst et al. 2018). We suggest that such large impacts produced an exosphere thick enough to self-shield against photodestruction; such self-shielding would relax these impact parameter restrictions, allowing for faster, more typical impact speeds and/or lower impact latitudes to more easily populate Mercury's cold traps with water ice.

Our results suggest that Callisto's detected $CO_2$ exosphere (*Carlson et al. 1999*) is unlikely to be the result of impact delivery of $CO_2$, but rather could be maintained by ~7 hectares of exposed $CO_2$ ice spread about the surface of Callisto; such exposures could be maintained by mass wasting. Our model finds that the bright elevations and spires of Callisto are not deposits of $CO_2$ ices, but rather are likely $H_2O$ frosts that are deposited by the $H_2O$ exosphere of Callisto. We find that the detected $CO_2$ in these bright terrains (*Hibbitts et al. 2002*) is consistent with thermodynamic equilibrium with the $CO_2$ exosphere, rather than primary deposits of $CO_2$ themselves. We use our model to predict the locations of peak density of Callisto's $CO_2$ exosphere, which could be tested by the upcoming ESA JUICE mission.

Our model also finds that Iapetus' two-toned appearance likely requires that its dark Cassini Regio have exposures of underlying water ice that cover ~0.06% of its surface, likely through small, unresolved impact craters. Although current conditions preclude Saturn's inner mid-sized moons from developing similar albedo dichotomies, we propose that, if current conditions are not representative of these moons' long term evolution, then they could have previously formed albedo dichotomies that were later erased by Enceladus' activity and the formation of the E-ring, which is presently resurfacing these moons. Similarly, $CO_2$ in the Uranian system has a similar response to surface albedo as $H_2O$ in the Saturnian system.

Although no Iapetus-like albedo dichotomies are found on the Uranian moons, the leading/trailing dichotomy of $CO_2$ deposits on Ariel, Umbriel, Titania, and Oberon (*Grundy et al., 2003, 2006; Cartwright et al., 2015*) are consistent with exosphere-mediated volatile transport, and may be seasonal features that form during the equinoxes, but are largely erased during the solstices. We find that ~2.4 – 6.4 mm of $CO_2$ frost can be mobilized on these moons in a seasonal cycle; our model is consistent with an endogenous origin or magnetospheric delivery of this $CO_2$, and inconsistent with impact delivery. However, albedo dichotomies that form on Uranus' moons are likely erased by seasonal pole-to-pole volatile transport, a direct result of the high spin pole obliquity in the Uranian system; the observed leading/trailing $CO_2$ asymmetry on these moons may form during the equinox, only to be erased in Summer.

## Acknowledgements

We wish to thank Will Grundy, who provided constructive discussions and suggestions that improved the quality of this work. We also wish to acknowledge the Texas Advanced Computing Center (TACC), who hosts, maintains, and supports the Stampede2 supercomputer, with which we ran the PLANET DSMC code to validate our numerical model. This research project was funded by NASA award 80NSSC17K0725. J.K.S. was also partially supported by NASA award 80NSSC19K0556.

## Appendix A: Computing Surface Temperature and Desorption Longitude

We compute a scaled desorption time by normalizing the time in sunlight by the residence time, where the residence time is also a function of time because surface temperature (see equations 2 and 3) varies during the day:

$$\hat{t}_{desorb\,(\theta,\phi)} = \int_0^t \frac{dt}{\tau_{res(\theta,\phi)}(t)}, \tag{A-1}$$

where

$$\tau_{res(\theta,\phi)}(t) = \tau_0 \exp\left(\frac{u_o}{k_B T_{(\theta,\phi)}(t)}\right). \tag{A-2}$$

A characteristic time $t^*_{(\theta)}$ is determined by the implicit equation

$$\hat{t}_{desorb(\theta,\phi)} = 1 = \int_0^{t^*_{(\theta)}} \frac{dt}{\tau_{res(\theta,\phi)}(t)}. \tag{A-3}$$

The corresponding desorption longitude $\Delta\phi_{desorb(\theta)}$ is determined by

$$\Delta\phi_{desorb(\theta)} = 360° \frac{t^*_{(\theta)}}{P_{rot}} \tag{A-4}$$

and represents the angular displacement from the current dawn terminator at which most desorption occurs.

## Appendix B: Computation of Fraction of Molecules Lost

We integrate over the portion of the surface emitting molecules (assume nightside to have negligible sublimation flux) to compute the fraction of molecules lost from across the illuminated hemisphere of the body. For the icy surface (infinite reservoir) case, we integrate over the entire illuminated surface of the object. Because warmer surfaces have greater sublimation mass fluxes and contribute more molecules to the exosphere, we weight our integration over the surface by the sublimation mass flux of each surface element. Thus,

$$f_{lost_global} = \frac{\dot{m}_{ejected}}{\dot{m}_{total}} = \iint \frac{\dot{m}_{(\theta,\phi)}}{\dot{m}_{total}} f_{lost\,(\theta,\phi)} \cos\theta \, d\theta \, d\phi \tag{A-5}$$

$$= \iint \frac{\dot{m}_{(\theta,\phi)}}{\dot{m}_{total}} \left(1 + \beta^2_{(\theta,\phi)} v_0^2\right) e^{-\beta^2_{(\theta,\phi)} v_0^2} \cos\theta \, d\theta \, d\phi. \tag{A-6}$$

We integrate this function numerically to compute the fraction of ejected molecules that are directly ejected on escape trajectories, along with the ejected mass flux. The unilluminated sides of airless bodies are sufficiently cold that a negligible fraction of condensed molecules on the surface sublimate into the exosphere, allowing such contributions to be neglected.

For the refractory surface (finite reservoir) case, we consider the temperatures along the line of the morning exosphere. Because adsorbed/condensed volatiles are not present in an infinitely thick layer (but rather in a thin, possibly monolayer), we do not normalize by mass flux, resulting in an expression that computes the global fraction of molecules that are directly ejected to space along the path of the morning exosphere

$$f_{lost_global} = \iint f_{lost\,(\theta,\phi)} \cos\theta \, d\theta \, d\phi \tag{A-7}$$

$$= \iint \left(1 + \beta^2_{(\theta,\phi)} v_0^2\right) e^{-\beta^2_{(\theta,\phi)} v_0^2} \cos\theta \, d\theta \, d\phi \tag{A-8}$$

where $\phi$ and $\theta$ are the set of latitudes and longitudes corresponding to location of the morning exosphere. This expression assumes that condensed/adsorbed molecules are uniformly distributed in latitude at the morning terminator, and emitted only at the location of the morning exosphere (itself a function of latitude). Since this exosphere forms a generally linear, banana-like feature, this can be expressed as a line integral, such that

$$f_{lost_global} = \int f_{lost\,(\ell)} d\ell \tag{A-9}$$

$$= \int \left(1 + \beta_{(\ell)}^2 v_0^2\right) e^{-\beta_{(\ell)}^2 v_0^2}\, d\ell \qquad \text{(A-10)}$$

where $\ell$ is the line of the morning terminator exosphere.

For cases of the refractory surface (finite reservoir) in which the volatile does not immediately sublimate during the day (the integrated residence time is not much shorter than a diurnal cycle), the resulting exosphere is tenuous, more widely distributed, and not banana-shaped, with molecules ejected from the illuminated hemisphere in a manner that depends on surface temperature. This case best resembles the icy surface (infinite reservoir) case, without weighting by mass flux due to the limited amount of volatile present on the surface, but with weighting related to the inverse of residence time

$$f_{lost_global} = \iint g_{(\theta,\phi)} f_{lost\,(\theta,\phi)} \cos\theta\, d\theta\, d\phi \qquad \text{(A-11)}$$

where $g_{(\theta,\phi)}$ is a weighing factor that accounts for the differences in residence time (i.e., probability of ejection) across the illuminated surface. The factor $g_{(\theta,\phi)}$ weights each part of the surface based on the inverse of its residence time, and normalizes the entire integral by the sum of this weighting factors ($G_{res}$)

$$f_{lost_global} = G_{res} \iint \frac{1}{\tau_{res\,(\theta,\phi)}} f_{lost\,(\theta,\phi)} \cos\theta\, d\theta\, d\phi \qquad \text{(A-12)}$$

where

$$G_{res} = \iint \tau_{res\,(\theta,\phi)} \cos\theta\, d\theta\, d\phi. \qquad \text{(A-13)}$$

Thus

$$f_{lost_global} = \iint \frac{G_{res}}{\tau_{res\,(\theta,\phi)}} \left(1 + \beta_{(\theta,\phi)}^2 v_0^2\right) e^{-\beta_{(\theta,\phi)}^2 v_0^2} \cos\theta\, d\theta\, d\phi. \qquad \text{(A-14)}$$

**Appendix C: Computing Ballistic Timescales From Velocity with Celestial Mechanics**

We use celestial mechanics to compute the duration of a ballistic hop, using relationships between the mean anomaly ($M$), eccentric anomaly ($E$), and true anomaly ($\upsilon$) to determine radial position ($r$) as a function of time ($t$):

$$M = n(t - t_p) \qquad \text{(A-15)}$$

$$E - e \sin \mathrm{E} = M \qquad \text{(A-16)}$$

$$\cos \upsilon = \frac{\cos E - e}{1 - e \cos E} \qquad \text{(A-17)}$$

$$r = \frac{a(1-e^2)}{1 + e \cos \upsilon} \qquad \text{(A-18)}$$

Where $t_p$ is the time of pericenter passage of the orbit, and mean orbital angular velocity ($n$) depends on orbital period ($P$):

$$n = \frac{2\pi}{P} \qquad \text{(A-19)}$$

(*Fitzpatrick, 2012*). As described in Kepler's Third Law, the orbital period itself depends only on the semi-major axis of the orbit ($a$) and mass ($M$) of the primary body:

$$P = 2\pi \sqrt{\frac{a^3}{GM}}. \qquad \text{(A-20)}$$

The duration of a ballistic hop is therefore the time elapsed between instances where the orbital radius ($r$) is equal to the radius of the object (or using the symmetry of the problem to compute half the ballistic hop duration as the time between apoapse and the instance where $r$ is equal to the object radius). Thus, the ballistic hop duration can be computed by numerically solving a system of equations that depends only on the mass ($M$) and radius ($R$) of the body, and the semi-major axis ($a$) and eccentricity ($e$) of the ejected molecule's orbit.

The semi-major axis and eccentricity themselves depend on the ejection speed ($v$) and ejection angle ($\gamma$; measured with respect to the surface normal direction) of a sublimated molecule. From the *vis-viva* equation of celestial mechanics

$$v^2 = GM\left(\frac{2}{r} - \frac{1}{a}\right) \quad \text{(A-21)}$$

we solve the equation for semimajor axis and plug in the conditions at launch

$$a = \frac{GMR}{2GM - Rv^2}. \quad \text{(A-22)}$$

The orbital eccentricity depends on the specific orbital energy ($\epsilon$) and specific orbital angular momentum ($h$) of the ejected molecule:

$$e = \sqrt{1 + \frac{2\epsilon h^2}{(GM)^2}} \quad \text{(A-23)}$$

$$\epsilon = -\frac{GM}{2a} \quad \text{(A-24)}$$

$$h = |\vec{r} \times \vec{v}| = Rv \sin\gamma. \quad \text{(A-25)}$$

Thus, a molecule's ejection velocity completely determines the duration of its ballistic flight.

As noted above, the equilibrium speed distribution function of molecules leaving a surface at a specified temperature $T$ is

$$F_{(v)}dv = \frac{1}{2}\left(\frac{m}{k_B T}\right)^2 v^3 e^{-\frac{mv^2}{2k_B T}} dv\ . \quad \text{(A-26)}$$

Any molecules moving faster than the speed sufficient to reach the Hill radius never return to the surface. The ejected molecules follow a Lambertian distribution, in which the probability of a molecule being ejected in a particular direction is proportional to the cosine of the angle with respect to the normal direction. We can describe this profile by breaking the velocity into components and considering their distributions. We use spherical coordinates, in which we describe the component in the normal direction ($v_{\text{norm}}$) and its associated component distribution ($f_{\text{norm}}$), and integrate over azimuthal and polar directions to describe the tangential component of speed ($v_{\text{tan}}$) and its associated component distribution ($f_{\text{tan}}$)

$$f_{\text{norm}\,(v_{\text{norm}})} = 2\beta^2 v_{\text{norm}} e^{-\beta^2 v_{\text{norm}}^2} \quad \text{(A-27)}$$

$$f_{\tan\,(v_{\tan})} = 2\beta^2 v_{\tan} e^{-\beta^2 v_{\tan}^2}. \quad \text{(A-28)}$$

where $\beta$ is the inverse of a characteristic molecular speed

$$\beta = \sqrt{\frac{m}{2k_B T}}, \quad \text{(A-29)}$$

From these component distributions, we compute the ejection angle ($\gamma$) and speed ($v$)

$$v = \sqrt{v_{\text{norm}}^2 + v_{\tan}^2} \quad \text{(A-30)}$$

$$\gamma = \text{atan}\left(\frac{v_{\tan}}{v_{\text{norm}}}\right). \quad \text{(A-31)}$$

**Appendix D: Validating Numerical Approach**

We implemented our analytic scheme in a MATLAB script, and used the PLANET DSMC code to validate this approach. PLANET is a fully 3D Direct Simulation Monte Carlo (DSMC) code, which can accurately model the exospheric behavior of water ice condensed on airless bodies, accounting for ballistic loss and photodestruction (*Prem et al., 2018; 2018; Prem 2017*). For the purposes of validating our approach, we turned off molecular collisions in PLANET DSMC, to allow for free molecular flow, and also turned off non-inertial forces (e.g.,

Coriolis forces), which allow PLANET to match the assumptions of our analytical model. Later, we explore whether turning off non-inertial forces introduces errors into our calculations.

For our validation, we restricted our comparison to large objects for which water readily forms an exosphere, which PLANET DSMC can readily handle without modification. This limits us to four objects, all in the inner Solar System: Mercury, the Moon, Ceres, and Vesta. For these objects, we coat their unilluminated surfaces with a uniform layer of adsorbed water, and start the simulation to record where these molecules desorb, where they land (if they land), and how long each molecule's ballistic flight lasted. We than compare these DSMC-derived values (longitude of desorption, ballistic hop distance, and ballistic hop duration) with comparable quantities from the analytic model.

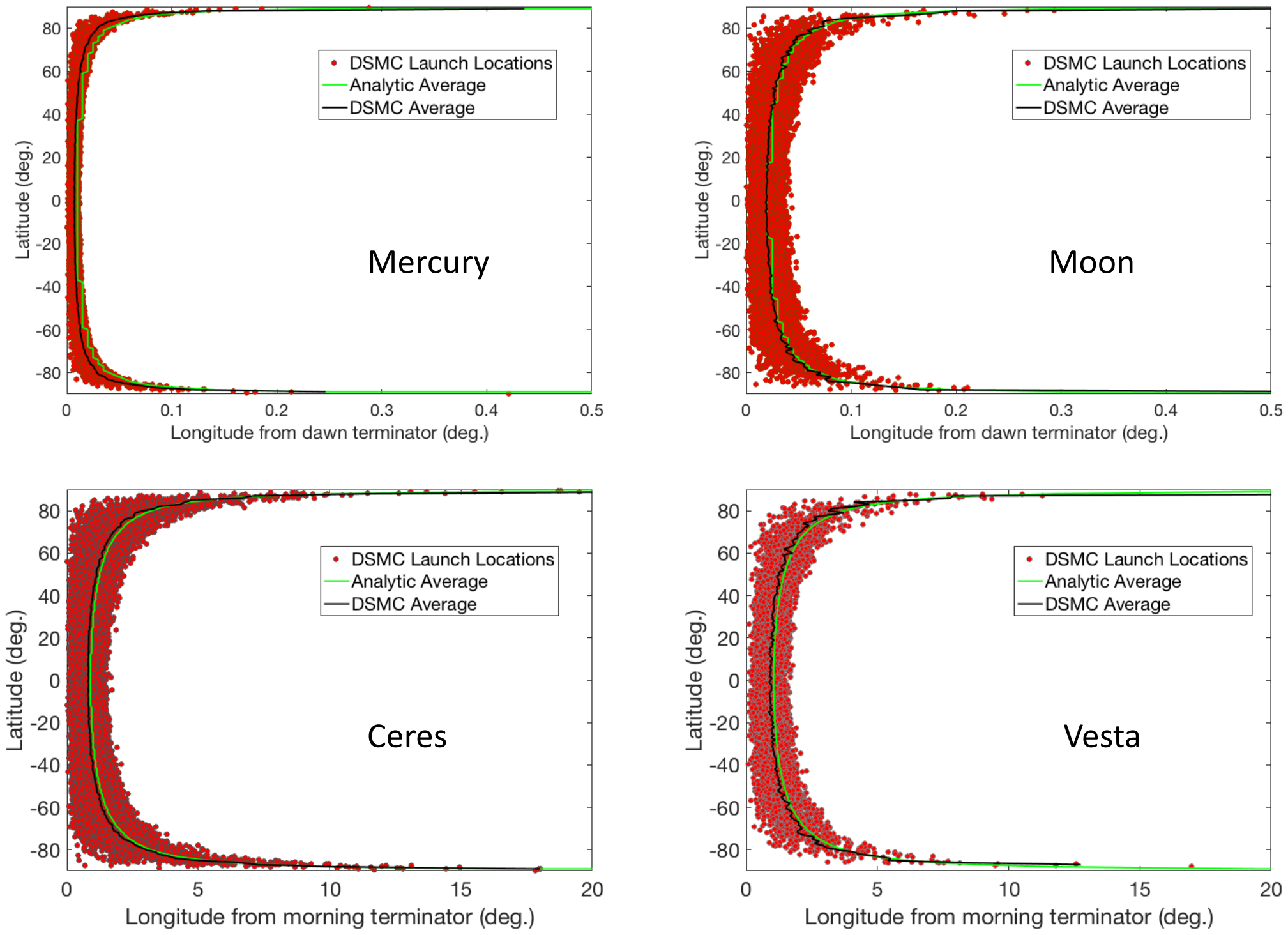

**Figure D-1:** *Comparison of desorption longitudes computed by DSMC simulations and Analytic code.* The analytic code accurately reproduces the DSMC desorption longitudes. The stepwise behavior in the analytic averages is due to the finite resolution in the analytic codes numerical integration from the dawn terminator.

*Desorption Longitude*

The desorption longitude describes the angular distance from the dawn terminator at which molecules desorb from the surface (Appendix A). Whereas PLANET DSMC is a stochastic model that results in a relative broad region over which molecules desorb, the analytic model produces a definite latitude-dependent longitude at which molecules are most likely to desorb from the surface. Successful validation should show that the analytic model produces a longitude of desorption that overlies the DSMC region of desorption, and matches its average behavior. We find that the analytic model successfully reproduces the desorption longitudes of the DSMC (See Figure D-1). The desorption longitudes in the analytic model round up to the next longitude step, resulting in desorption longitudes that may exceed the DSMC code by up to one longitude step (0.005° for Mercury). Nevertheless, such errors are negligible with a sufficiently high resolution grid.

*Ballistic Hop Distance and Duration*

We similarly compare the average ballistic hop distance and duration with the DSMC model, to ensure accurate function of the analytic code. We use the launch and landing time in the DSMC code to compute the mean ballistic hop duration for each particle that returns to the surface of the body. We find that the Analytic approach accurately reproduces the DSMC mean to within a few percent (see Fig. D-2). Similarly, we use the launch and landing positions of each particle that returns to the surface of each body in the DSMC code to compute the great-circle distance between the launch and landing position (the net displacement). We then use these values to compute the mean hop distance in the DSMC code, and compare with the analytic average (see figure D-3). We find that the analytic average accurately reproduces the DSMC average for each of these bodies to within a few percent.

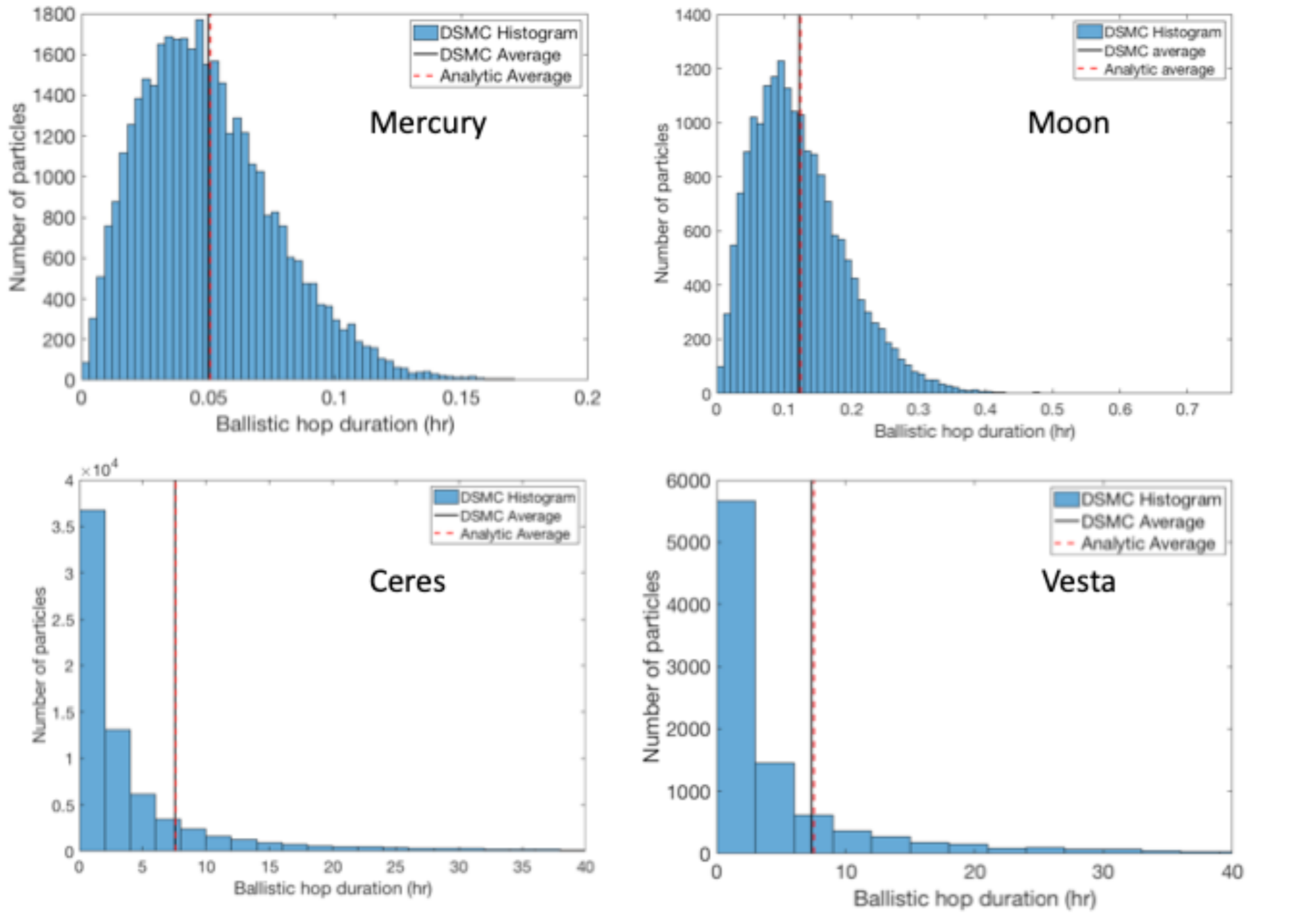

**Figure D-2:** *Comparison of Ballistic Hop Durations between DSMC simulations and Analytic calculation.* We find that the Analytic code accurately matches the ballistic hop duration of the DSMC simulations, to within a few percent.

These bodies of comparison span the range of parameters for this analytic approach is valid. Mercury and the Moon are relatively warm, very slowly rotating bodies. Such slow rotation challenges the analytic code to accurately reproduce the desorption longitudes.

Similarly, their large bodies and warm surfaces result in loss dominated by photodestruction rather than ballistic escape. Conversely, Ceres and Vesta are small, relatively cold, rapidly rotating bodies. This leads Ceres and Vesta to lose molecules predominantly through ballistic escape, rather than photodestruction. Thus, the underlying mechanisms included in the analytic model presented here can reliably reproduce the behavior of exospheres on other airless bodies.

We compare these models with previous calculations of ballistic transport of water on some of the bodies. *Schorghofer et al., (2017*) computes properties of water molecules on the surface of Ceres using a half-Maxwellian velocity distribution for particles desorbed from a surface, rather than the more appropriate Armand distribution of desorption velocities (*Armand, 1977; Hodges & Mahaffy, 2016*). As a result, *Schorghofer et al., (2017*) derive an escape fraction for a 200K surface of 41% (instead of 58.9%), and compute the exospheric residence time on Ceres to be ~7 hours (instead of 8.17 hours). If we modify our code to use this half-Maxwellian distribution, we obtain an escape fraction of 42% and an exospheric residence time of 7.6 hours. Thus, our code is consistent with *Schorghofer et al., (2017)* if we adopt their assumptions.

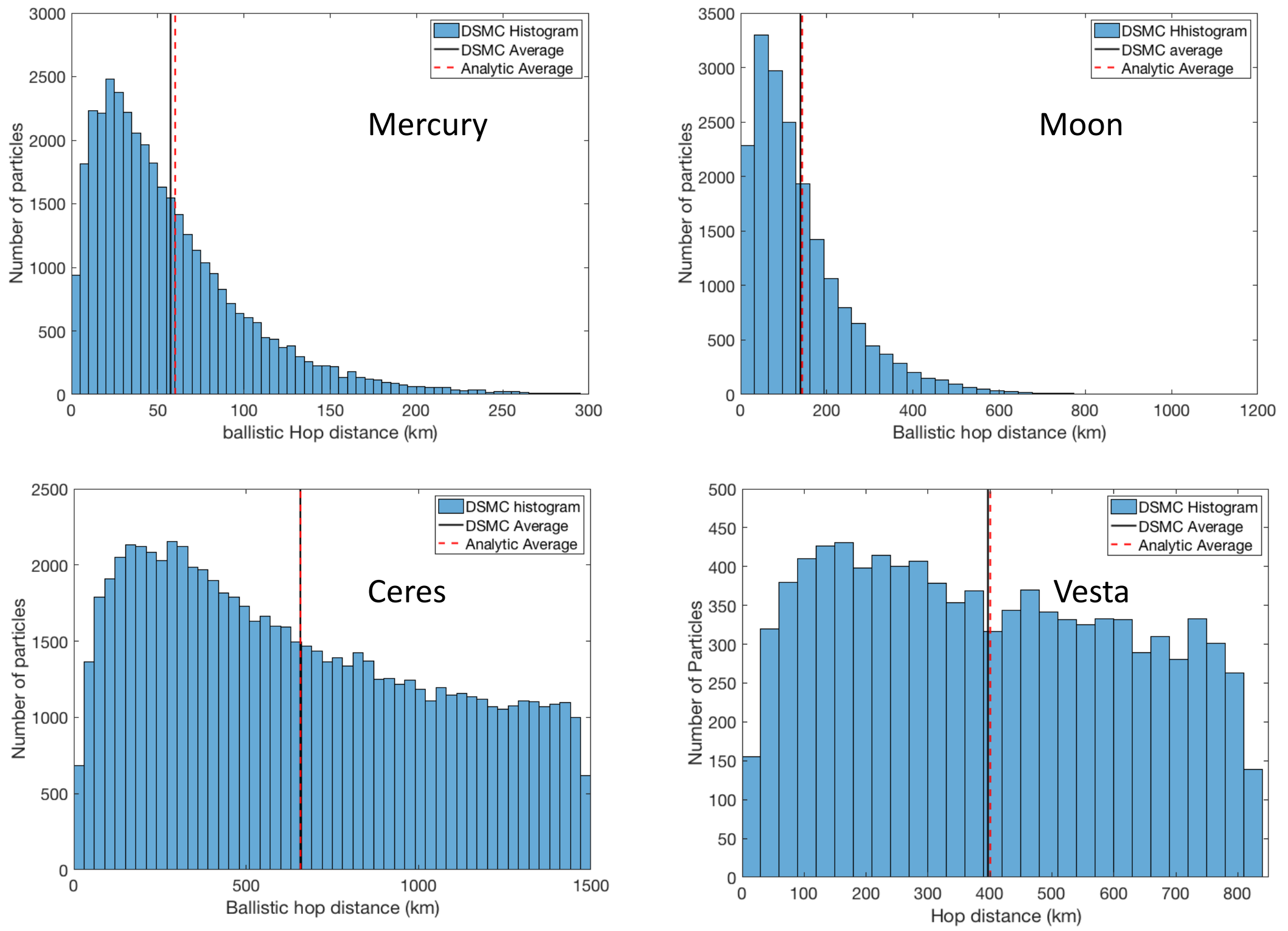

**Figure D-3:** *Comparison of Ballistic Hop Distances between DSMC simulations and Analytic calculation.* We find that the Analytic code very accurately matches the ballistic hop distance of the DSMC simulations, to within a few percent. The largest errors are associated with the analytic computation for Mercury, which the analytic calculation tends to overestimate the longitude of desorption, resulting in slightly warmer surface temperatures, and therefore slightly faster escape speeds.

*Arnold (1979)* used the Fredholm equation model from *Hodges (1973)* to compute the ballistic hop distance and flight time of water as a function of a set surface temperature. For this, we set both our DSMC code and analytic model to have a uniform surface temperature, to replicate the surface conditions for this calculation. At 200 K, we compute that a typical water molecule will travel 199 km in 542 s, which is significantly greater than the estimate in *Arnold (1979)* of 119 km in 277 s. Similarly, at 400 K, we compute a ballistic hop distance and time of 450 km in 932 s, which is also much greater than the estimate in *Arnold (1979)* of 250 km in 411 s. We think this discrepancy due to the assumption of a half-Maxwellian velocity distribution (rather than the Armand distribution for desorbing molecules). Additionally, the assumption of an isotropic emission from the surface, which overcounts shallow ejection angles (less than 45° from horizontal), dramatically reduces expected hop distance and flight time. However, even when we adopt their assumptions, we were unable to closely recreate their published results. It is unclear why the model results diverge.

*Errors From Neglecting Non-Inertial Forces*

Finally, in Fig. D-4 we compare with DSMC simulations that include non-inertial forces, to understand the errors associated with neglecting them. We focus on Ceres and Vesta, which are small, rapid rotators; hence, non-inertial forces should have a maximum impact on the behavior of their exospheres.

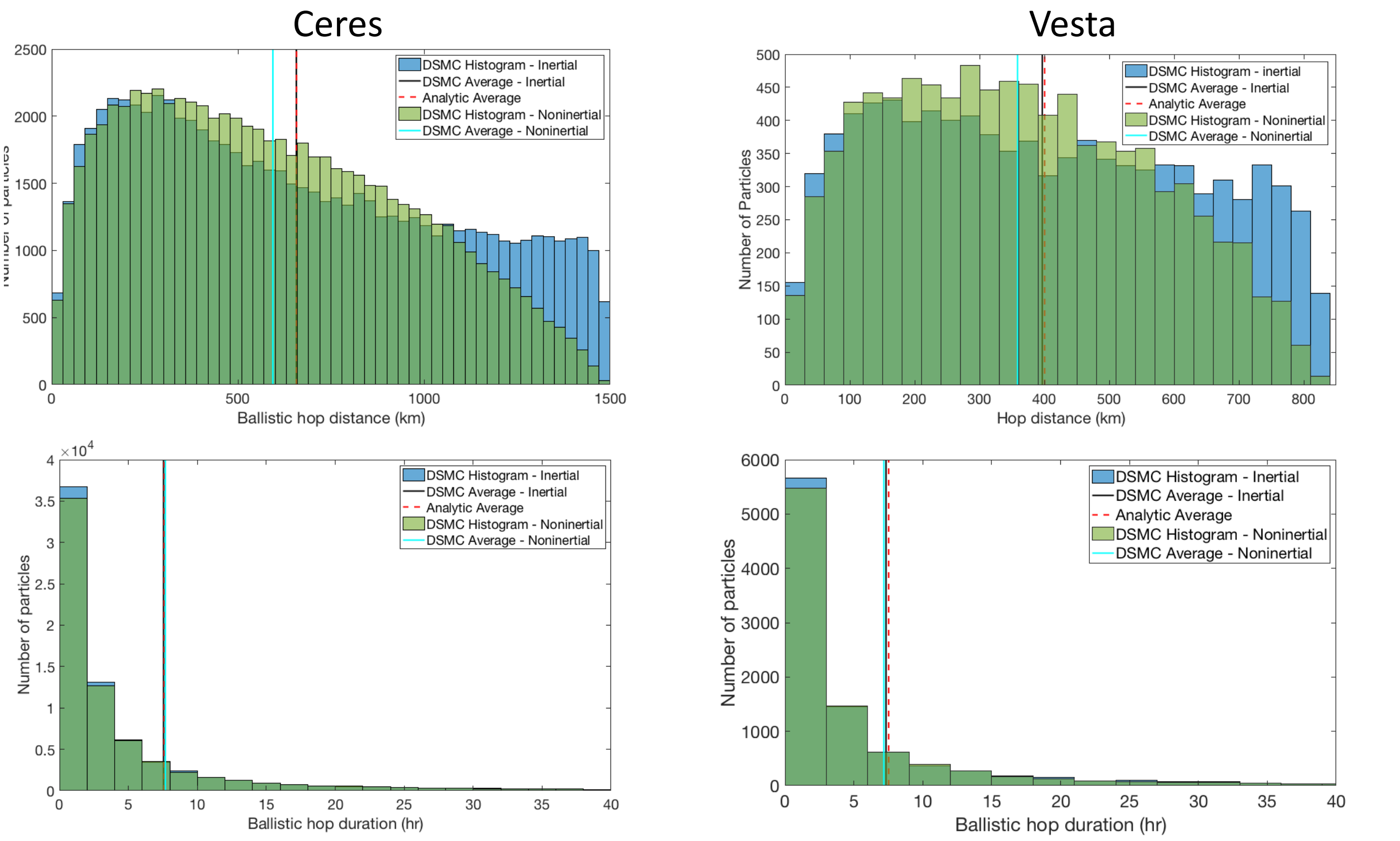

**Figure D-4:** *Comparison of Inertial and Non-inertial DSMC simulations with Analytic calculation.* We plot the ballistic hop distance and duration histograms of our DSMC simulations (both inertial and non-inertial reference frames) and their average values against the average value from the analytic code. We find that the non-inertial (rotating) reference frame alters the hop distance histogram to favor shorter hop distances. However, the hop durations are essentially unaltered.

We find that the errors associated with neglecting non-inertial forces remain small and acceptable even for these "worst-case" bodies. The average hop duration changes by only 1.7% and 2.0% for Ceres and Vesta respectively. Hop distances have significantly larger errors, of 9.59% and 9.57% for Ceres and Vesta respectively. These differences are not surprising; hop duration depends only on launch speed and direction, which is largely unaffected by non-inertial forces. However, launch distance is measured by the difference in landing and launch points, which can change significantly when non-inertial forces (e.g., Coriolis forces) are turned on in the model. For example, a prograde hop will experience the body rotating in the same direction as the traveling molecule, reducing the distance between launch and landing points. This is somewhat offset by molecules that hop in the opposite direction, and thus may experience a lengthened hop distance due to the addition of non-inertial forces. However, if a retrograde particle travels greater than halfway around the body, the net distance traveled will be reduced, since hop distance is computed along a great circle arc. Thus, the net effect is for non-inertial hop distances to be shorter than those in an inertial reference frame. Nevertheless, the errors associated with this are still small (less than 10%) even for these worst-case scenarios of a fast rotating body with little gravity. Thus, the analytic model can be considered to be reasonably accurate for all airless bodies, even though it neglects the effects of body rotation.

Table D-1 summarizes the thermal parameters used for this validation, along with the above results for all the bodies comparing the analytic model results with those of obtained via DSMC, including showing the impact of including non-inertial reference frame effects, i.e. Coriolis forces, on the computed mean ballistic time, hop distance, and the mean fraction of molecules lost due to ballistic escape and photodissociation during a single hop. We find that in all cases, the analytic model accurately replicates the results of the DSMC code with only minor errors.

**Table D-1**: *Comparison of Results from Analytic Model with those from DSMC simulations.*

| | | Mercury | Moon | Ceres | Vesta |
|---|---|---|---|---|---|
| **Heliocentric distance (AU)** | | 0.39 | 1.0 | 2.77 | 2.36 |
| **Peak Surface Temperature (K)** | | 700 [a] | 392[b] | 241 (est.) | 250 (est.) |
| **Minimum Surface Temperature (K)** | | 100 [a] | 110 | 100 | 100 |
| **Mean Ballistic Time** | **Analytic** | 183 s | 448 s | 7.59 hours | 7.52 hours |
| | **DSMC - Inertial** | 180 s | 442 s | 7.55 hours | 7.33 hours |
| | **DSMC – Non-inertial** | 180 s | 439 s | 7.68 hours | 7.18 hours |
| **Mean Ballistic Hop Distance** | **Analytic** | 60 km | 143 km | 658 km | 400 km |
| | **DSMC - Inertial** | 57 km | 139 km | 657 km | 397 km |
| | **DSMC – Non-inertial** | 58 km | 138 km | 594 km | 359 km |
| **Fraction of molecules lost (ballistic + photo)** | **Analytic** | 1.47% | 0.55% | 48.16% | 78.89% |
| | **DSMC - Inertial** | 1.42% | 0.48% | 48.27% | 78.20% |
| | **DSMC – Non-inertial** | 1.43% | 0.48% | 49.84% | 78.82% |

[a] Vasavada et al., 1999 [b] Hurley et al., 2015

## Appendix E Sensitivity of Results to Errors in Surface Temperature Model

In this model, we assume the canonical thermal model used in *Stewart et al.* (*2011*) and *Prem et al.* (*2015*)

$$T_{(\theta,\phi)}(t) = \begin{cases} T_{min} + \left(T_{peak} - T_{min}\right)\left[\cos\theta\cos(\phi - \phi_o(t))\right]^{\frac{1}{4}} & \text{if } |\phi - \phi_o| \le 90^{\circ} \\ T_{min} & \text{if } |\phi - \phi_o| > 90^{\circ} \end{cases} \quad \text{(A-32)}$$

where $T_{peak}$ and $T_{min}$ are the peak and minimum surface temperatures respectively, and $\phi_o$ is the subsolar longitude. This model is consistent with instantaneous radiative equilibrium with sunlight in the limit that $T_{min} = 0$ K, but always consistent at the subsolar point; although such minimum temperatures are unknown on any object due to nonzero thermal inertia of their surface. Nevertheless, other surface temperature profiles have been proposed/modeled for other various objects.

*The Moon*

Owning to its proximity to Earth, numerous spacecraft missions, and even surface temperature probes installed by the Apollo astronauts, the Moon has among the most well understood surface temperature distributions of all celestial bodies. This has allowed highly accurate empirical surface temperature models to be constructed, which accurately describe the surface temperature structure of the moon. Among these, the empirical model of Hurley et al. (2015) fit *Diviner* data to construct a quasi-empirical surface temperature model. The dayside temperatures use a similar $\cos^{0.25}$ model, except with $T_{peak}$ set to 392 and $T_{min}$ set to zero; with a lower temperature boundary of 130 K specified. The temperature on the night side of the moon is specified by an empirical fit that accounts for the decrease in the lunar surface temperature as the night progresses and as the surface increases in latitude. The Hurley et al. (*2015*) model generally provides a better fit to observations than our model, although both are reasonably close (Figure E-1).

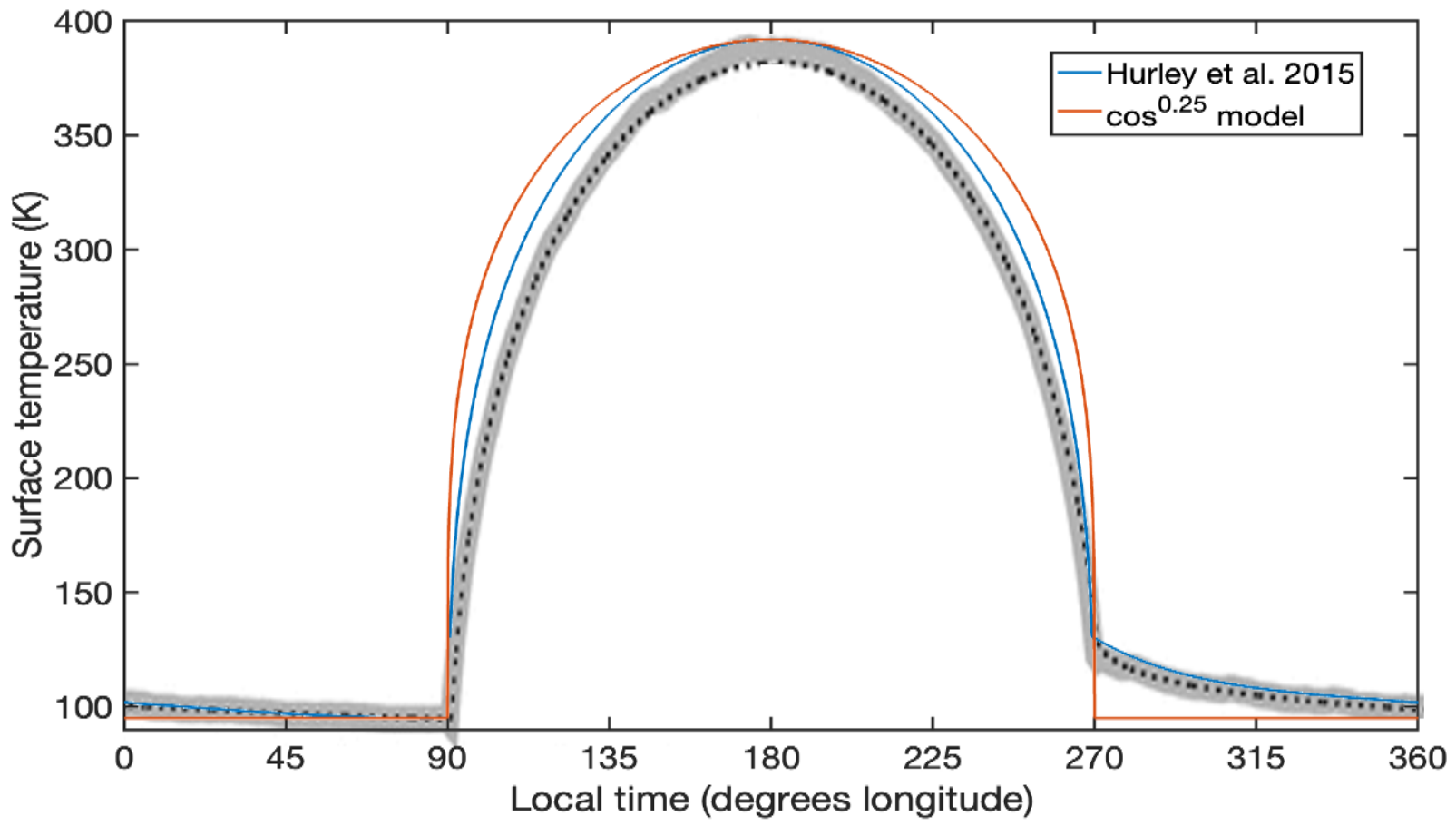

**Figure E-1:** *Comparison of surface temperature models of the Moon with observations.* The $\cos^{0.25}$ model and the quasi-empirical model of Hurley et al. (2015) are plotted here, along with Lunar *Diviner* measurements of the equatorial surface temperatures from Vasavada et al. (*2012*; modified from figure 9); dotted curve is the best fit surface thermal model of Vasavada et al. (2012). Although neither model perfectly recreates the observed temperature profile along the Moon's equator, we find that Hurley et al. (2015) provides a closer fit.

Here we compare the results of our thermal model's effects on exospheric properties with those that would result from the model of Hurley et al. (*2015*). We find that both models generally agree pretty closely with one another (to within a few percent) on all properties tracked, except for the fraction of molecules lost ballistically (see Table E-1). The higher morning temperatures in our model cause the thermal speeds in the high-velocity tail of thermal velocity distribution to be much higher than those in the tail of the distribution using the Hurley et al. (*2015*) model, resulting in orders of magnitude greater losses from ballistic escape in our model. However, ballistic loss of water from the Moon is negligible compared to photodestruction, producing no appreciable error in the behavior of water molecules on the Moon. Furthermore, the Hurley et al. (2015) model is specific to the Moon, and therefore cannot be readily generalized to other airless bodies in the Solar System.

| Temperature Model | $Cos^{0.25}$ | Hurley et al. (2015) |
|---|---|---|
| **Object** | Moon | |
| **Object Radius** | 1737 km | |
| **Avg. Bond Albedo** | 0.123±0.002[a] | |
| **Heliocentric distance (AU)** | 1.0 | |
| **Peak Surface Temperature (K)** | 392[b] | |
| **Minimum Surface Temperature (K)** | 95[c] | |
| **Surface escape speed (m/s)** | 2,380 | |
| **Fraction of molecules lost (ballistically, per hop)** | $2.23\times10^{-15}$ % | $3.65\times10^{-17}$ % |
| **Fraction of molecules lost (photodestruction, per hop)** | 0.53% | 0.50% |
| **Fraction of molecules lost (total, per hop)** | 0.53% | 0.50% |
| **Expected Number of hops** | 187 | 198 |
| **Ballistic timescale ($H_2O$)** | 443 s | 418 s |
| **Expected Exospheric Residence Time ($H_2O$)** | 23.0 hours | 23.0 hours |
| **Ballistic hop distance ($H_2O$)** | 140 km | 126 km |
| **Expected Migration distance (RMS)** | 1910 km | 1770 km |
| **Photodissociation timescale at heliocentric distance ($H_2O$)** | ~25 hours | |

**Table E-1:** *Comparing the exosphere properties of the Moon using our model and the more accurate Hurley et al. (2015) model.* We find that these properties generally agree to within a few percent, demonstrating that the general properties of these exospheres are largely independent of the specific functional form of the chosen thermal model, so long as the model provides a reasonably good fit to observations. [a]Lane and Irvine, 1973 [b]Hurley et al., 2015 [c]Williams et al. 2017

*Callisto*

Hartkorn et al. (*2017*) use a sinusoidal approximation for the temperature of Callisto of

$$\frac{1}{2}\left(T_{peak} + T_{min}\right) + \frac{1}{2}\left(T_{peak} - T_{min}\right)\cos\delta \qquad \text{(A-33)}$$

where $\delta$ is the angular distance to the subsolar point, $T_{peak}$ is 155 K, and $T_{min}$ is 80 K. Moore et al. (*2004*) compiled *Galileo* spacecraft measurments of Callisto's effective temperature along the equator, which we compare to our canonical model (using $T_{peak}$ is 167 K, and $T_{min}$ is 80 K) and the Hartkorn et al. (*2017*) model. We see that our canonical model provides a good approximation to the measured effective surface temperatures (see Figure E-2).

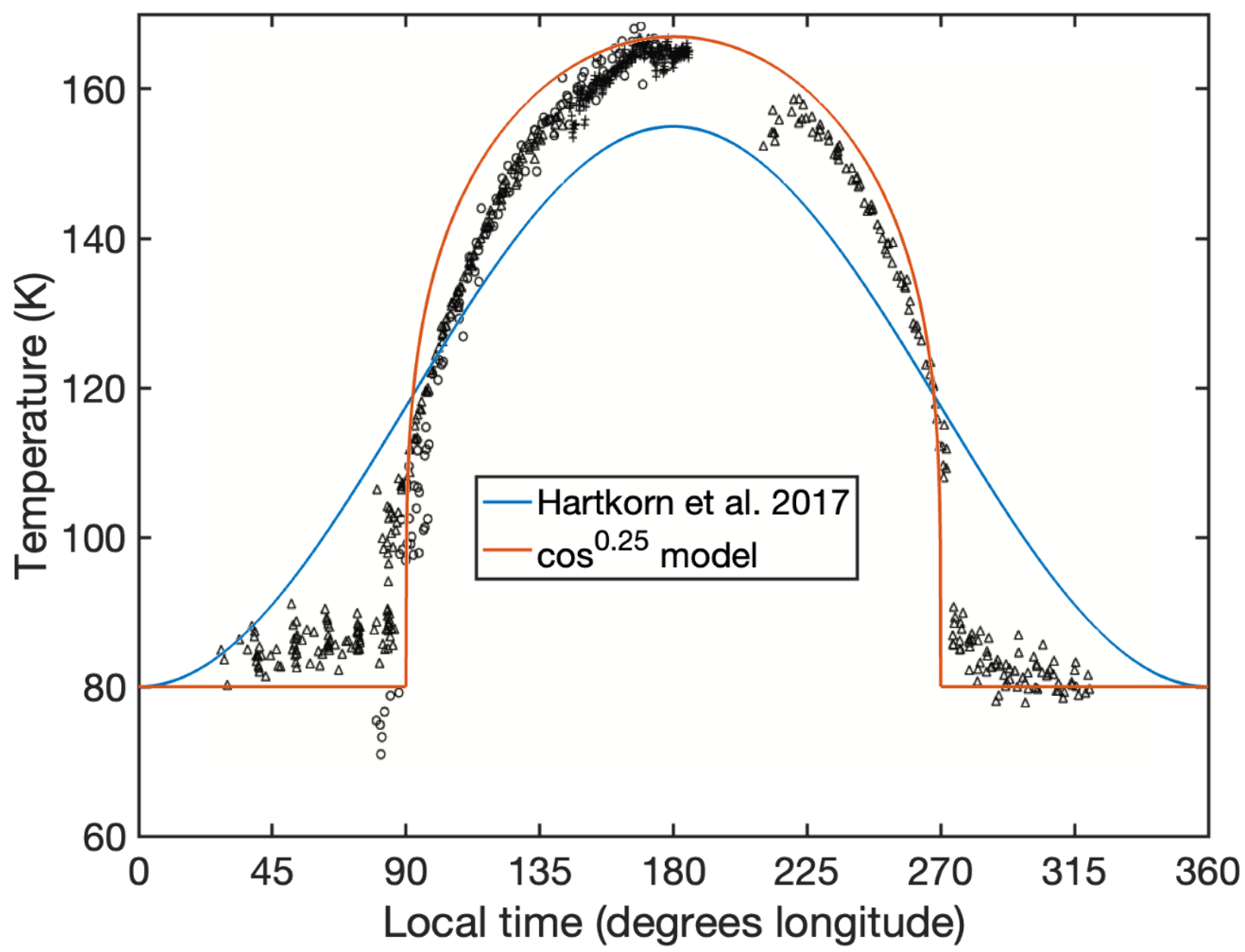

**Figure E-2:** *Comparison of surface temperature models of Callisto with observations.* The models are plotted over the compiled Galileo spacecraft data of figure 17.7 from Moore et al. (*2004*). Different data symbols represent different Callisto flybys and different wavelength brightness temperatures, which were then converted to equatorial effective temperatures (see *Moore et al. 2004* for details). From inspection, we find that our canonical $\cos^{0.25}$ model is a good approximation to observed temperatures on Callisto, better matching the measured diurnal temperature profile than Hartkorn et al. (2017).

*Rhea*

The icy moons of the outer solar system demonstrate sizeable departures from our canonical surface temperature model. Howett et al. (*2010*) used *Cassini* CIRS data to create best fit thermal properties of these surface, and generate temperature profiles of Mimas, Enceladus, Tethys, Dione, Rhea, Iapetus, and Phoebe. Aside from a peak temperature that is offset from the subsolar point, we find that a $\cos^{0.50}$ model or a $\cos^{0.25}$ model with a minimum temperature of zero (on the dayside) fits the modeled data much better than our canonical $\cos^{0.25}$ model (see figure E-3).

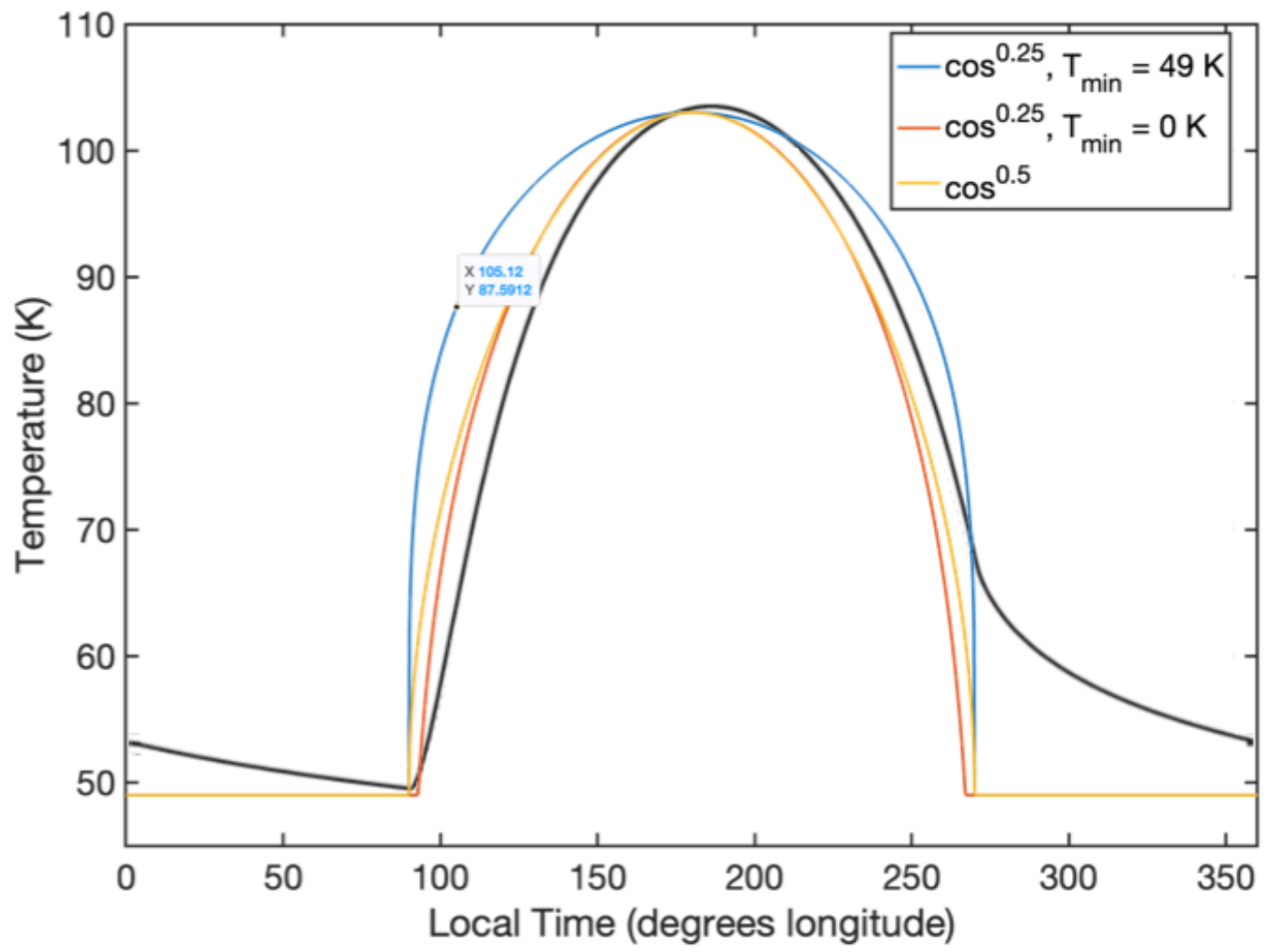

**Figure E-3:** *Comparison of surface temperature models with observation-constrained surface temperature model.* The models are plotted over best fit model to Rhea's surface temperature (black curve) from Howett et al. (*2010*; replotted from figure 16). We find that a $\cos^{0.25}$ with a minimum temperature of 0K (rather than our 49K minimum) fits best (albeit with a longitude offset), with a $\cos^{0.5}$ model also providing a good fit. Our canonical $\cos^{0.25}$ model with a minimum temperature of 49K provides the largest departure from the Howett et al. (*2010*) model that was fit to *Cassini* CIRS observations of the surface.

Nevertheless, we find that, when we compute the physical properties of the resulting exosphere, there is little variation between these three different models (Table E-2). Indeed, all computed properties agree to within a few percent, except for the computed sublimative mass-loss rate. We find that our canonical $\cos^{0.25}$ model overestimates the sublimative mass loss rate by nearly a factor of two. This demonstrates that the accuracy of our results depends on the quantity desired. For sublimative mass flux, our results are accurate to within half an order-of-magnitude, but for other properties, we find that the exosphere is generally insensitive to the functional form of these surface temperature models, and is accurate to within a few percent.

| | | $Cos^{0.25}$ Min=49K | $Cos^{0.25}$ Min=0K | $Cos^{0.50}$ |
|---|---|---|---|---|
| | **Object** | Rhea | | |
| | **Object Radius** | 764.3 km | | |
| | **Avg. Bond Albedo** | 0.42–0.48[a] | | |
| | **Heliocentric distance (AU)** | 9.6 | | |
| | **Peak Surface Temperature (K)** | ~103[f] | | |
| | **Minimum Surface Temperature (K)** | ~49[f] | 0 | ~49[f] |
| | **Surface escape speed (m/s)** | 635 | | |
| **$H_2O$** | **Mass loss rate (kg/s)** | $1.21\times10^{-4}$ | $6.59\times10^{-5}$ | $6.38\times10^{-5}$ |
| | **Fraction of molecules lost (ballistic, per hop)** | 7.2% | 7.2% | 7.2% |
| | **Fraction of molecules lost (photo, per hop)** | 0.096 % | 0.096% | 0.096% |
| | **Fraction of molecules lost (ballistic + photo, per hop)** | 7.3 % | 7.3 % | 7.3 % |
| | **Expected number of hops** | 12.7 | 12.7 | 12.7 |
| | **Ballistic timescale** | 2.29 hrs | 2.28 hrs | 2.29 hrs |
| | **Expected Exospheric Residence Time** | 29.0 hrs | 29.0 hrs | 29.1 hrs |
| | **Ballistic hop distance** | 758 km | 757 km | 757 km |
| | **Expected Migration distance (RMS)** | 2,700 km | 2,700 km | 2,700 km |
| | **Photodissociation timescale** | ~97 days | | |
| **$CO_2$** | **Fraction of molecules lost (ballistic, per hop)** | 0.00192 % | 0.00162 % | 0.00155 % |
| | **Fraction of molecules lost (photo, per hop)** | 0.00385 % | 0.00163 % | 0.00377 % |
| | **Fraction of molecules lost (ballistic + photo, per hop)** | 0.00577 % | 0.00542 % | 0.00532 % |
| | **Expected number of hops** | 17,300 | 18,500 | 18,800 |
| | **Ballistic timescale** | 29.2 min | 28.8 min | 28.7 min |
| | **Expected Exospheric Residence Time** | 352 days | 370 days | 375 days |
| | **Ballistic hop distance** | 232.9 km | 229.0 km | 227.8 km |
| | **Expected Migration distance (RMS)** | 30,700 km | 31,100 km | 31,200 km |
| | **Photodissociation timescale** | 528 days | | |

**Table E-2:** *Comparing the exosphere properties of Rhea using three different surface temperature models.* We compare our canonical $\cos^{0.25}$ surface temperature model with a variation in which the minimum temperature is set to zero, and a model in which the surface temperature follows a $\cos^{0.50}$ profile. We find that, in general, the computed properties are largely unaffected by this choice of surface temperature model.

**References**

A'Hearn, M.F.; Belton, M.J.S.; Delamere, W.A.; Feaga, L.M.; Hampton, D. et al., 2011. EPOXI at Comet Hartley 2. *Science* 332, 1396 - 1400

A'Hearn, M.F.; Feldman, P.D. (1992) Water vaporization on Ceres. *Icarus* 98, 54 - 60

Armand, G. (1977) Classical theory of desorption rate velocity distribution of desorbed atoms; possibility of a compensation effect. *Surface Science* **66**, 321 – 345

Arnold, J.R. 1979. Ice in the lunar polar regions. *JGR* 84, 5659 - 5668

Benna, M.; Hurley, D.M.; Stubbs, T.J.; Mahaffy, P.R.; Elphic, R.C. 2019. Lunar soil hydration constrained by exospheric water liberated by meteoroid impacts. *Nature Geoscience* 12, 333 - 338

Berg, J.J.; Goldstein, D.B.; Varghese, P.L.; Trafton, L.M. (2016) DSMC simulation of Europa water vapor plumes. *Icarus* 277, 370 – 380

Bockelée-Morvan, D.; Crovisier, J.; Mumma, M.J.; Weaver, H.A. (2004) The composition of cometary volatiles. in *Comets II* Editors: Festou, M.; Keller, H.U.; Weaver, H.A. University of Arizona Press – Tucson, AZ, 391 - 423

Bridge, H.S.; Belcher, J.W.; Coppi, B.; Lazarus, A.J.; McNutt Jr., R.L. et al., (1986) Plasma observations near Uranus: Initial results from Voyager 2. *Science* 233, 89–93

B. J. Buratti, J. A. Mosher, P. D. Nicholson, C. A. McGhee, R. French, Icarus 136, 223 (1998).

Bussey, D.B.J.; Spudis, P.D; Robinson, M.S. 1999. Illumination conditions at the lunar south pole. *GRL* 26(9), 1187 - 1190

Bussey, D.B.J.; Lucey, P.G.; Steutel, D.; Robinson, M.S.; Spudis, P.D.; Edwards, K.D. 2003. Permanent shadow in simple craters near the lunar poles. *GRL* 30(6), 1278 (4 pp)

Butler, B.J. (1997) The migration of volatiles on the surfaces of Mercury and the Moon. *JGR* **102**, 19283 - 19291

Campbell, B.A.; Campbell, D.B.; Chandler, J.F.; Hine. A.A.; Nolan, M.C.; Perillat, P.J. 2003. Radar images of the lunar poles. *Nature* 426, 137 – 138

Campbell, D.B.; Campbell, B.A.; Carter, L.M.; Margot, J.-L.; Stacy, N.J.S. 2006. No evidence for thick deposits of ice at the lunar south pole. *Nature* 443, 835 - 837

Canup, R.M. (2010) Origin of Saturn's rings and inner moons by mass removal from a lost Titan-sized satellite. *Nature* 468, 943 - 946

Carlson, R.W. (1999) A Tenuous Carbon Dioxide Atmosphere on Jupiter's Moon Callisto. *Science* **253**, 820 - 821

Carlson, R.; Smythe, W.; Baines, K.; Barbinis, E.; Becker, K. et al., (1996) Near-infrared spectroscopy and spectral mapping of Jupiter and the Galilean satellites: results from Galileo's initial orbit. *Science* 274, 385 - 388

Cartwright, R.J.; Emery, J.P.; Rivkin, A.S.; Trilling, D.E.; Pinilla-Alonso, N. (2015) Distribution of $CO_2$ ice on the large moons of Uranus and evidence for compositional stratification of their near-surfaces. *Icarus* 257, 428 – 456

Cartwright, R.J.; Emery, J.P.; Pinilla-Alonso, N.; Lucas, M.P.; Rivkin, A.S.; Trilling, D.E. (2018) Red material on the large moons of Uranus: Dust from the irregular satellites? *Icarus* 314, 210 – 231

Cassini, J.D. (1677) Some new observations made by Dig. Cassini and deliver'd in the journal des scavans, concerning the two Planets about Saturn, formerly discover'd by the same, as appears in N. 92. Of these tracts. *Philosophical Transactions of the Royal Society* **12**(133), 831 - 833

Chabot, N.L.; Ernst, C.M.; Harmon, J.K.; Murchie, S.L.; Solomon, S.C. et al., 2013. Craters hosting radar-bright deposits in Mercury's north polar region: Areas of persistent shadow determined from MESSENGER images. *JGR Planets* 118, 26-36

Chabot, N.L.; Ernst, C.M.; Denevi, B.W.; Nair, H.; Deutsch, A.N. et al., 2014. Images of surface volatiles in Mercury's polar craters acquired by the MESSENGER spacecraft. *Geology* 42(12), 1051 - 1054

Chabot, N.L.; Ernst, C.M.; Paige, D.A.; Nair, H.; Denevi, B.W. et al., 2016. Imaging Mercury's polar deposits during MESSENGER's low-altitude campaign. *GRL* 43, 9461 – 9468

Chabot, N.L., Shread, E.E.; Harmon, J.K. 2018. Investigating Mercury's south polar deposits: Arecibo radar observations and high-resolution determination of illumination conditions. *JRL:Planets* 123, 666 – 681

Charnoz, S.; Crida, A.; Castillo-Rogez, J.C.; Lainey, V.; Dones, L. et al., (2011) Accretion of Saturn's mid-sized moons during the viscous spreading of young massive rings: Solving the paradox of silicate-poor rings versus silicate-rich moons. *Icarus* 216, 535 - 550

Choukroun, M.; Altwegg, K.; Kührt, E.; Biver, N.; Bockelée-Morvan, D.; Drążkowska, J.; Hérique, A.; Hilchenbach, M.; Marschall, R.; Pätzold, M.; Taylor, M.G.G.T.; Thomas, N. 2020. Dust-to-Gas and Refractory-to-ice Mass Ratios of Comet 67P/Churyumov-Gerasimenko from Rosetta Observations. *Space Sci. Rev.* 216:44 (38 pp.)

Clark, R.N. 2009. Detection of adsorbed water and hydroxyl on the Moon. *Science* 326, 562 – 564

Clark, R.N.; Curchin, J.M.; Jaumann, R.; Cruikshank, D.P.; Brown, R.H.; Hoefen, T.M. et al., (2008) Compositional mapping of Saturn's satellite Dione with Cassini VIMS and implications of dark material in the Saturn system. *Icarus* **193**, 372 – 386

Colaprete, A.; Schultz, P.; Heldmann, J.; Wooden, D.; Shirley, M. et al., 2010. Detection of water in the LCROSS ejecta plume. *Science* 330, 463 - 468

Connerney, J.E.P.; Acuña, M.H.; Ness, N.F. (1987) The magnetic field of Uranus. *JGR* **92**, 15329 – 15336

Costello, E.S.; Phillips, C.B.; Lucey, P.G.; Ghent, R.R (2021) Impact gardening on Europa and repercussions for possible biosignatures. *Nature Astronomy* 5, 951-956

Crider, D.H. and Vondrak, R.R. (2003) Space weathering effects on lunar cold trap deposits. *JGR* **108**, 5079 (15 pp.)

Cruikshank, D.P.; Dalton, J.B.; Dalle Ore, C.M.; Bauer, J.; Stephan, K. et al., (2007) Surface composition of Hyperion. *Nature* 448, 54 - 56

Ćuk, M.; Dones, L.; Nesvorny, D. (2016) Dynamical evidence for a late formation of Saturn's moons. *ApJ* 820:97 (16pp)

Deutsch, A.N.; Chabot. N.L.; Mazarico, E.; Ernst, C.M.; Head, J.W. et al., 2016. Comparison of areas in shadow from imaging and altimetry in the north polar region of Mercury and implications for polar ice deposits. *Icarus* 280, 158 – 171

Deutsch, A.N.; Head, J.W.; Neumann, G.A. (2019) Age constraints of Mercury's polar deposits suggest recent delivery of ice. *EPSL* 520, 26 - 33

Ernst, C.M.; Chabot, N.L.; Barnouin, O.S. (2018) Examining the potential contribution of the Hokusai impact to water ice on Mercury. *JGR:Planets* 123, 2628 - 2646

Feldman, W.C.; Maurice, S.; Binder, A.B.; Barraclough, B.L.; Elphic, R.C.; Lawrence, D.J. 1998. Fluxes of fast and epithermal neutrons from Lunar Prospector: Evidence for water ice at the lunar poles. *Science* 281, 1496 - 1500

Feldman, W.C.; Lawrence, D.J.; Elphic, R.C.; Barraclough, B.L.; Maurice, S.; Genetay, I.; Binder, A.B. 2000. Polar hydrogen deposits on the Moon. *JGR* 105, 4175 – 4195

Feldman, W.C; Maurice, S.; Lawrence, D.J.; Little, R.C.; Lawson, S.L. et al., 2001. Evidence for water ice near the lunar poles. *JGR* 106, 23231 - 23251

Fisher, E.A.; Lucey, P.G.; Lemelin, M.; Greenhagen, B.T.; Siegler, m>A.; Mazarico, E.; Ahronson, O.; Williams, J.-P.; Hayne, P.O.; Neumann, G.A.; Paige, D.A.; Smith, D.E.; Zuber, M.T. 2017. Evidence for surface water ice in the lunar polar regions using reflectance measurements from the Lunar Orbiter Laser Altimeter and temperature measurements from the Diviner Lunar Radiometer Experiment. *Icarus* 292, 74 - 85

Fitzpatrick, R. 2012. An introduction to celestial mechanics. Cambridge University Press – New York, NY. p. 45

Flasar, F.M.; Achterberg, R.K.; Conrath, B.J.; Pearl, J.C.; Bjoraker, G.L. et al., (2005) Temperatures, winds, and composition in the Saturnian system. *Science* 307, 1247 – 1251

Frantseva, K.; Nesvorný, D.; Mueller, M.; van der Tak, F.F.S; ten Kate, I.L.; Pokorný, P. (2022) Exogeneous delivery of water to Mercury. *Icarus*. 114980. In Press

Frenkel, J. 1924. Theorie der Adsorption und verwandter Erscheinungen. *Zeitschrift für Physik* 26(1), 117 – 138

Le Gall, A.; West, R.D.; Bonnefoy, L.E. (2019) Dust and Snow Cover on Saturn's Icy Moons. *GRL* 46, 11,747 – 11,755

Greeley, R.; Klemaszewski, J.E.; Wagner, R.; the Galileo Imaging Team (2000) Galileo views of the geology of Callisto. *Planetary and Space Science* 48, 829 - 853

Grundy, W.H.; Buratti, B.J.; Cheng, A.F.; Emery, J.P.; Lunsford, A. et al., (2007) New Horizons mapping of Europa and Ganymede. *Science* 318, 234 – 237

Grundy, W.M.; Young, L.A.; Spencer, J.R.; Johnson, R.E.; Young, E.F.; Buie, M.W (2006) Distributions of $H_2O$ and $CO_2$ ices on Ariel, Umbriel, Titania, and Oberon from IRTF/SpeX observations. *Icarus* 184, 543 – 555

Grundy, W.M.; Young, L.A.; Young, E.F. (2003) Discovery of $CO_2$ ice and leading-trailing spectral asymmetry on the uranian satellite Ariel. *Icarus* 162, 223–230

Gundlach, B.; Skorov, Y.; Blum, J. (2011) Outgassing of icy bodies in the Solar System – I. The sublimation of hexagonal water ice through dust layers. *Icarus* 213, 710 - 719

Halekas, J.S.; Benna, M.; Mahaffy, P.R.; Elphic, R.C.; Poppe, A.R.; Delory, G.T. 2015. Detections of lunar exospheric ions by the LADEE neutral mass spectrometer. *GRL* 42, 5162 – 5169

Hanel, R.; Conrath, B.; Flasar, F.M.; Kunde, V.; Maguire, W.; Pearl, J. et al., (1986) Infrared Observations of the Uranian System. *Science* **233**, 70 - 74

Harmon, J.K.; Slade, M.A. 1992. Radar mapping of Mercury: Full-disk images and polar anomalies. *Science* 258, 640 – 643

Harmon, J.K.; Slade, M.A.; Rice, M.S. 2011. Radar imagery of Mercury's putative polar ice: 1999 – 2005 Arecibo results. *Icarus* 211(1), 37 – 50

Hartkorn, O.; Saur, J.; Strobel, D.F. (2017) Structure and density of Callisto's atmosphere from a fluid-kinetic model of its ionosphere: Comparison with Hubble Space Telescope and Galileo observations. *Icarus* 282, 237 - 259

Hayne, P.O.; Aharonson, O. 2015b. Thermal stability of ice on Ceres with rough topography. *JGR – Planets* 120, 1567 – 1584

Hayne, P.O.; Hendrix, A.; Sefton-Nash, E.; Siegler, M.A.; Lucey, P.G. et al., 2015a. Evidence for exposed water ice in the Moon's south polar regions from Lunar Reconnaissance Orbiter ultraviolet albedo and temperature measurements. *Icarus* 255, 58 – 69

Hedman, M.M.; Burns, J.A.; Hamilton, D.P.; Showalter, M.R. (2012) The three-dimensional structure of Saturn's E ring. *Icarus* **217**, 322 - 338

Hendrix, A.R.; Hurley, D.M.; Farrell, W.M.; Greenhagen, B.T.; Hayne, P.O.; Retherford, K.D. et al. (2019) Diurnally Migrating Lunar Water: Evidence from Ultraviolet Data. *GRL* 46, 2417-2424

Hibbitts, C.A.; Klemaszewski, J.E.; McCord, T.B.; Hansen, G.B.; Greeley, R. (2002) $CO_2$-rich impact craters on Callisto. *JGR:Planets* **107**, 5084

Hodges, R.R. (1973) Helium and Hydrogen in the Lunar Atmosphere. *JGR* **78**(34), 8055 - 8064

Hodges, R.R. (1975) Formation of the Lunar atmosphere. *The Moon* 14, 139 – 157

Hodges, R.R. 2002. Ice in the lunar polar regions revisited. *JGR* 107, 5011 (7pp)

Hodges, R.R.; Mahaffy, P.R. (2016) Synodic and semiannual oscillations of argon-40 in the lunar exosphere. *GRL* **43**, 22 - 27

Hoey, W.A.; Yeoh, S.K.; Trafton, L.M.; Goldstein, D.B.; Varghese, P.L. (2017a) Rarefied gay dynamic simulation of transfer and escape in the Pluto-Charon system. *Icarus* 287, 87 – 102

Hoey, W.A. (2017b) Rarefied gas dynamic simulations of planetary atmospheric systems. PhD dissertation. University of Texas at Austin, Austin, TX

Howard, A.D.; Moore, J.M. (2008) Sublimation-driven erosion on Callisto: A landform simulation model test. *GRL* **35**, L03203 (4 pp.) doi:10.1029/2007GL032618

Howard, A.D.; Moore, J.M.; Schenk. P.M.; White, O.L.; Spencer, J. (2012) Sublimation-driven erosion on Hyperion: Topographic analysis and landform simulation model tests. *Icarus* 220, 268 – 276

Howett, C.J.A.; Spencer, J.R.; Pearl, J.; Segura, M. (2010) Thermal inertia and bolometric Bond albedo values for Mimas, Enceladus, Tethys, Dione, Rhea, and Iapetus as derived from Cassini/CIRS measurements. *Icarus* 206, 573 - 593

Huebner, W.F.; Keady, J.J.; Lyon, S.P. (1992) Solar photo rates for planetary atmospheres and atmospheric pollutants. *Astrophysics and Space Science* 195, 1-294

Huebner, W.F.; Mukherjee, J. (2015) Photoionization and photodissociation rates in solar and blackbody radiation fields. *Planetary & Space Science* 106, 11-45

Hurley, D.M.; Sarantos, M.; Grava, C.; Williams, J.-P.; Retherford, K.D.; Siegler, M.; Greenhagen, B.; Paige, D. (2015) An analytic function of lunar surface temperature for exospheric modeling. *Icarus* 255, 159 - 163

Jacobson, R.A.; Campbell, J.K.; Taylor, A.H.; Synnott, S.P. (1992) The masses of Uranus and its major satellites from Voyager tracking data and Earth-based Uranian satellite data. *AJ* 103(6), 2068 – 2078

Jia, Y.-D.; Villarreal, M.B.; Russell, C.T. (2017) Possible Ceres bow shock surfaces based on fluid models, *JGR: Space Physics* 122, 4976 – 4987. DOI: 10.1002/2016JA023712

Johnson, R.E.; Lanzerotti, L.J.; Brown, W.L.; Armstrong, T.P. (1981) Erosion of Galilean satellite surfaces by jovian magnetosphere particles. *Science* 212, 1027 - 1030

Johnston, W. R. (2018) TNO AND CENTAUR DIAMETERS, ALBEDOS, AND DENSITIES V1.0. urn:nasa:pds:tno-centaur_diam-albedo-density::1.0. NASA Planetary Data System

Jones, B.M.; Sarantos, M.; Orlando, T.M. (2020) A New In Situ Quasi-continuous Solar-wind Source of Molecular Water on Mercury. *ApJL* 891:L43 (8pp) doi:10.3847/2041-8213/ab6bda

Karkoschka, E. (2001) Comprehensive photometry of the rinds and 16 satellites of Uranus with the Hubble Space Telescope. *Icarus* 151, 51 - 68

Karkoschka, E. (2003) Sizes, shapes, and albedos of the inner satellites of Neptune. *Icarus* 162, 400 – 407

Krimigis, S.M.; Armstrong, T.P.; Axford, W.I.; Cheng, A.F.; Gloeckler, G. et al., (1986) The magnetosphere of Uranus: Hot plasma and radiation environment. *Science* 233, 97–102

Küppers, M.; O'Rourke, L.; Bockelée-Morvan, D.; Zakharov, V.; Lee, S. et al., (2014) Localized sources of water vapour on the dwarf planet (1) Ceres. *Nature* 505, 523 – 527

Lainey, V.; Jacobson, R.A.; Tajeddine, R.; Cooper, N.J.; Murray, C. et al., (2017) New constraints on Saturn's interior from Cassini astrometric data. *Icarus* 281, 286 - 296

Landis, M.E.; Byrne, S.; Combe, J.-Ph.; Marchi, S.; Castillo-Rogez, J. et al., (2019) Water vapor contribution to Ceres' exosphere from observed surface ice and postulated ice-exposing impacts. *JGR:Planets* 124, 61 - 75

Lane, A.P.; Irvine, W.M. 1973. Monochromatic phase curves and albedos for the lunar disk. *AJ* 78, 267 – 277

Langmuir, I. 1913. The vapor pressure of metallic tungsten. *Phys. Rev.* 2, 329 - 342

Langmuir, I. 1916. The evaporation, condensation and reflection of molecules and mechanism of adsorption. *Phys. Rev.* 8, 149 – 176

Lawrence, D.J.; Hurley, D.M.; Feldman, W.C.; Elphic, R.C.; Maurice, S.; Miller, R.S.; Prettyman, T.H. 2011. Sensitivity of orbital neutron measurements to the thickness and abundance of surficial lunar water. *JGR* 116, E01002

Lawrence, D.J.; Feldman, W.C.; Goldsten, J.O.; Maurice, S.; Peplowski, P.N. 2012. Evidence for water ice near Mercury's north pole from MESSENGER neutron spectrometer measurements. *Science* 339, 292 - 296

Li, J.-Y.; A'Hearn, M.F.; McFadden, L.A.; Belton, M.J.S. 2007. Photometric analysis and disk-resolved thermal modeling of Comet 19P/Borrelly from Deep Space 1 data. *Icarus* 188, 195 – 211

Li, J.-Y.; A'Hearn, M.F.; Farnham, T.L.; McFadden, L.A. 2009. Photometric analysis of the nucleus of Comet 81P/Wild 2 from Stardust images. *Icarus* 204, 209 - 226

Li, J.-Y.; A'Hearn, M.F.; Belton, M.J.S.; Farnham, T.L.; Klaasen, K.P. et al., 2013a. Photometry of the nucleus of Comet 9P/Tempel 1 from Stardust-NExT flyby and the implications. *Icarus* 222, 467 - 476

Li, J.-Y.; Le Corre, L.; Schröder, S.E.; Reddy, V.; Denevi, B.W. et al., 2013b. Global photometric properties of Asteroid (4) Vesta observed with Dawn Framing Camera. *Icarus* 226, 1252 - 1274

Li, J.-Y.; Reddy, V.; Natheus, A.; Le Corre, L.; Izawa, M.R.M. et al., 2016. Surface albedo and spectral variability of Ceres. *ApJ* 817:L22 (7pp)

Li, S.; Lucey, P.G.; Miliken, R.E.; Hayne, P.O.; Fisher, E. et al., 2018. Direct evidence of surface exposed water ice in the lunar polar regions. *PNAS* 115(36), 8907 – 8912

López-Oquendo, A.J.; Rivera-Valentín, E.G.; Dalle Ore, C.M.; Kirchoff, M.R.; Nichols-Fleming, F. *et al.,* (2019) Constraints on Crater Formation Ages on Dione from Cassini VIMS and ISS. *50th Lunar and Planetary Science Conference*. Abstract #2435.

Marconi, M.L. 2007. A kinetic model of Ganymede. *Icarus* 190, 155 – 174

Margot, J.L.; Campbell, D.B.; Jurgens, R.F.; Slade, M.A. 1999. Topography of the lunar poles from radar interferometry: A survey of cold trap locations. *Science* 284, 1658 – 1660

McCord, T.B.; Carlson, R.W.; Smythe, W.D.; Hansen, G.B.; Clark, R.N.; Hibbitts, C.A.; Fanale, F.P.; Granahan, J.C.; Segura, M.; Matson, D.L.; Johnson, T.V.; Martin, P.D. (1997) Organics and other molecules in the surfaces of Callisto and Ganymede. *Science* **278**, 271 – 275

McCord, T.B.; Hansen, G.B.; Clark, R.N.; Martin, P.D.; Hibbitts, C.A.; Fanale, F.P.; Granahan, J.C.; et al., (1998) Non-water-ice constituents in the surface material of the icy Galilean satellites from the Galileo near-infrared mapping spectrometer investigation. *JGR* **103**, 8603 - 8626

McDoniel, W.J.; Goldstein, D.B.; Varghese, P.L.; Trafton, L.M. (2017) The interaction of Io's plumes and sublimation atmosphere. *Icarus* 294, 81 - 97

Mitrofanov, I.; Litvak, M.; Sanin, A.; Malakhov, A.; Golovin, D.; Boynton, W.; Droege, G.; Chin, G.; Evans, L. et al., (2012) Testing polar spots of water-rich permafrost on the Moon: LEND observations onboard LRO. *JGR* 117, E00H27

Moore, J.M.; Asphaug, E.; Morrision, D.; Spencer, J.R.; Chapman, C.R. et al., (1999) Mass movement and landform degradation on the icy Galilean satellites: results of the Galileo nominal mission. *Icarus* 140, 294 – 312

Moore, J.M.; Chapman, C.R.; Bierhaus, E.B.; Greeley, R.; Chuang, F.C.; Klemaszewski, J.; Clark, R.N.; Dalton, J.B.; Hibbitts, C..; Schenk, P.M.; Spencer, J.R.; Wagner, R. (2004) Callisto. In *Jupiter: the Planet, Satellites, and Magnetosphere.* Eds: Bagenal, F.; Dowling, T.E.; McKinnon, W.B. Cambridge University Press

Nakajima, A.; Ida, S.; Kimura, J.; Brasser, R. (2019) Orbital evolution of Saturn's mid-sized moons and the tidal heating of Enceladus. *Icarus* 317, 570 - 582

Needham, D.H.; Kring, D.A., 2017. Lunar volcanism produced a transient atmosphere around the ancient Moon. *EPSL* 478, 175 - 178

Neish, C.D.; Bussey, B.J.; Spudis, P.; Marshall, W.; Thomson, B.J. et al., 2011. The nature of lunar volatiles as revealed by Mini-RF observations of the LCROSS impact site. *JGR*, 116, E01005(8 pp)

Nelson, R.M.; Buratti, B.J.; Wallis, B.D.; Smythe, W.D.; Horn, L.J. et al., (1990) Spectral geometric albedo and bolometric bond albedo of Neptune's satellite Triton from Voyager observations. *GRL* 17(10), 1761 – 1764

Ness, N.F.; Acuña, M.H.; Behannon, K.W.; Burlaga, L.F.; Connerney, J.E.P. et al., (1986) Magnetic fields at Uranus. *Science* **233**, 85 - 89

Neumann, G.A.; Cavanaugh, J.F.; Sun, X.; Mazarico, E.M.; Smith, D.E. et al., 2012. Bright and dark polar deposits on Mercury: evidence for surface volatiles. *Science* 339(6117), 296-300

Nozette, S.; Lichtenberg, C.L.; Spudis, P.; Bonner, R.; Ort, W. et al., 1996. The Clementine bistatic radar experiment. *Science* 274, 1495 – 1498

Nozette, S.; Spudis, P.D.; Robinson, M.S.; Bussey, D.B.J.; Lichtenberg, C.; Bonner, R. 2001. Integration of lunar polar remote-sensing data sets: Evidence for ice at the lunar south pole. *JGR* 106, 23253 – 23266

O'Neill, C.; Nimmo, F. (2010) The role of episodic overturn in generating the surface geology and heat flow on Enceladus. *Nature Geoscience* 3, 88 - 91

Ong, L.; Asphaug, E.I.; Korycansky, D.; Coker, R.F. (2010) Volatile retention from cometary impacts on the Moon. *Icarus* 207, 578 - 589

Owen, T.C.; Cruikshank, D.P.; Dalle Ore, C.M.; Geballe, T.R.; Roush, T.L.; de Bergh, C. (1999) Detection of water ice on Saturn's satellite Phoebe. *Icarus* 140, 379 - 382

Paige, D.A.; Siegler, M.A.; Harmon, J.K.; Neumann, G.A.; Mazarico, E.M. et al., 2013. Thermal stability of volatiles in the north polar region of Mercury. *Science* 339, 300 - 303

Palomba, E.; Longobardo, A.; De Sanctis, M.C.; Stein, N.T.; Ehlmann, B. et al., (2019) Compositional differences among Bright Spots on the Ceres surface. *Icarus* 320, 202 - 212

Pieters, C.M.; Goswami, J.M.; Clark, R.N.; Annadurai, M.; Boardman, J. et al., 2009. Character and spatial distribution of OH/$H_2O$ on the surface of the Moon seen by $M^3$ on Chandrayaan-1. *Science* 326, 568 - 572

Pitman, K.M.; Buratti, B.J.; Mosher, J.A. 2010. Disk-integrated bolometric Bond albedos and rotational light curves of Saturnian satellites from *Cassini* Visual and Infrared Mapping Spectrometer. *Icarus* 206, 537 – 560

Plainaki, C.; Milillo, A.; Massetti, S.; Mura, A.; Jia, X.; Orsini, S. et al. (2015) The $H_2O$ and $O_2$ exospheres of Ganymede: The result of a complex interaction between the jovian magnetospheric ions and the icy moon. *Icarus* 245, 306 - 319

Platz, T.; Nathues, A.; Schorghoger, N.; Preusker, F.; Mazarico, E. et al., 2016. Surface water-ice deposits in the northern shadowed regions of Ceres. *Nature Astronomy* 1(1). 6pp

Porco, C.C.; West, R.A.; McEwen, A.; Del Genio, A.D.; Ingersoll, A.P. et al., (2003) Cassini Imaging of Jupiter's Atmosphere, Satellites, and Rings. *Science* 299, 1541 - 1547

Porco, C.C.; Baker, E.; Barbara, J.; Beurle, K.; Brahic, A. et al., (2005) Cassini imaging science: Initial results on Phoebe and Iapetus. *Science* 307, 1237 – 1242

Porco, C.C.; Helfenstein, P.; Thomas, P.C.; Ingersoll, A.P.; Wisdom, J.; West, R. et al., (2006) Cassini observed the active south pole of Enceladus. *Science* **311**, 1393 - 1401

Prem, P. (2017) DSMC simulations of volatile transport in a transient lunar atmosphere and ice deposition in cold traps after a comet impact. (doctoral dissertation). University of Texas at Austin, Austin, Texas

Prem, P.; Artemieva, N.A.; Goldstein, D.B.; Varghese, P.L.; Trafton, L.M. 2015. Transport of water in a transient impact-generated lunar atmosphere. *Icarus* 255, 148 – 158

Prem, P.; Goldstein, D.B.; Varghese, P.L.; Trafton, L.M. 2018. The influence of surface roughness on volatile transport on the Moon. *Icarus* 299, 31 – 45

Prem, P.; Hurley, D.; McFarland, E.L.; Chabot, N.L.; Ernst, C.M.; Goldstein, D.B. (2019) Modeling the Transport, Loss and Deposition of Water on Mercury. *AGU Fall Meeting 2019* Abstract P11B-04. Oral Presentation.

Prettyman, T.H.; Yamashita, N.; Toplis, M.J.; McSween, H.Y.; Schorghofer, N. et al., 2017. Extensive water ice within Ceres' aqueously altered regolith: Evidence from nuclear spectroscopy. *Science* 355, 55 - 59

Raponi, A.; De Sanctis, M.C.; Carrozzo, F.G.; Ciarniello, M.; Castillo-Rogez, J.C. et al., (2019) Mineralogy of Occator crater on Ceres and insights into its evolution from the properties of carbonates, phyllosilicates, and chlorides. *Icarus* 320, 83 – 96

Roth, L.; Ivchenko, N.; Gladstone, G.R.; Saur, J.; Grodent, D. et al., (2021) A sublimated water atmosphere on Ganymede detected from Hubble Space Telescope observations. *Nature Astronomy* https://doi.org/10.1038/s41550-021-01426-9

Rubin, M.; Fougere, N.; Altwegg, K.; Combi, M.R.; Le Roy, L.; Tenishev, V.M.; Thomas, N. 2014. Mass transport around comets and its impact on the seasonal differences in water production rates. *ApJ* 788:168 (8 pp.)

Safrit, T.K.; Steckloff, J.K.; Bosh, A.S.; Nesvorny, D.; Walsh, K.; Brasser, R.; Minton, D.A. (2020) The Formation of Bilobate Comet Shapes through Sublimation Torques. *PSJ* 2:14 (10 pp).

Sandford, S.A.; Allamandola, L.J. 1988. The condensation and vaporization behavior of $H_2O$:CO ices and implications for interstellar grains and cometary activity. *Icarus* 76, 201 – 224

Sandford, S.A.; Allamandola, L.J. (1990) The physical and infrared spectral properties of $CO_2$ in astrophysical ice analogues. *ApJ* 355, 357 – 372

Schenk, P.; Hamilton, D.P.; Johnson, R.E.; McKinnon, W.B.; Paranicas, C. et al., (2011) Plasma, plumes, and rings: Saturn system dynamics as recorded in global color patterns on its midsize satellites. *Icarus* 211, 740 – 757

Schenk, P.; Sizemore, H.; Schmidt, B.; Castillo-Rogez, J.; De Sanctis, M. et al., (2019) The central pit and dome at Cerealia Facula bright deposits and floor deposits in Occator crater, Ceres: Morphology, comparisons and formation. *Icarus* 320, 159 - 187

Schorghofer, N. 2014. Migration calculations for water in the exosphere of the Moon: Dusk-dawn asymmetry, heterogeneous trapping, and D/H fractionation. *GRL* 41, 4888 - 4893

Schorghofer, N.; Mazarico, E.; Platz, T.; Preusker, F.; Schröder, S.E. et al., 2016. The permanently shadowed regions of dwarf planet Ceres. *GRL* 43, 6783 – 6789

Schorghofer, N.; Byrne, S.; Landis, M.E.; Mazarico, E.; Prettyman, T.H. et al., (2017) The putative Cerean exosphere. *ApJ* 850:85 (7pp.)

Shi, M; Baragiola, R.A.; Grosjean, D.E.; Johnson, R.E.; Jurac, S.; Schou, J. (1995) Sputtering of water ice surfaces and the production of extended neutral atmospheres. *JGR* 100, 26387 - 26395

Shoemaker, E.M.; Wolfe, R.A., (1982) Cratering timescales for the Galilean satellites, in: Morrison, D. (Ed.), *Satellites of Jupiter*, Univ. of Arizona Press, Tucson, pp. 277–339

Sierks, H.; Barbieri, C.; Lamy, P.L.; Rodrigo, R.; Koschny, D. et al., 2015. On the nucleus structure and activity of comet 67P/Churyumov-Gerasimenko. *Science* 347 (5 pp.)

Simonelli, D.P.; Rossier, L.; Thomas, P.C.; Veverka. J.; Burns, J.A. (2000) Leading/trailing albedo asymmetries of Thebe, Amalthea, and Metis. *Icarus* 147, 353 – 365

Slade, M.A.; Butler, B.J.; Muhleman, D.O. 1992. Mercury radar imaging: Evidence for polar ice. *Science* 258, 635 – 640

Smith, B.A.; Soderblom, L.A.; Beebe, R.; Bliss, D.; Boyce, J.M. et al., (1986) Voyager 2 in the Uranian System: Imaging science results. *Science* 233, 43 - 64

Sone, Y. (2007) Flows induced by temperature fields. In *Molecular Gas Dynamics.* Birkhäuser – Boston, 233 - 280

Sori, M.M.; Bapst, J.; Bramson, A.M.; Byrne, Landis, M.E. (2017) A Wunda-ful world? Carbon dioxide ice deposits on Umbriel and other Uranian moons. *Icarus* **290**, 1 – 13

Spencer, J.R. (1987a) The surfaces of Europa, Ganymede, and Callisto: An investigation using Voyager IRIS thermal infrared spectra. Ph.D. thesis, University of Arizona.

Spencer, J.R. (1987b) Thermal Segregation of Water Ice on the Galilean Satellites. *Icarus* **69**, 297 - 313

Spencer, J.R.; Denk, T. (2010) Formation of Iapetus' extreme albedo dichotomy by exogenically triggered thermal ice migration. *Science* 327, 432 – 435

Spencer, J.R.; Nimmo, F. (2013) Enceladus: An Active Ice World in the Saturn System. *Annual Rev. of Earth and Planetary Sciences* 41, 693 - 717

Squyres, S.W. (1980) Surface temperatures and retention of $H_2O$ frost on Ganymede and Callisto. *Icarus* 44, 502 - 510

Stacy, N.J.S.; Campbell, D.B.; Ford, P.G. 1997. Arecibo radar mapping of the lunar poles: A search for ice deposits. *Science* 276, 1527 – 1530

Steckloff, J.K.; Goldstein, D.; Prem, P.; Varghese, P.; Trafton, L. 2018. The migration of impact-delivered water to the cold traps of airless bodies. *American Astronomical Society – Division of Planetary Science 50$^{th}$ meeting.* Abstract #103.01

Steckloff, J.K.; Jacobson, S.A. (2016) The formation of striae within cometary dust tails by a sublimation-driven YORP-like effect. *Icarus* 264, 160 – 171

Steckloff, J.K.; Graves, K.; Hirabayashi, M.; Melosh, H.J.; Richardson, J.E. (2016) Rotationally induced surface slope-instabilities and the activation of $CO_2$ activity on comet 103P/Hartley 2. *Icarus* 272, 60-69

Steckloff, J.K.; Johnson, B.C.; Bowling, T.; Melosh, H.J.; Minton, D. et al., (2015) Dynamic sublimation pressure and the catastrophic breakup of Comet ISON. *Icarus* 258, 430 – 437

Steckloff, J.K.; Samarasinha, N.H. (2018) The sublimative torques of Jupiter Family Comets and mass wasting events on their nuclei. *Icarus* 312, 172 - 180

Steckloff; J.K.; Lisse, C.M.; Safrit, T.K.; Bosh, A.S.; Lyra, W.; Sarid, G. (2021) The sublimative evolution of (486958) Arrokoth. *Icarus* 356, 113998

Stern, S.A. (1999) The lunar atmosphere: History, status, current problems, and context. *Rev. or Geophys.* 37, 453 - 491

Stewart, B.D.; Pierazzo, E.; Goldstein, D.B.; Varghese, P.L.; Trafton, L.M. 2011. Simulations of a comet impact on the Moon and associated ice deposition in polar cold traps. *Icarus* 215, 1-16

Stubbs, T.J.; Wang, Y. 2012. Illumination conditions at the asteroid 4 Vesta: Implications for the presence of water ice. *Icarus* 217, 272 - 276

Sunshine, J.M.; Farnham, T.L.; Feaga, L.M.; Groussin, O.; Merlin, F. et al., 2009. Temporal and spatial variability of lunar hydration as observed by the Deep Impact spacecraft. *Science* 326, 565 – 568

Sunshine, J.M.; Thomas, N.; El-Maarry, R.; Farnham, T. 2016. Evidence for geologic processes on comets. *JGR:Planets* 121, 2194 - 2210

Teolis, B.D.; Jones, G.H.; Miles, P.F.; Tokar, R.L.; Magee, B.A. et al., (2010) Cassini finds an oxygen-carbon dioxide atmosphere at Saturn's icy moon Rhea. *Science* 330, 1813 - 1815

Teolis, B.D.; Waite, J.H. (2016) Dione and Rhea seasonal exospheres revealed by Cassini CAPS and INMS. *Icarus* 272, 277 - 289

Teolis, B.D.; Wyrick, D.Y.; Bouquet, A.; Magee, B.A.; Waite, J.H. (2017) Plume and surface feature structure and compositional effects on Europa's global exosphere: Preliminary Europa mission predictions

Thomas, E.C.; Vu, T.H.; Hodyss, R.; Johnson, P.V.; Choukroun, M. (2019) Kinetic effect on the freezing of ammonium-sodium-carbonate-chloride brines and implications for the origin of Ceres' bright spots. *Icarus* 320, 150 - 158

Thomas, G.E. 1974. Mercury: Does its atmosphere contain water? *Science* 183, 1197 -1198

Thomas, P.C.; Armstrong, J.W.; Asmar, S.W.; Burns, J.A.; Denk, T. et al., 2007. Hyperion's sponge-like appearance. *Nature* 448, 50 - 53

Thomson, B.J.; Bussey, D.B.J.; Neish. C.D.; Cahill, J.T.S.; Heggy, E. et al., 2012. An upper limit for ice in Shackleton crater as revealed by LRO Mini-RF orbital radar. *GRL* 39, L14201 (4pp)

Tokar, R.L.; Johnson, R.E.; Thomsen, M.F.; Sittler, E.C.; Coates, A.J. et al., (2012) Detection of exospheric $O_2^+$ at Saturn's moon Dione. *GRL* 39, L03105 (7pp.)

Trafton, L. (1984) Large seasonal variations in Triton's atmosphere. *Icarus* 58, 312 – 324

Trafton, L.M.; Whipple, A.L.; Stern, S.A. (1987) The Extended Atmosphere of the Pluto-Charon System. *Bull. Amer. Astron. Soc.* 19. 1071 - 1072

Tu, L.; Ip, W.-H.; Wang, Y.-C. (2014) A sublimation-driven exospheric model of Ceres. *Planetary and Space Science* 104B, 157 - 162

Tucker, O.J.; Johnson, R.E.; Young, L.A. (2014) Gas transfer in the Pluto-Charon system: A Charon atmosphere. *Icarus* 246, 291 - 297

Tu, L.; Ip, W.-H.; Wang, Y.-C. (2014) A Sublimation-driven Exospheric Model of Ceres. *Planetary and Space Science* **104**, 157 - 162

Vasavada, A.R.; Paige, D.A.; Wood, S.E. (1999) Near-Surface Temperatures on Mercury and the Moon and the Stability of Polar Ice Deposits. *Icarus* 141, 179 – 193

Vasavada, A.R.; Bandfield, J.L.; Greenhagen, B.T.; Hayne, P.O.; Siegler, M.A.; Williams, J.-P.; Paige, D.A. (2012) Lunar equatorial surface temperatures and regolith properties from the Diviner Lunar Radiometer Experiment. *JGR* 117, E00H18. doi:10.1029/2011JE003987

Veverka, J.; Thomas, P.; Davies, M.; Morrison, D. (1981) Amalthea: Voyager imaging results. *JGR* 86, 8675 - 8692

Veverka, J.; Helfenstein, P.; Hapke, B.; Goguen, J.D. 1988. Photometry and polarimetry of Mercury. In *Mercury.* University of Arizona Press – Tucson, AZ, 37 – 58

Vincent, J.-B.; Hviid, S.F.; Mottola, S. et al. 2017. Constraints on cometary surface evolution derived from a statistical analysis of 67P's topography. MNRAS 469, S329 – S338

Vorburger, A.; Wurz, P.; Lammer, h.; Barabash, S.; Mousis, O. (2015) Monte-carlo simulation of Callisto's exosphere. *Icarus* 262, 14 – 29

Vorburger, A.; Fatemi, S.; Galli, A.; Liuzzo, l.; Poppe, A.R.; Wurz, P. (2022) 3D Monte-Carlo simulation of Ganymede's water exosphere. *Icarus* 375, 114810

Vu, T.H.; Hodyss, R.; Johnson, P.V.; Choukroun, M. (2017) Preferential formation of sodium salts from frozen sodium-ammonium-choloride-carbonate brines – Implications for Ceres' bright spots. *P&SS*

Walker, A.C.; Moore, C.H.; Goldstein, D.B.; Varghese, P.L.; Trafton, L.M. 2012. A parametric study of Io's thermophysical surface properties and subsequent numerical atmosphere simulations based on the best fit parameters. *Icarus* 220, 225 - 253

Watson, K.; Murray, B.C.; Brown, H. 1961. The behavior of volatiles on the lunar surface. *JGR* 66(9), 3033 – 3045

Whipple, A.L.; Trafton, L.M.; Stern, S.A. (1989) A Gravitational Restricted Three-Body Model of Pluto's Upper Atmosphere. *Bull. Amer. Astron. Soc.* 21, 982

Williams, J.-P.; Paige, D.A.; Greenhagen, B.T.; Sefton-Nash, E. (2017) The global surface temperatures of the Moon as measured by the Diviner Lunar Radiometer Experiment. *Icarus* 283, 300 - 325

Wilmoth, R.G. (1973) Measurement of surface stay times for physical adsorption of gases. (doctoral dissertation). University of Virginia, Charlottesville, Viriginia

Wisdom, J.; Peale, S.J.; Mignard, F. (1984) The chaotic rotation of Hyperion. *Icarus* **58**, 137 - 152

Zahnle, K.; Dones, L.; Levison, H. (1998) Cratering rates on the Galilean satellites. *Icarus* 136, 202 - 222

Zahnle, K.; Schenk, P.; Sobieszczyk, S.; Dones, L.; Levison, H.F. (2001) Differential cratering of synchronously rotating satellites by ecliptic comets. *Icarus* 153, 111 – 129

Zahnle, K.; Schenk, P.; Levison, H.; Dones, L. (2003) Cratering rates in the outer Solar System. *Icarus* 163, 263 - 289

Zurbuchen, T.H.; Raines, J.M.; Gloeckler, G.; Krimigis, S.M.; Slavin, J.A. et al., 2008. MESSENGER observations of the composition of Mercury’s ionized exosphere and plasma environment. *Science* 321, 90 – 92